\pdfoutput=1

\documentclass[11pt,twoside,a4paper,cmspaper,final,collab]{cms-tdr}
\def\svnVersion{485615P}\def\cmsCernNoTag{CERN-EP-2018-253}\def\cmsCernDate{\today}\def\cmsMessage{Published in the Journal of High Energy Physics as \href{http://dx.doi.org/10.1007/JHEP12(2018)117}{\doi{10.1007/JHEP12(2018)117.}}}

\begin{document}\cmsNoteHeader{SMP-17-003}

\hyphenation{had-ron-i-za-tion}
\hyphenation{cal-or-i-me-ter}
\hyphenation{de-vices}
\newcommand{\PYTHIAeight}{{\PYTHIA}8\xspace}
\newcommand{\PYTHIAsix}{{\PYTHIA}6\xspace}
\newcommand{\pythiaC}   {{\PYTHIA}8\xspace CUETP8M1\xspace}
\newcommand{\pythiaM}   {{\PYTHIA}8\xspace Monash\xspace}
\newcommand{\bitem}     {\begin{itemize}}
\newcommand{\eitem}     {\end{itemize}}
\newcommand{\btab}      {\begin{tabular}}
\newcommand{\etab}      {\end{tabular}}
\newcommand{\beqn}      {\begin{equation}}
\newcommand{\eeqn}      {\end{equation}}
\newcommand{\h}         {\ensuremath{H_{\mathrm{T},2}}\xspace}
\newcommand{\nt} 	{\ensuremath{{\hat n}_{\mathrm{T}}}\xspace}
\newcommand{\ptiv} 	{\ensuremath{\vec{p}_{\mathrm{T},i}}\xspace}
\newcommand{\pti}       {\ensuremath{p_{\mathrm{T},i}}\xspace}
\newcommand{\taup}	{\ensuremath{\tau_{\perp}}\xspace}
\newcommand{\bt}        {\ensuremath{B_{\mathrm{Tot}}}\xspace}
\newcommand{\rhoTot}   	{\ensuremath{\rho_{\mathrm{Tot}}}\xspace}
\newcommand{\rhoPerp}  	{\ensuremath{\rho_{\mathrm{Tot}}^{\mathrm{T}}}\xspace}
\newcommand{\rhoh}      {\ensuremath{\rho_{\mathrm{H}}}\xspace}
\newcommand{\thrust}    {\ensuremath{T_{\perp}}\xspace}
\newcommand{\HTzer}     {\ensuremath{73 <\h < 93}\xspace}
\newcommand{\HTone}     {\ensuremath{93 <\h < 165}\xspace}
\newcommand{\HTtwo}     {\ensuremath{165 <\h < 225}\xspace}
\newcommand{\HTthr}     {\ensuremath{225 <\h < 298}\xspace}
\newcommand{\HTfou}     {\ensuremath{298 <\h < 365}\xspace}
\newcommand{\HTfiv}     {\ensuremath{365 <\h < 452}\xspace}
\newcommand{\HTsix}     {\ensuremath{452 <\h < 557}\xspace}
\newcommand{\HTsev}     {\ensuremath{\h > 557}\xspace}

\cmsNoteHeader{SMP-17-003}
\title{Event shape variables measured using multijet final states in proton-proton collisions
at \texorpdfstring{$\sqrt{s} = 13\TeV$}{sqrt(s) = 13 TeV}}

\date{\today}
\abstract{
The study of global event shape variables can provide sensitive tests of predictions for multijet
production in proton-proton collisions. This paper presents a study of several event shape variables
calculated using jet four momenta in proton-proton collisions at a centre-of-mass energy of 13\TeV and
uses data recorded with the CMS detector at the LHC corresponding to an integrated luminosity of
$2.2\fbinv$. After correcting for detector effects, the resulting distributions are compared with
several theoretical predictions. The agreement generally improves as the energy,
represented by the average transverse momentum of the two leading jets, increases.
}
\hypersetup{%
pdfauthor={CMS Collaboration},%
pdftitle={Hadronic event shape variables in proton-proton collisions at sqrt(s)=13 TeV},%
pdfsubject={CMS},%
pdfkeywords={CMS, physics, shape variables, multijet events}}
\maketitle
\section{Introduction}
The production of quarks and gluons in hadron collisions and the process of hadron formation are subject to
in-depth theoretical and experimental studies. The experiments at the CERN LHC have studied production of
hadronic jets by measuring differential cross-sections, ratios of numbers of jets, angular distributions,
etc., to deepen the understanding of quantum chromodynamics (QCD). While the production of quarks and gluons
with large transverse momentum (\pt) is well described by calculations based on perturbative QCD, the
hadronization process probes energy scales where perturbative calculations are not applicable.
Instead, phenomenological models inspired by QCD are used to predict the experimental results.

Event shape variables (ESVs) are sensitive to the flow of energy in hadronic final states. These variables
are safe from collinear and infrared divergences and have reduced experimental uncertainties~\cite{Banfi:2004nk}.
Some distributions of ESVs are sensitive to the details of the hadronization
process~\cite{Banfi:2010xy, Dasgupta:2003iq, Rubin:2010xp}, so they can be used to
tune parameters of Monte Carlo (MC) event generators,
determine the strong coupling \alpS~\cite{Jones:2003yv, Dissertori:2007xa, Dissertori:2009ik}, and
to search for new physics phenomena~\cite{Chatterjee:2012qt, Datta:2011vg, Konar:2005bd}.

Various ESVs have been studied in electron-positron collisions at the CERN LEP collider
to determine \alpS~\cite{Heister:2003aj, Abreu:1996na, Acciarri:1997dn, Achard:2004sv, Acton:1993zh}.
ESVs have also been studied in electron-proton collisions at the DESY HERA collider~\cite{Aktas:2005tz}
and in proton-antiproton collisions at the FNAL Tevatron collider~\cite{Aaltonen:2011et},
where they were compared with next-to-leading-order (NLO) calculations and with various tunes
of the PYTHIA6 event generator~\cite{Sjostrand:2006za}. At the CERN LHC collider studies by the
ALICE, ATLAS, and CMS Collaborations have exploited proton-proton collisions at centre-of-mass energies
of $\sqrt{s} = 0.9$, 2.76, and 7\TeV to evaluate
ESVs~\cite{Khachatryan:2011dx, Chatrchyan:2013tna, Chatrchyan:2013ala, Khachatryan:2014ika, Aad:2012np,
Aad:2012fza, Aad:2016ria, Abelev:2012sk}.

This paper reports a measurement of ESVs by the CMS Collaboration using hadronic jets in \Pp\Pp{} collisions at
$\sqrt{s} = 13\TeV$ corresponding to an integrated luminosity ($\Lumi_{\text{int}}$) of 2.2$\fbinv$.
The following variables are studied: the complement of transverse thrust, total jet broadening, total
jet mass, and total transverse jet mass. The theoretical uncertainties in the predictions of these ESVs
can be reduced by careful choice of the quantity used to classify the energy scale of the events.
Following Ref.~\cite{Rubin:2010xp}, we use $\h = (p_{\mathrm{T,jet1}}+p_{\mathrm{T,jet2}})/2$,
where $p_{\mathrm{T,jet1}}$ and $p_{\mathrm{T,jet2}}$ refer to the transverse momenta
of the highest and second highest \pt jets.The measured distributions are corrected for
detector effects and compared with the predictions of QCD models implemented in
the \PYTHIAeight~\cite{Sjostrand:2014zea}, \MGvATNLO{}+\PYTHIAeight~\cite{Alwall:2014hca}, 
and \HERWIGpp~\cite{Bellm:2015jjp} event generators.

The paper is organized as follows. The ESVs are discussed in Section~\ref{sec:esvs}.
After briefly describing the elements of the CMS detector in Section~\ref{sec:cmsdet}, the jet
reconstruction relevant to this analysis is described in Section~\ref{sec:jetr}.
The data sample and event selection criteria are described in Section~\ref{sec:data}.
Sections~\ref{sec:unfold} and~\ref{sec:syst} present the unfolding technique and the
systematic uncertainties, respectively. Section~\ref{sec:results} contains comparisons
between CMS data and theoretical predictions, and the results are summarized in
Section~\ref{sec:summary}.
\section{Event shape variables}\label{sec:esvs}
The four ESVs studied in this analysis are defined using the four-momenta of hadronic jets.

\textit{The complement of transverse thrust:}
The complement of thrust is defined as:
\beqn
	\taup \equiv 1 - \thrust,	\label{eqn:Complement}
\eeqn
where the thrust in the transverse plane is:
\begin{equation}
\thrust \equiv \max_{\nt} \frac{\sum_i \abs{\ptiv \cdot \nt}} {\sum_i \pti}. \label{eqn:Thrust}
\end{equation}
Here, \ptiv is the component of momentum of the $i^{th}$ jet perpendicular to the beam direction and
thrust direction \nt is the unit vector that maximizes the projection and defines the transverse thrust axis.
The $\taup$ is zero for a perfectly balanced two-jet event and is $1-2/\pi$ for an isotropic multijet event.

\textit{Total jet broadening:}
For each event, the transverse thrust axis is used to divide the event into upper (U) and lower (L) regions.
The jets in U satisfy $\ptiv.\nt > 0$ and those in L have $\ptiv.\nt < 0$. For these two regions,
the $p_{T}$-weighted pseudorapidities and azimuthal angles are
\begin{equation}
\eta_X \equiv \frac{\sum_{i\in{X}} \pti\eta_i}
                        {\sum_{i\in{X}} \pti},
\phi_X \equiv \frac{\sum_{i\in{X}} \pti\phi_i}
                        {\sum_{i\in{X}} \pti},
\end{equation}
where X refers to the U or L regions. The jet broadening variable in each region is defined as
\begin{equation}
B_{X} \equiv {\frac{1}{2\,P_{T}}
	\sum_{i\in{X}} \pti \sqrt{(\eta_i -\eta_X)^2 + (\phi_i - \phi_X)^2}},
\end{equation}
where $P_{\mathrm{T}}$ is the scalar \pt sum of all the jets in the event. The total jet
broadening is then defined as
\begin{equation}
	\bt \equiv  B_{\mathrm{U}} + B_{\mathrm{L}}.
\end{equation}

\textit{Total jet mass:} The normalized squared invariant mass of the jets in the U and L regions
of the event is defined by
\begin{equation}
	\rho_X \equiv \frac{M^2_X}{P^2},  	
\end{equation}
where $M_\mathrm{X}$ is the invariant mass of the jets in the region X, and P is the scalar sum of
the momenta of all central jets. The total jet mass is defined as the sum of the
masses in the U and L regions,
\begin{equation}
	\rhoTot \equiv \rho_{\mathrm{U}} + \rho_{\mathrm{L}}.	\label{eqn:JetMass}
\end{equation}

\textit{Total transverse jet mass:} The quantity corresponding to \rhoTot in the transverse plane,
the total transverse jet mass (\rhoPerp), is similarly calculated using $\ptiv$ of jets.

These four ESVs probe different aspects of QCD~\cite{Banfi:2010xy} and are designed to
have higher values for multijet, spherical events and lower values for back-to-back dijet
events. While \taup is sensitive to the hard-scattering process, the jet masses and jet
broadening depend more on the nonperturbative aspects of QCD, responsible for
hadronisation process.

\section{The CMS detector}\label{sec:cmsdet}
The central feature of the CMS apparatus is a superconducting solenoid of 6\unit{m} internal diameter,
providing a magnetic field of 3.8\unit{T}. The solenoid volume holds a silicon pixel and strip
tracker, a lead tungstate crystal electromagnetic calorimeter (ECAL), and a brass and scintillator
hadron calorimeter (HCAL), each composed of a barrel and two endcap sections. Steel and quartz-fibre
Cherenkov hadron forward calorimeters extend the pseudorapidity ($\eta$) coverage provided by
the barrel and endcap detectors to the region $3.0 < \abs{\eta} < 5.2$. Muons are measured in
gas-ionisation detectors embedded in the steel flux-return yoke outside the solenoid. In the
region $\abs{\eta} < 1.74$,  the HCAL cells have widths of 0.087 in $\eta$ and 0.087 radians in
azimuthal angle ($\phi$). For $\abs{\eta} < 1.48$, the HCAL cells map onto $5\!\times\!5$ ECAL crystals
arrays in the $\eta$-$\phi$ plane to form calorimeter towers projecting radially outwards
from close to the nominal interaction point. At larger values of $\eta$, the size in $\eta$ of
the towers increases and the matching ECAL arrays contain fewer crystals. CMS uses a two stage
online trigger to select events for offline analysis. In the first stage, a hardware-based
level-1 (L1) trigger uses information from calorimeter and muon subsystems and selects
event at a rate of about 100\unit{kHz}.
In the second stage, a
software-based high-level trigger (HLT), running on computer farms, uses full event information
and reduces the event rate to about 1\unit{KHz} before data storage. A more detailed description
of the CMS detector can be found in Ref.~\cite{Chatrchyan:2008zzk}.

\section{Jet reconstruction}\label{sec:jetr}
The particle-flow (PF) event algorithm~\cite{Sirunyan:2017ulk} reconstructs photons, electrons,
charged and neutral hadrons, and muons with an optimised combination of information
from the various elements of the CMS detector. The energy of a photon is directly obtained from the
ECAL measurement. The energy of an electron is determined from a combination of the electron momentum
at the primary interaction vertex as determined by the tracker, the energy of the corresponding
ECAL cluster, and the energy sum of all bremsstrahlung photons spatially compatible with originating
from the electron track. The momentum of a muon is obtained from the curvature of the corresponding
track. The energy of a charged hadron is determined from a combination of its momentum measured in
the tracker and the matching ECAL and HCAL energy deposits, corrected for zero-suppression effects
and for the response function of the calorimeters to hadronic showers. Finally, the energy of a
neutral hadron is obtained from the corresponding energy deposits in ECAL and HCAL.

Jets are reconstructed from photons, electrons, charged and neutral hadrons, and muons using
the anti-\kt clustering algorithm~\cite{Cacciari:2008gp, Cacciari:2011ma} with a distance
parameter R = 0.4. Measurement of jet energy is  affected by contamination from
additional \Pp\Pp\xspace interactions in the same bunch crossing (pileup), as well as by the
nonuniform and nonlinear response of the CMS calorimeters. The technique of charged-hadron
subtraction~\cite{Sirunyan:2017ulk} is used to reduce the contribution of particles that
originate from pileup interactions to the jet energy measurement.
The jet four-momentum is corrected for
the difference observed in simulation between jets built from reconstructed particles and
generator-level particles. The jet mass and direction are kept constant for the corrections,
which are functions of the $\eta$ and \pt of the jet, as well as the energy density and jet area
quantities defined in Ref.~\cite {Chatrchyan:2011ds}. The latter
are used to correct the energy offset introduced by the pileup interactions.
The energy of the jets is further corrected using dijet, \cPZ+jet, and $\gamma$+jet events,
where the \pt-balance of the event is exploited.
The jet energy resolution typically amounts to $15\%$ at 10\GeV,
$8\%$ at 100\GeV, and $4\%$ at 1\TeV.

\section{Data set and event selection}\label{sec:data}
\subsection{Collision data}

This analysis uses \Pp\Pp\xspace collision data collected in 2015 at $\sqrt{s} = 13\TeV$,
corresponding to $\Lumi_{\text{int}} = 2.2\fbinv$. Events are selected at L1 and
HLT that have jet \pt or \h thresholds, respectively, as shown in Table \ref{tab:trigger}.
The turn-on point for each trigger, offline \h at which the trigger is 99\% efficient,
is used to define the \h ranges for events.

Collision and simulated events are required to have at least three jets with $\pt>30\GeV$
within the coverage of the tracker $\abs{\eta} < 2.4$. For each event, three jets are used for
the calculation of the ESVs. The jets with the highest and the second-highest \pt are
selected. From the remaining jets, the one with the highest recoil term is selected as
the third jet. The recoil term for jet $k$ is
$$\mathcal{R}_{\mathrm{\perp,k}} =
\frac{\abs{\vec{p}_{\mathrm{T,jet1}}+\vec{p}_{\mathrm{T,jet2}}+\vec{p}_{\mathrm{T,jetk}}}}
	{\abs{\vec{p}_{\mathrm{T,jet1}}}+\abs{\vec{p}_{\mathrm{T,jet2}}}+\abs{\vec{p}_{\mathrm{T,jetk}}}}.$$
The data sample is divided into eight \h ranges such that the uncertainty due to the trigger
inefficiency is negligible. The ranges (in \GeVns{}) are: 73--93, 93--165, 165--225, 225--298,
298--365, 365--452, 452--557 and $>$557, as shown in Table~\ref{tab:trigger}, with the number
of events in each range.

\begin{table}[htbp]
\centering
\topcaption{L1 trigger thresholds, HLT thresholds, $\h$ range and number of events used in the
	analysis.}
\label{tab:trigger}
\btab{lccc}
\hline
L1 threshold for    & HLT threshold for & $\h$ range & Number of   \\
$p_{\mathrm{T,jet}}$ (\GeVns{}) & $\h$ (\GeVns{}) & (\GeVns{})  & events               \\
\hline
ZeroBias   & 60    & 73--93   & 222\,184    \\
52  	   & 80    & 93--165  & 36\,452   \\
92  	   & 140   & 165--225 & 81\,932   \\
128 	   & 200   & 225--298 & 363\,294   \\
128 or 176 & 260   & 298--365 & 134\,320   \\
128 or 176 & 320   & 365--452 & 354\,140   \\
128 or 176 & 400   & 452--557 & 443\,361   \\
128 or 176 & 500   & $>$557  & 295\,578  \\
\hline
\etab
\end{table}
\subsection{Simulated events}
Events are simulated using \PYTHIA~v8.212, \MGvATNLO~V5~2.2.2+\PYTHIAeight, and
\HERWIGpp~v2.7.1. The NNPDF3.0~\cite{Ball:2014uwa} parton distribution function (PDF) set is used.
The \PYTHIAeight and \HERWIGpp event generators use leading order 2$\rightarrow$2 matrix
element (ME) calculations and parton shower (PS) for generation of multijet topologies. The
\PYTHIAeight event generator uses a \pt -ordered PS, and the underlying event description is
based on the multiple parton interaction (MPI) model. Events are generated with two
\PYTHIAeight tunes: CUETP8M1~\cite{Khachatryan:2015pea} and Monash~\cite{Skands:2014pea}.
Minimum bias data collected by the CMS experiment were used to derive the \pythiaC
tune, which is based on the Monash tune. The \MGvATNLO generator uses ME calculations
to generate hard-scattering events with two to four partons and \pythiaC for subsequent
fragmentation and hadronization. The MLM~\cite{Alwall:2007fs} matching procedure is used
to avoid double counting of jets between the ME calculation and the PS description. The
\HERWIGpp generator uses an angular-ordered PS. For simulated events, particle-level jets
are obtained by applying the anti-\kt clustering algorithm to all generated stable particles,
excluding neutrinos, with $R = 0.4$.

The simulation events are passed through a complete and detailed reconstruction
in the CMS detector using the same reconstruction as the collision events.

\section{Unfolding of distributions}\label{sec:unfold}
A reconstructed collision event differs from the true event because of
finite resolution of the detector, detector acceptances, and uncertainties
and efficiencies of measurement. Hence, the detector-level distributions
obtained from data are unfolded to estimate the underlying particle-level
distributions, which can be compared with predictions from theoretical models
as well as with results obtained by other experiments.

Simulated events passing through the complete detector simulation, event reconstruction, and selection
chain are used to construct the response matrix for an ESV, which relates its particle-level
distribution with that at detector level. The response matrix incorporates all the experimental
effects and is subsequently used as input for the unfolding of the observed distribution in data.
Some events that satisfy the selection criteria at the particle level might not at
the detector level, leading to an inefficiency. The reverse may also happen, leading to
misidentification. Further, an event may migrate from one $\h$ range to
another. The corresponding efficiency and misidentification
rates are also incorporated in the unfolding process, and they contribute to the related
uncertainty of the unfolding process.

To investigate possible bias due to the choice of an MC generator to
construct the response matrices, we generate event samples from three different generators:
\pythiaC, \MGvATNLO, and \HERWIGpp. Each detector level distribution is unfolded
using these three response matrices and the corresponding particle-level
distributions are compared. No evidence for significant bias is observed.

Two different methods, which are implemented in RooUnfold~\cite{Adye:2011gm}, are used
for unfolding the observed distributions: D'Agostini iteration with
early stopping~\cite{DAgostini:1994fjx}, and Singular Value
Decomposition (SVD)~\cite{Hocker:1995kb}.
The difference between the unfolded distributions produced
with these two methods is much smaller than 1\%. Our unfolding is done using the D'Agostini iteration and
\pythiaC is used for constructing the response matrix. The SVD method is used as a cross-check.
\section{Systematic uncertainties}\label{sec:syst}
\begin{figure}[hbt]
\centering
\includegraphics[width=1\textwidth]{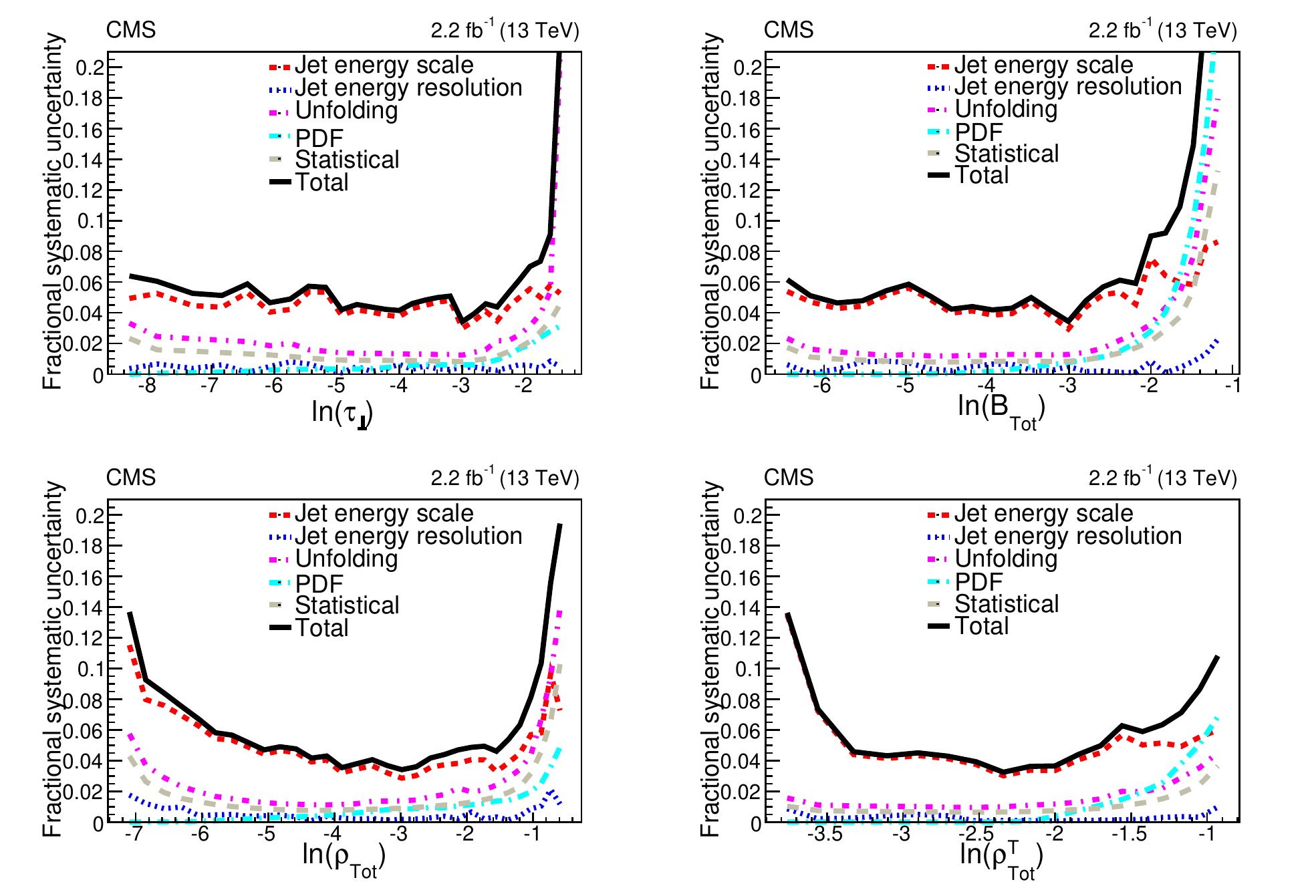}
  \caption{Total uncertainty (black line) for the four event shape  variables:
	the complement of transverse thrust (\taup) (upper left),
	total jet broadening (\bt) (upper right),
	total jet mass (\rhoTot) (lower left) and
	total transverse jet mass (\rhoPerp) (lower right) evaluated with jets for $\HTthr\GeV$.
	The contributions from different sources are also shown in each plot:
	JES (red dashed line), JER (blue dotted line), unfolding (pink dash-dotted line),
	PDF (light-blue dash-dotted line) and statistics (grey dashed line).}
  \label{fig:SystematicComponents}
\end{figure}

There are multiple sources of uncertainties in the unfolding process, and the contributions
from each individual source are added in quadrature to obtain the total uncertainty.
Figure~\ref{fig:SystematicComponents} shows the total uncertainty and the contributions
from various sources as a function of each ESV for the specific range $\HTthr\GeV$.

\bitem
  \item \textit{Jet energy scale (JES)}:
	CMS considers 26 different sources of uncertainties in the JES~\cite{Khachatryan:2016kdb}.
	To estimate the effect of each source, the four-momentum of each jet is scaled up and down by
	the corresponding uncertainty, the ESV is calculated, and the response matrix obtained
	with the nominal JES is used to unfold the distributions obtained with the nominal,
	scaled up, and scaled down JES values. For each bin of the unfolded distribution,
	the larger of the differences between the nominal,
	and the varied ones is taken as the systematic
	uncertainty. The systematic uncertainties due to different sources are then added in
	quadrature. For most bins in the distribution of an ESV, the uncertainty is 4--6\%.
	However, it reaches about 12\% for the highest and lowest bins of \rhoTot,
	lowest bins of \rhoPerp, and about 8\% for the highest bins of \bt.
	Typically JES is the largest source of systematic uncertainty in the ESVs.
  \item \textit{Jet energy resolution (JER)}: The JER is obtained from the ratio of \pt of the two jets
	in dijet events as a function of \pt and $\eta$~\cite{Khachatryan:2016kdb}.
	It has been observed that the JER is worse in data compared to simulation. Hence, extra
	smearing is applied to the simulated events, and different response matrices are constructed.
	The detector-level distribution of an ESV is unfolded with the different response matrices
	incorporating the uncertainty due to JER. The estimated uncertainties in the ESVs are
	of the order of 1\%.
  \item \textit{Unfolding}: The detector-level distribution of an ESV obtained from simulated events
	of \pythiaC is unfolded with two response matrices derived from \MGvATNLO and \HERWIGpp,
	and compared with the corresponding particle-level distribution in the same sample. Similar
	exercises are carried out for the \MGvATNLO sample using \pythiaC and \HERWIGpp response
	matrices, and for the \HERWIGpp sample using \pythiaC and \MGvATNLO response matrices.
	Out of these six differences for each bin, the largest is taken as the systematic
	uncertainty. In the closure tests of the individual response matrices, if, for a particular
	bin, the difference in the unfolded and generated values is larger than the uncertainty
	already assigned, the larger one is taken as the uncertainty due to the unfolding for that
	bin. The bias inherent in the D’Agostini method is estimated by using different generators.
	The difference in the unfolded
	results is included as an unfolding uncertainty.
	The uncertainty due to unfolding is of the order of 2\%, except for a few lowest, and
	highest bins where it dominates the total uncertainty.
  \item	\textit{Parton distribution function}: The uncertainty due to the PDFs in the particle-level
	distribution of an ESV is estimated using the 100 sets of NNPDF3.0 replicas. The standard
	deviation of the 100 values thus obtained for a bin is taken as the
	uncertainty due to PDFs for that bin. For most bins, the uncertainty due
	to the PDFs is less than 1\%, but increases for higher values of the
	variables. For \bt the uncertainty due to the PDFs increases very
	rapidly ($>$20\%) and dominates for the last few bins.
\eitem
The contribution of other sources of systematic uncertainty, \ie, pileup, and trigger efficiency
are negligible.
\section{Results}\label{sec:results}
\begin{figure}[hbtp]
\centering
\includegraphics[width=0.49\textwidth]{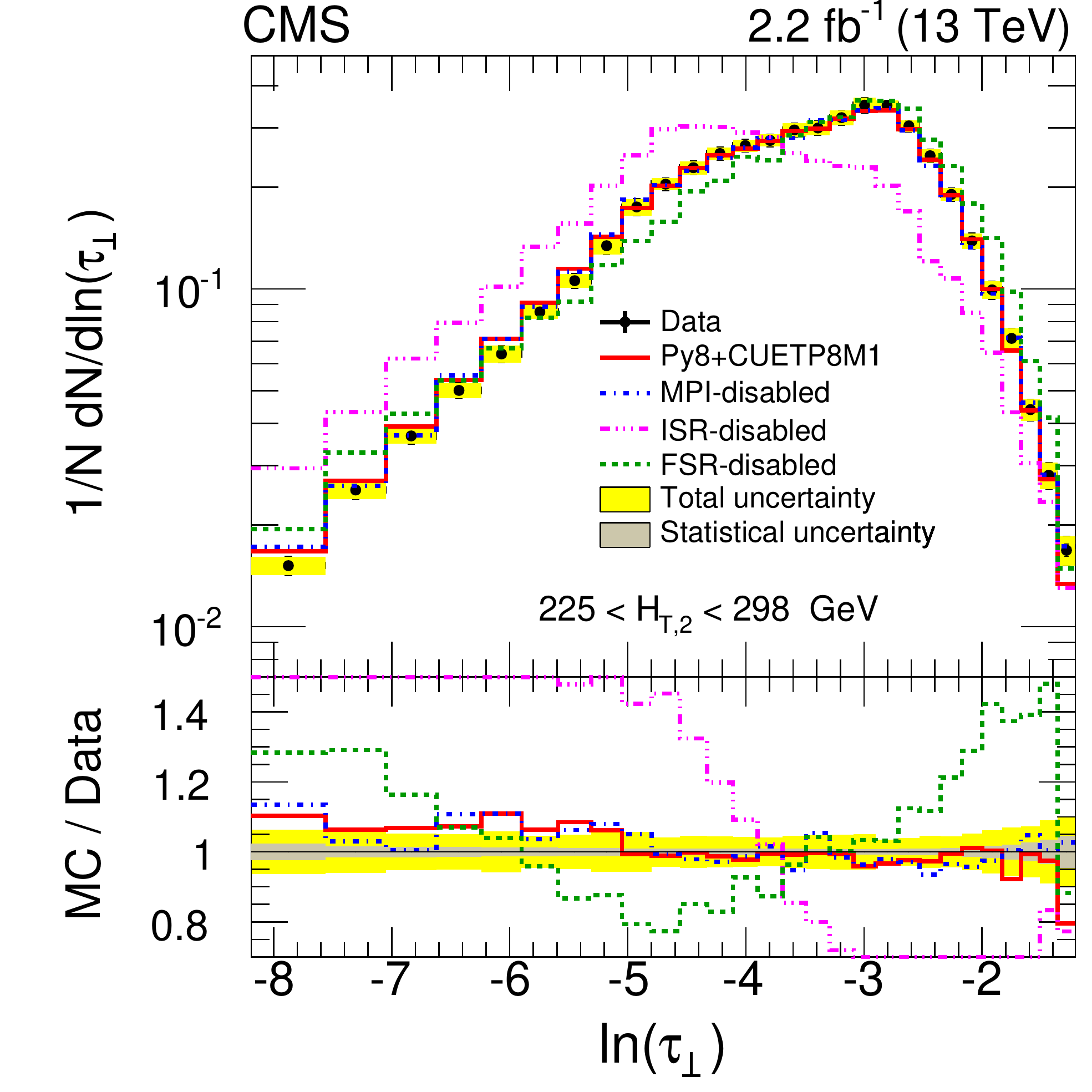}
\includegraphics[width=0.49\textwidth]{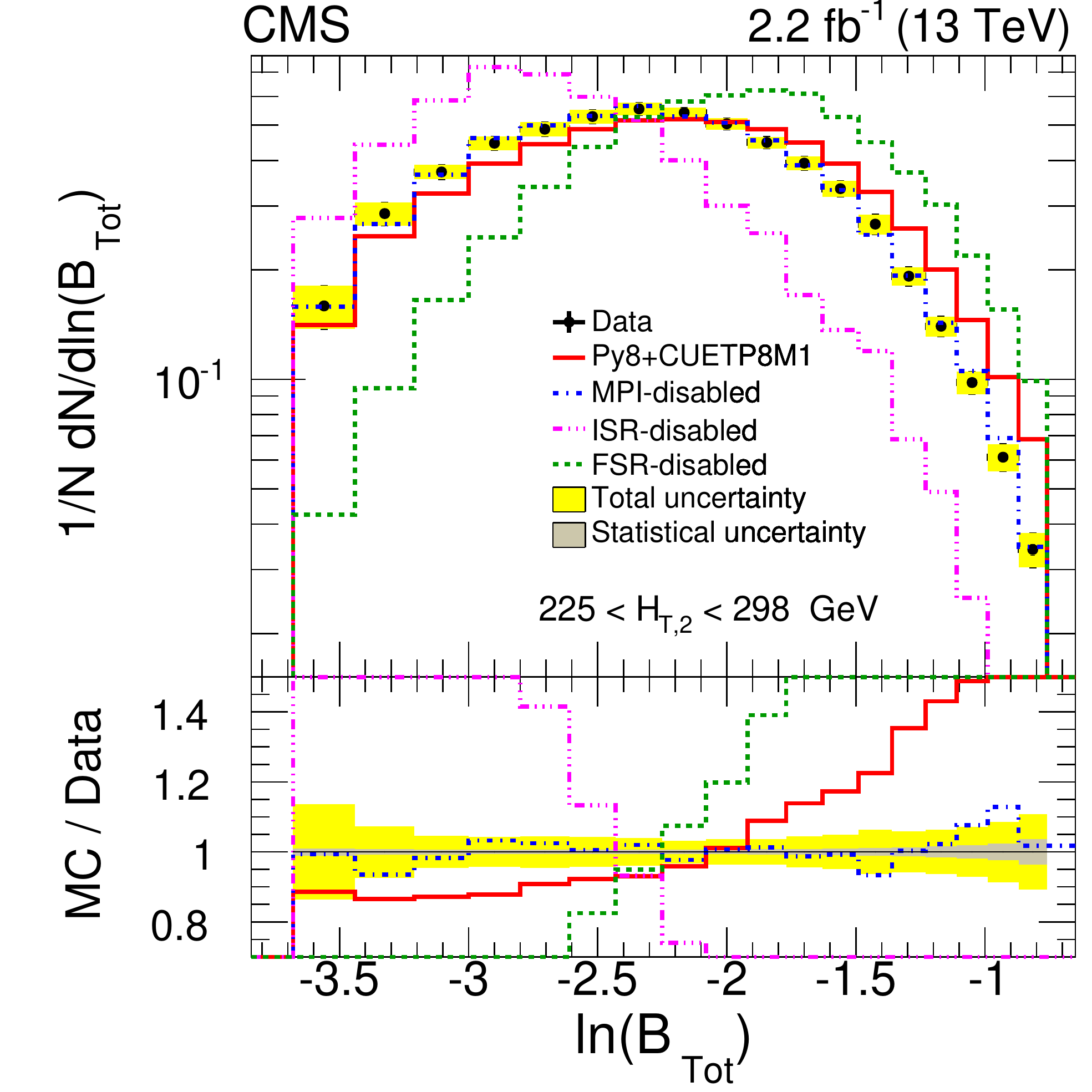}
\includegraphics[width=0.49\textwidth]{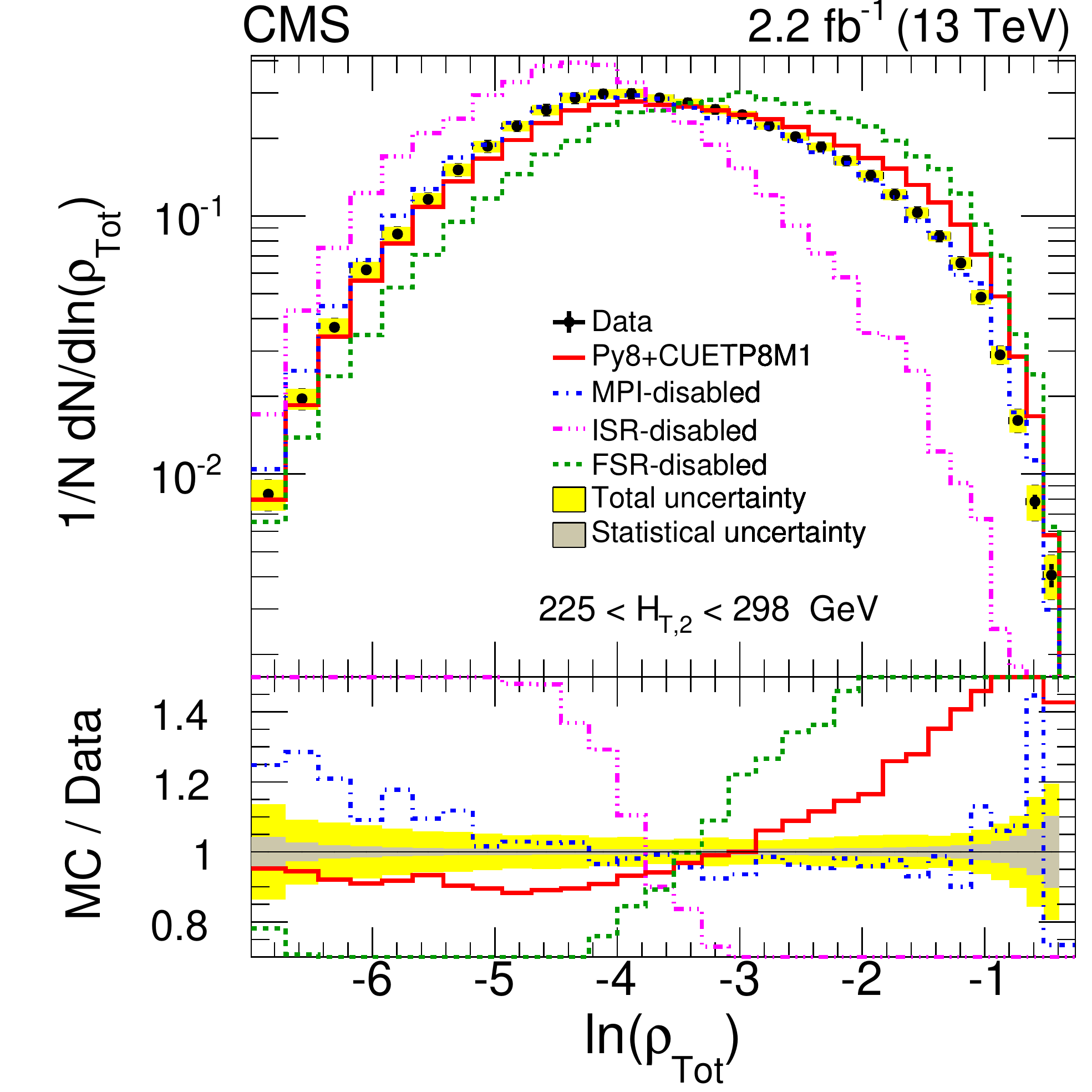}
\includegraphics[width=0.49\textwidth]{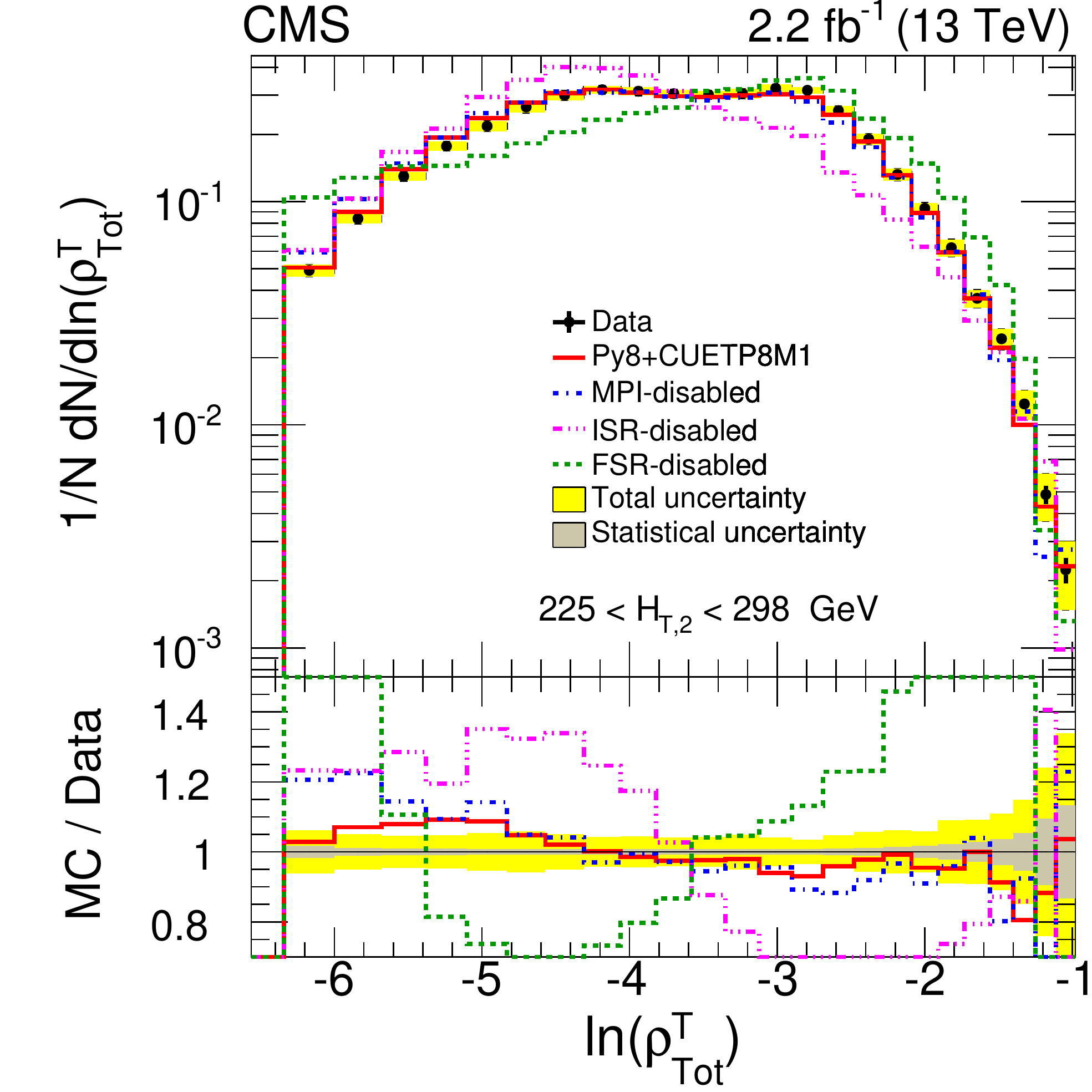}
  \caption{The effects of MPI, ISR, and FSR in \pythiaC on \taup (upper left), \bt (upper right),
	\rhoTot	(lower left) and \rhoPerp (lower right) for a typical range $\HTthr\GeV$.
	The ratio plots for simulation (MC) with respect to data
	are shown in the lower panel of each plot. The inner gray band represents the statistical
	uncertainty and the yellow band represents the total uncertainty (systematic + statistical)
	in each plot.}
  \label{fig:Switch}
\end{figure}

The modelling of initial-state radiation (ISR), final-state radiation (FSR) of gluons, and MPI in \pythiaC
is tested by studying each aspect individually, via the comparison of simulated ESV distributions with data,
as shown in Figure~\ref{fig:Switch}. This study shows that the effect of disabling ISR results in a very
large shift of the ESVs to lower values, i.e., reducing the spherical nature of the multijet events.
The effect of disabling the FSR is small compared to the ISR, and the effect of MPI is even smaller.

\begin{figure}[hbtp]
\centering
\includegraphics[width=0.49\textwidth]{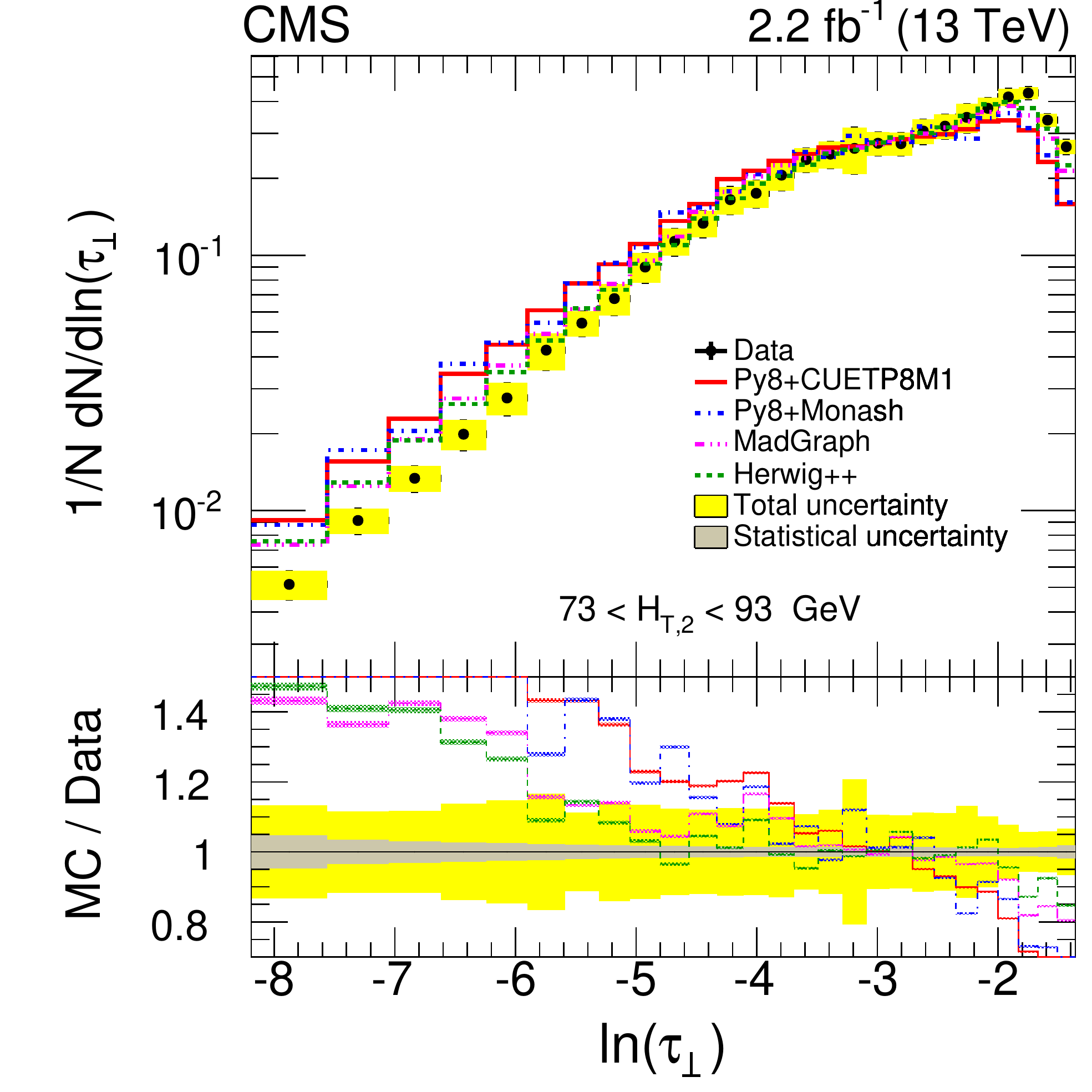}
  \includegraphics[width=0.49\textwidth]{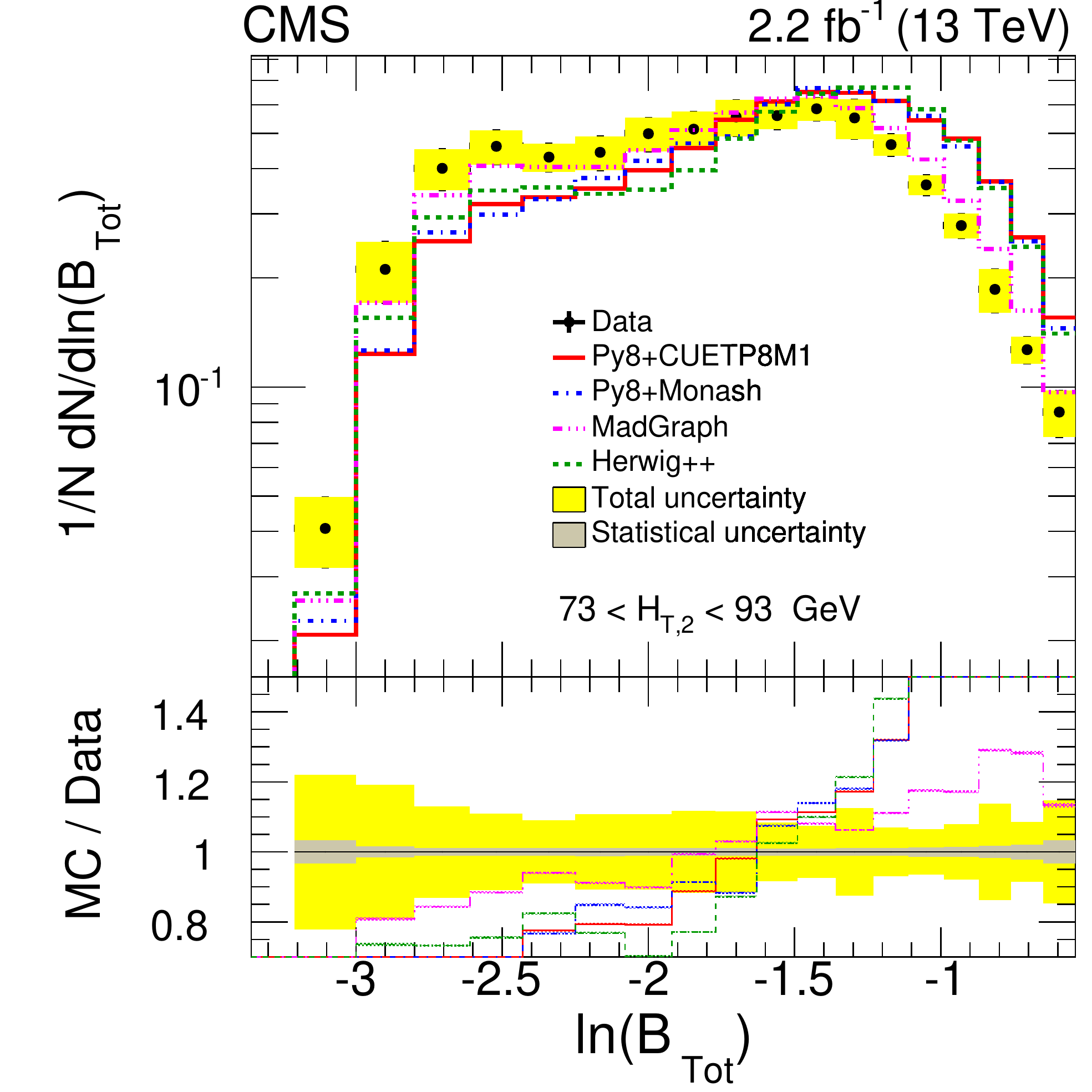}
  \includegraphics[width=0.49\textwidth]{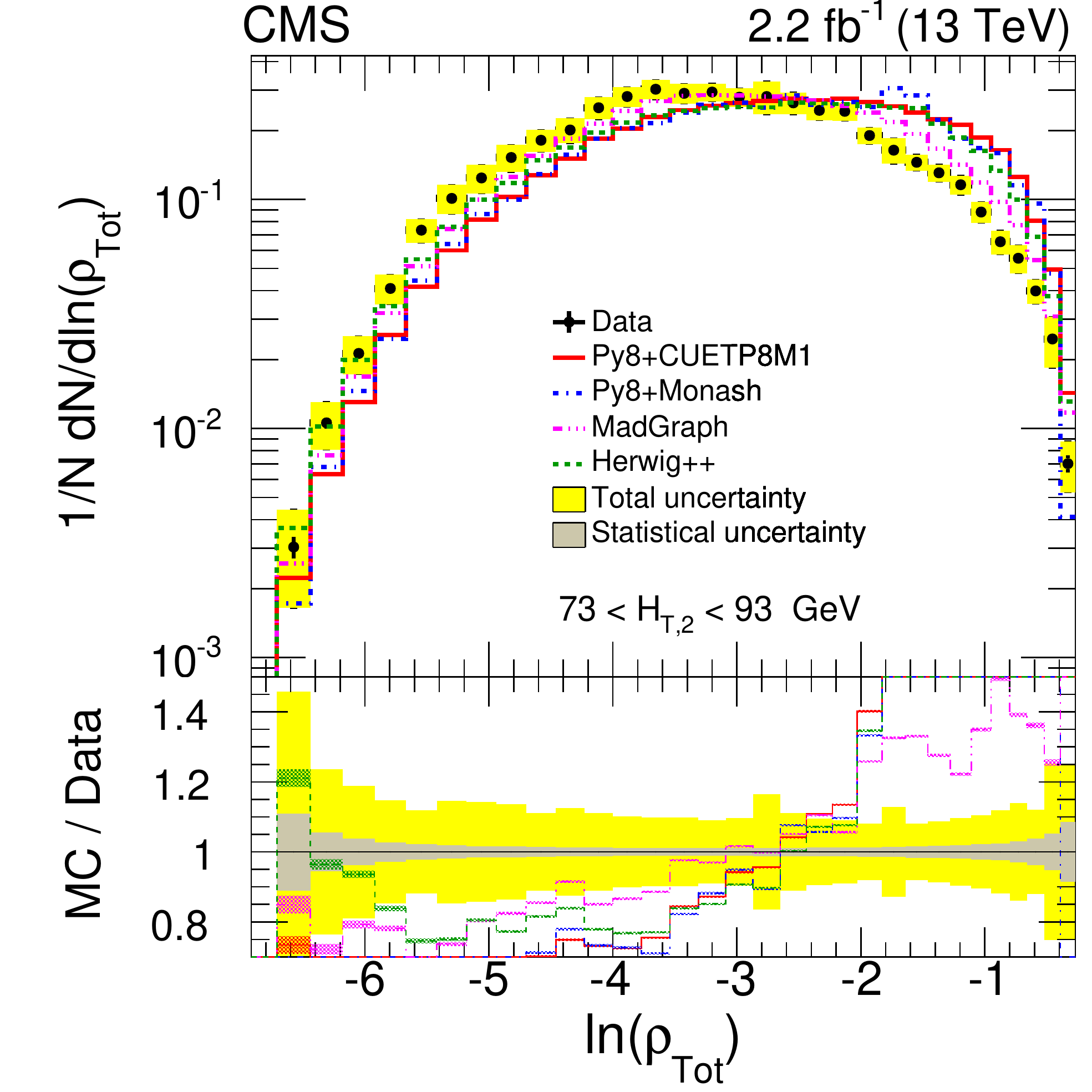}
  \includegraphics[width=0.49\textwidth]{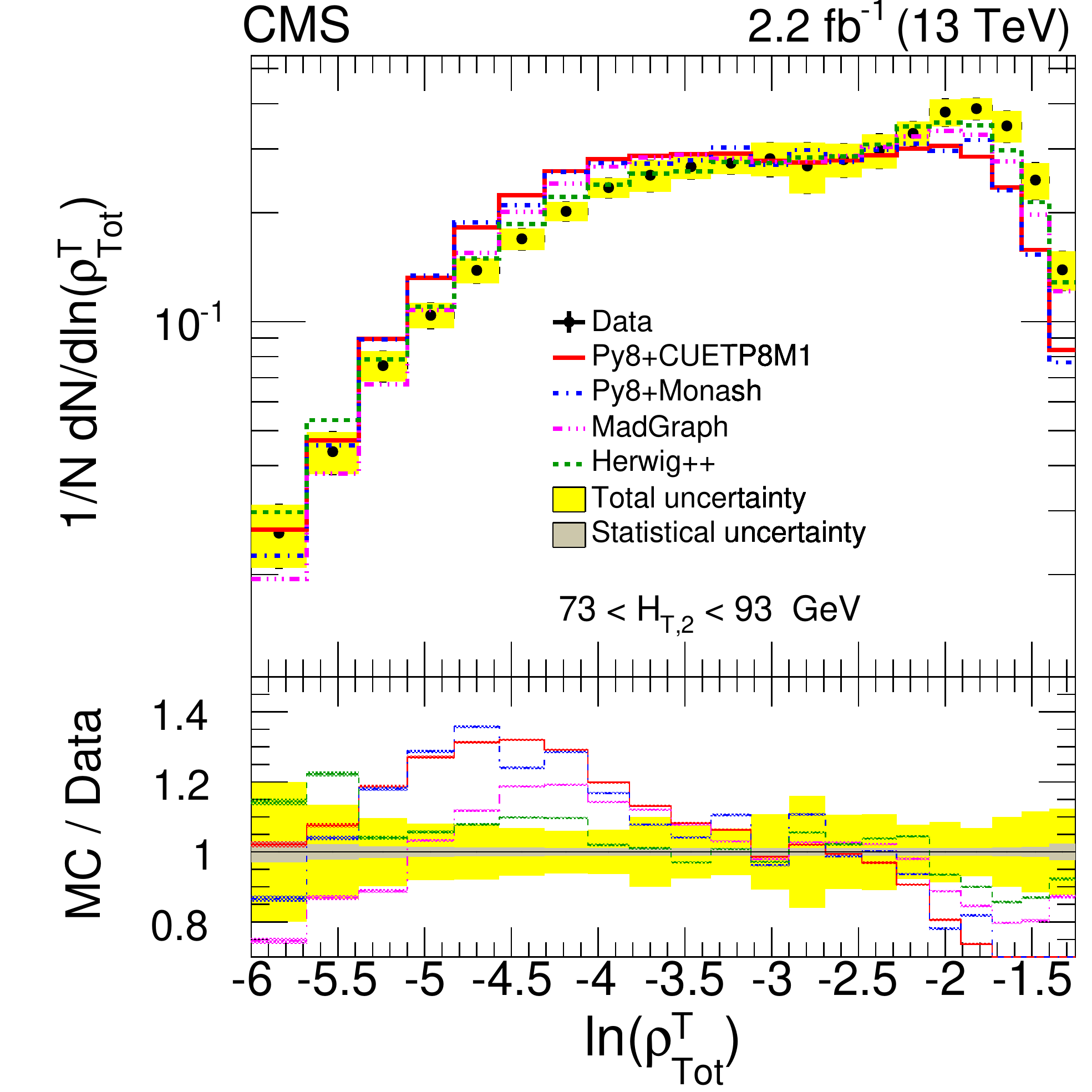}
  \caption{Normalized differential distributions of unfolded data compared with theoretical (MC)
	predictions of \pythiaC (red line), \pythiaM (blue dash-dotted line),
	\MGvATNLO (pink dash-dot-dotted line) and
        \HERWIGpp (brown dash-dot-dotted line)  as a function of ESV:
        complement of transverse thrust (\taup) (upper left),
        total jet broadening (\bt) (upper right),
        total jet mass (\rhoTot) (lower left) and
        total transverse jet mass (\rhoPerp) (lower right) for $\HTzer\GeV$.
        In each ratio plot, the inner gray band represents statistical uncertainty and the yellow band represents
	the total uncertainty (systematic and statistical components added in quadrature) on data and the MC
	predictions include only statistical uncertainty.}
\label{fig:Comparison-DataMC-zer}
\end{figure}

\begin{figure}[hbtp]
\centering
  \includegraphics[width=0.49\textwidth]{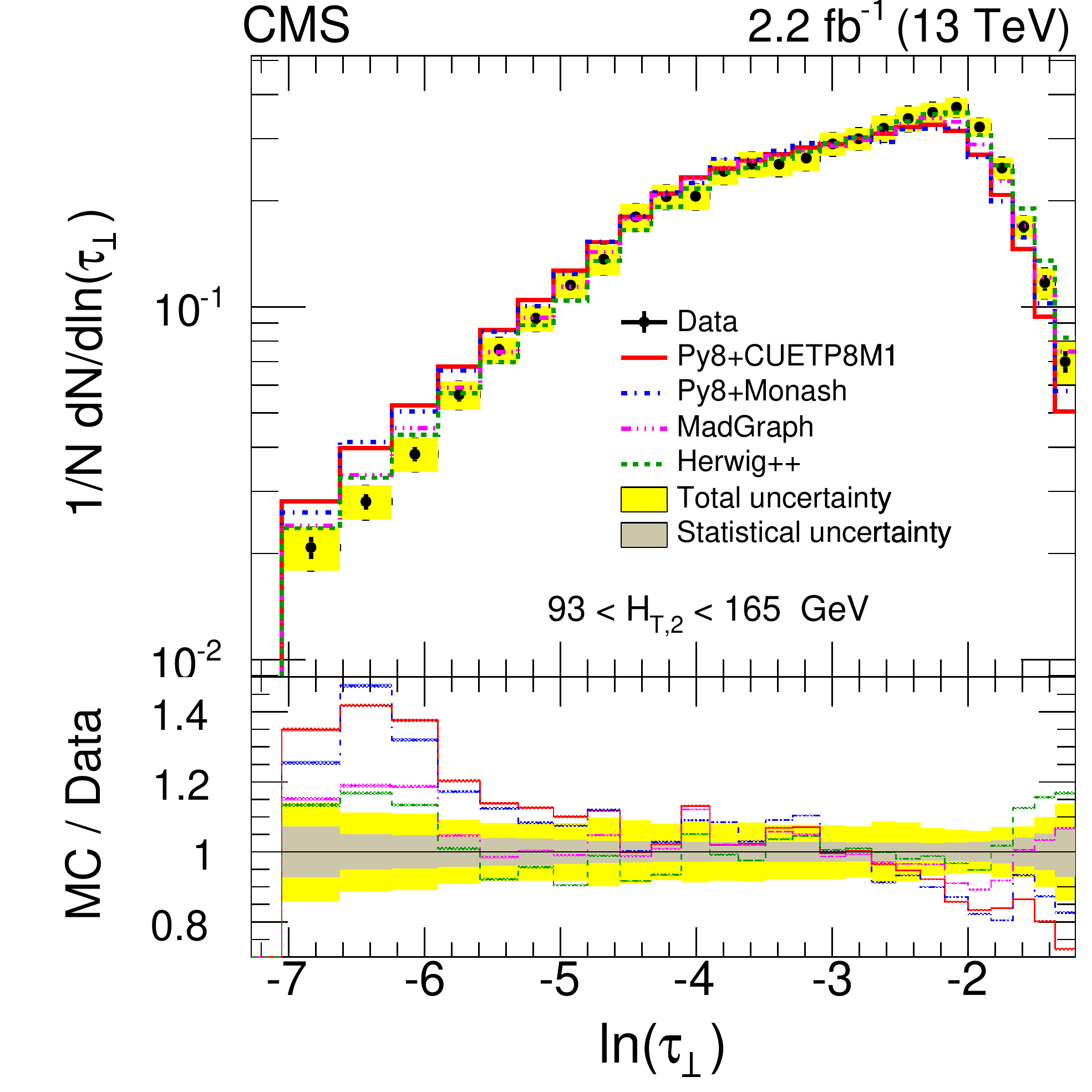}
  \includegraphics[width=0.49\textwidth]{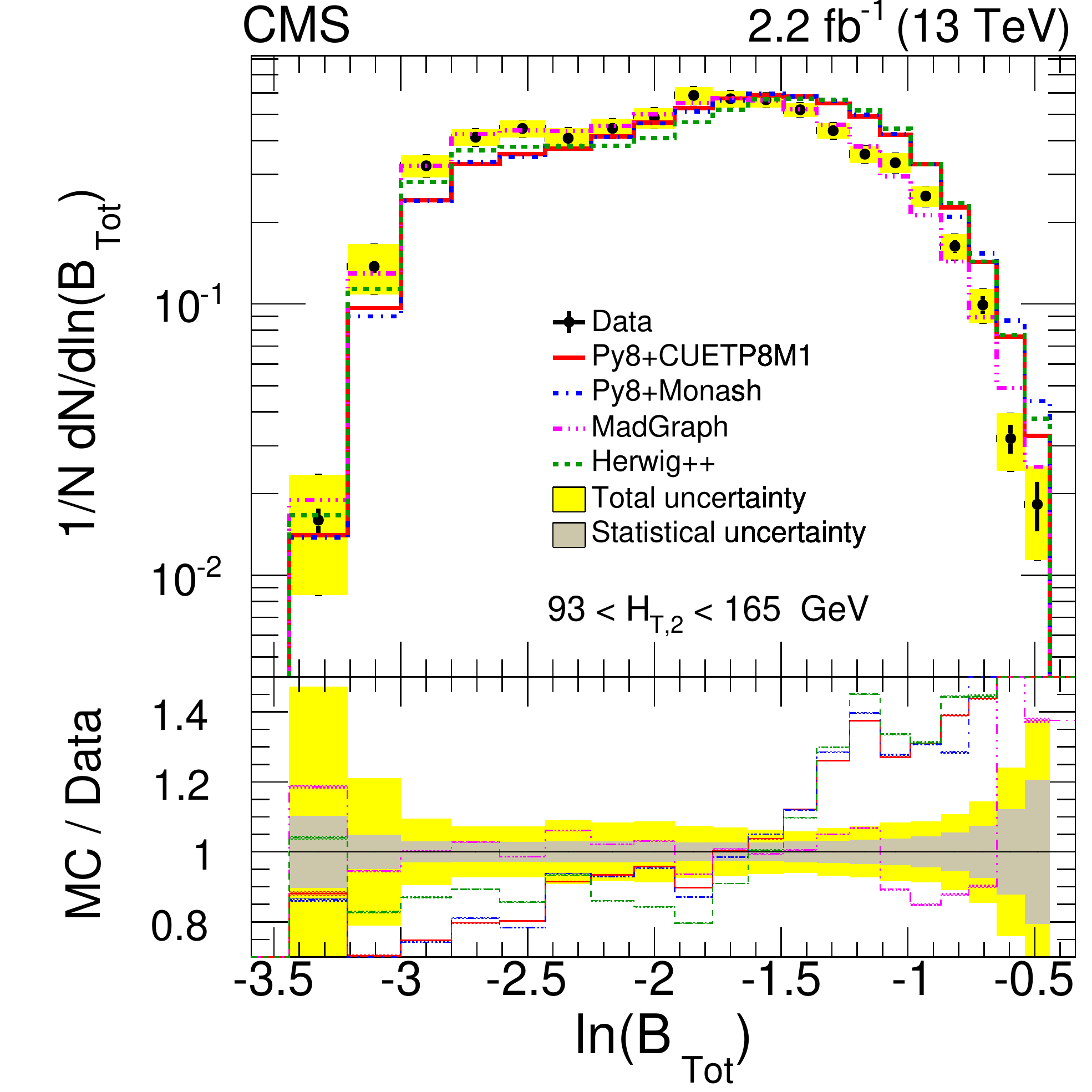}
  \includegraphics[width=0.49\textwidth]{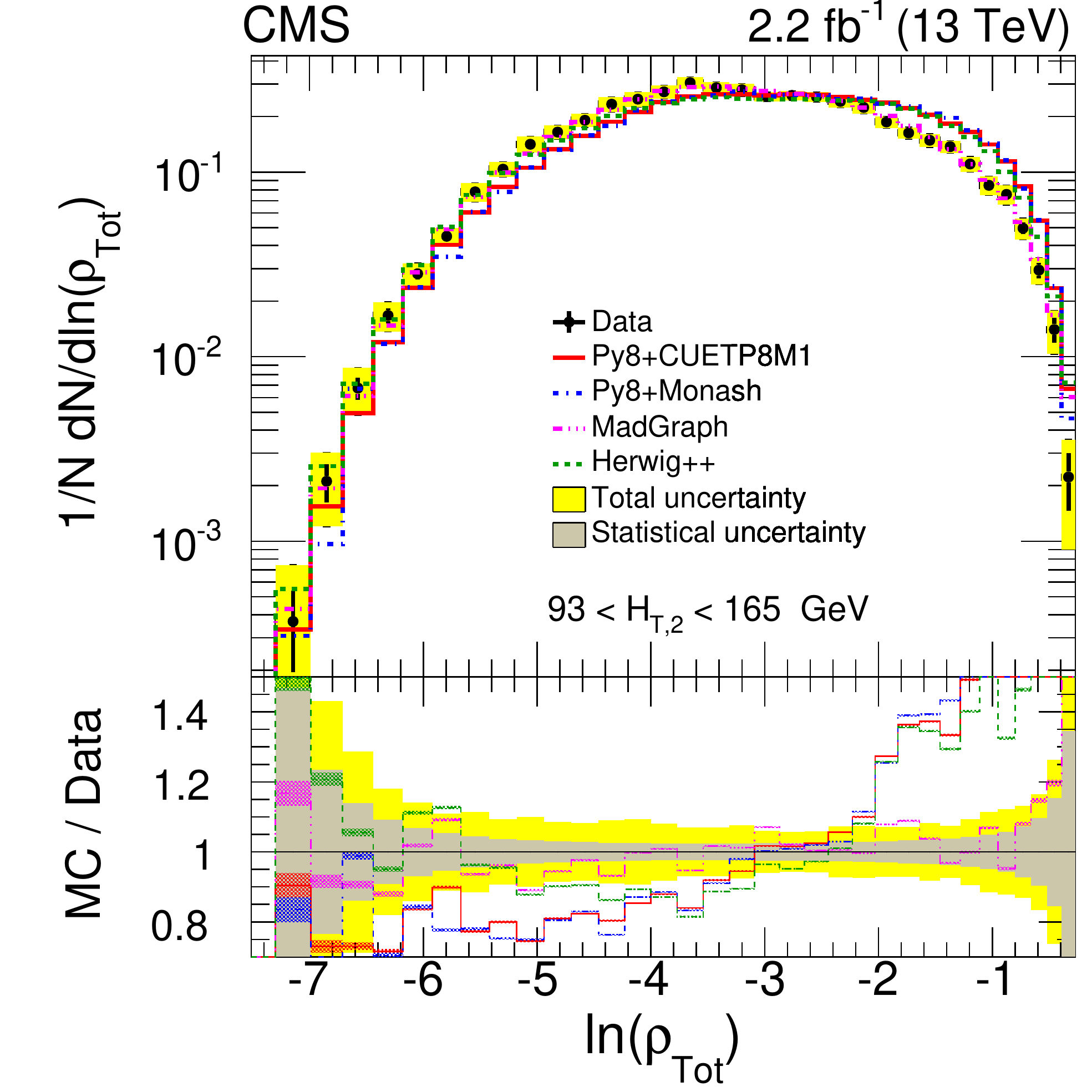}
  \includegraphics[width=0.49\textwidth]{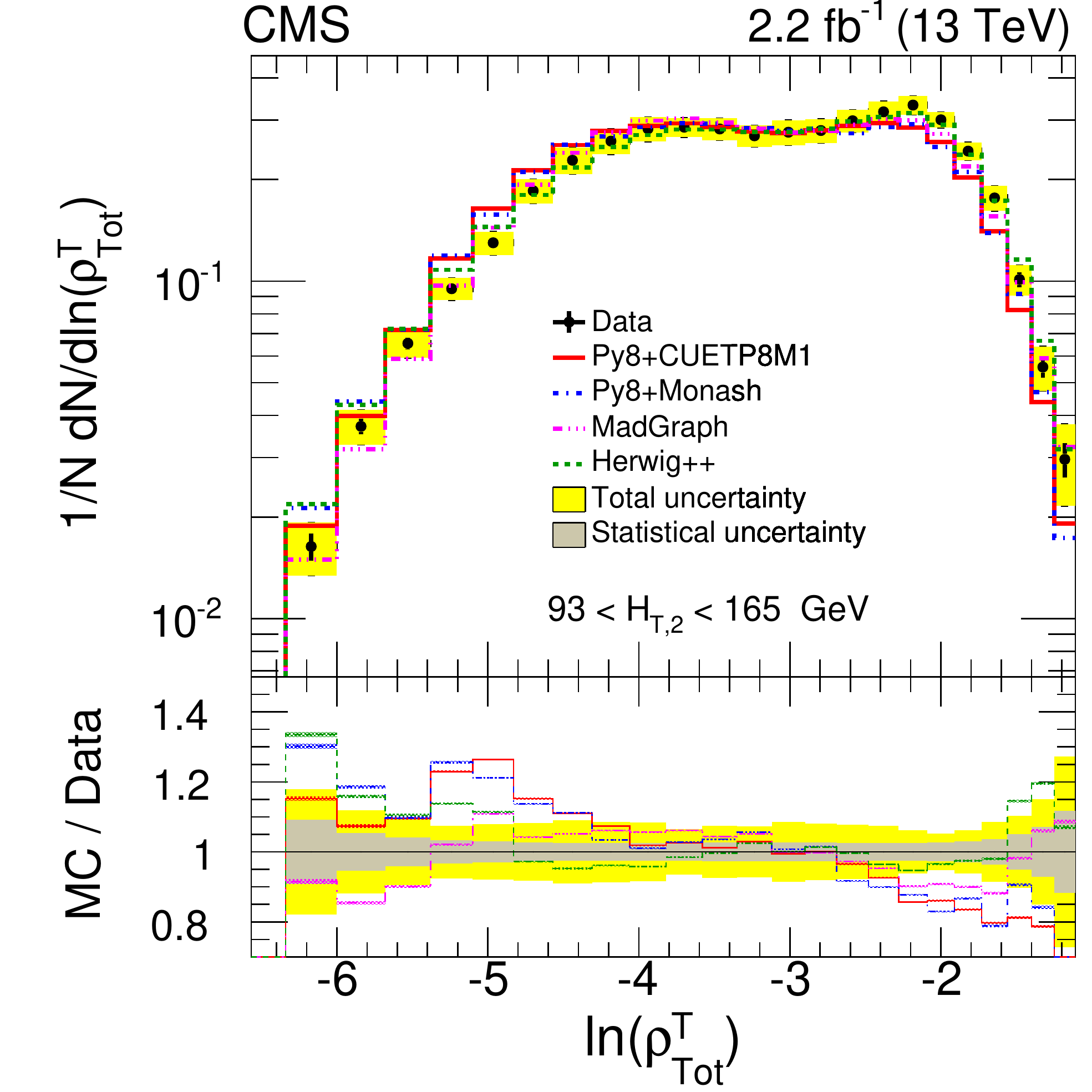}
  \caption{Normalized differential distributions of unfolded data compared with theoretical (MC)
	predictions of \pythiaC (red line), \pythiaM (blue dash-dotted line),
	\MGvATNLO (pink dash-dot-dotted line) and
        \HERWIGpp (brown dash-dot-dotted line)  as a function of ESV:
        complement of transverse thrust (\taup) (upper left),
        total jet broadening (\bt) (upper right),
        total jet mass (\rhoTot) (lower left) and
        total transverse jet mass (\rhoPerp) (lower right) for $\HTone\GeV$.
        In each ratio plot, the inner gray band represents statistical uncertainty and the yellow band represents
	the total uncertainty (systematic and statistical components added in quadrature) on data and the MC
        predictions include only statistical uncertainty.}
  \label{fig:Comparison-DataMC-one}
\end{figure}

\begin{figure}[hbtp]
\centering
  \includegraphics[width=0.49\textwidth]{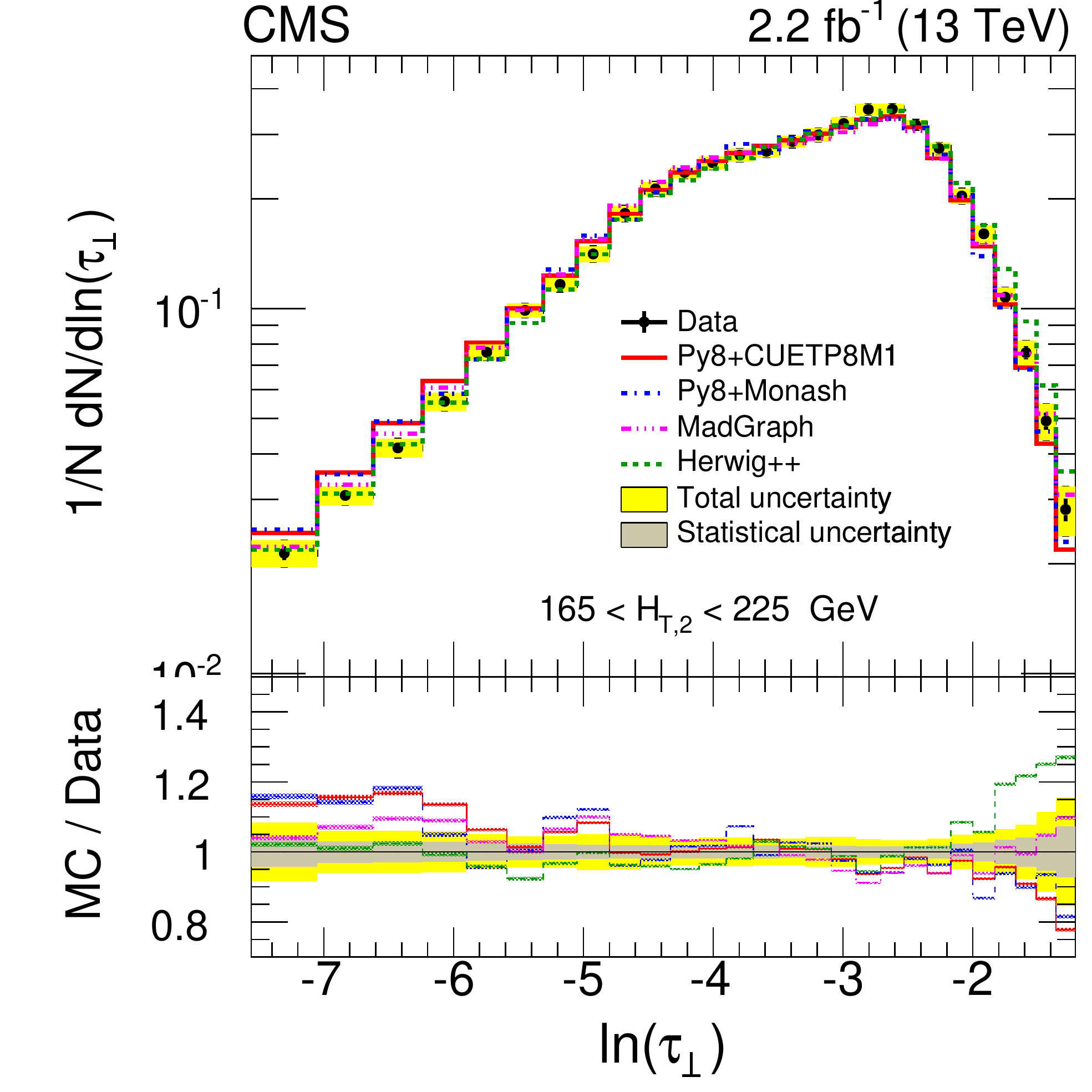}
  \includegraphics[width=0.49\textwidth]{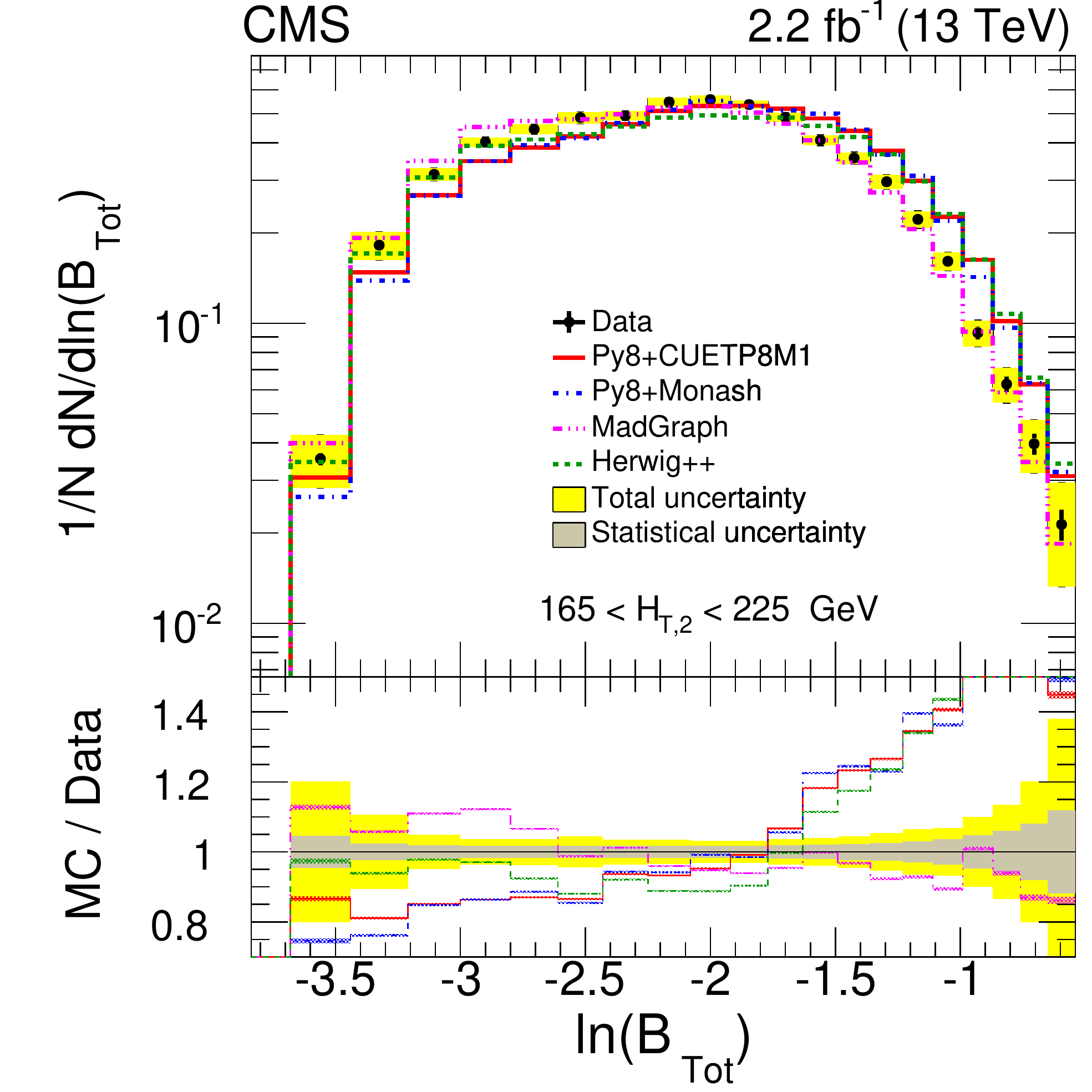}
  \includegraphics[width=0.49\textwidth]{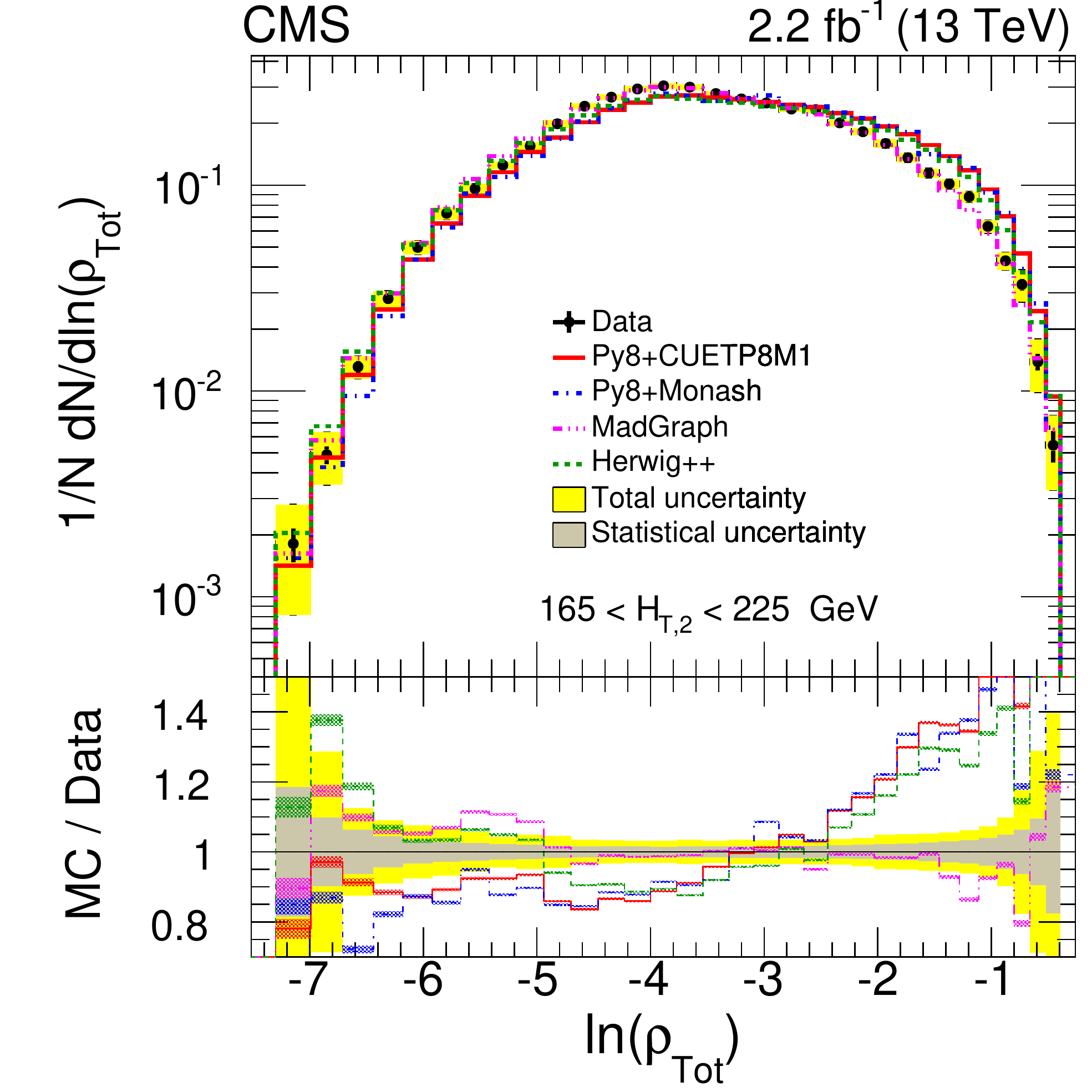}
  \includegraphics[width=0.49\textwidth]{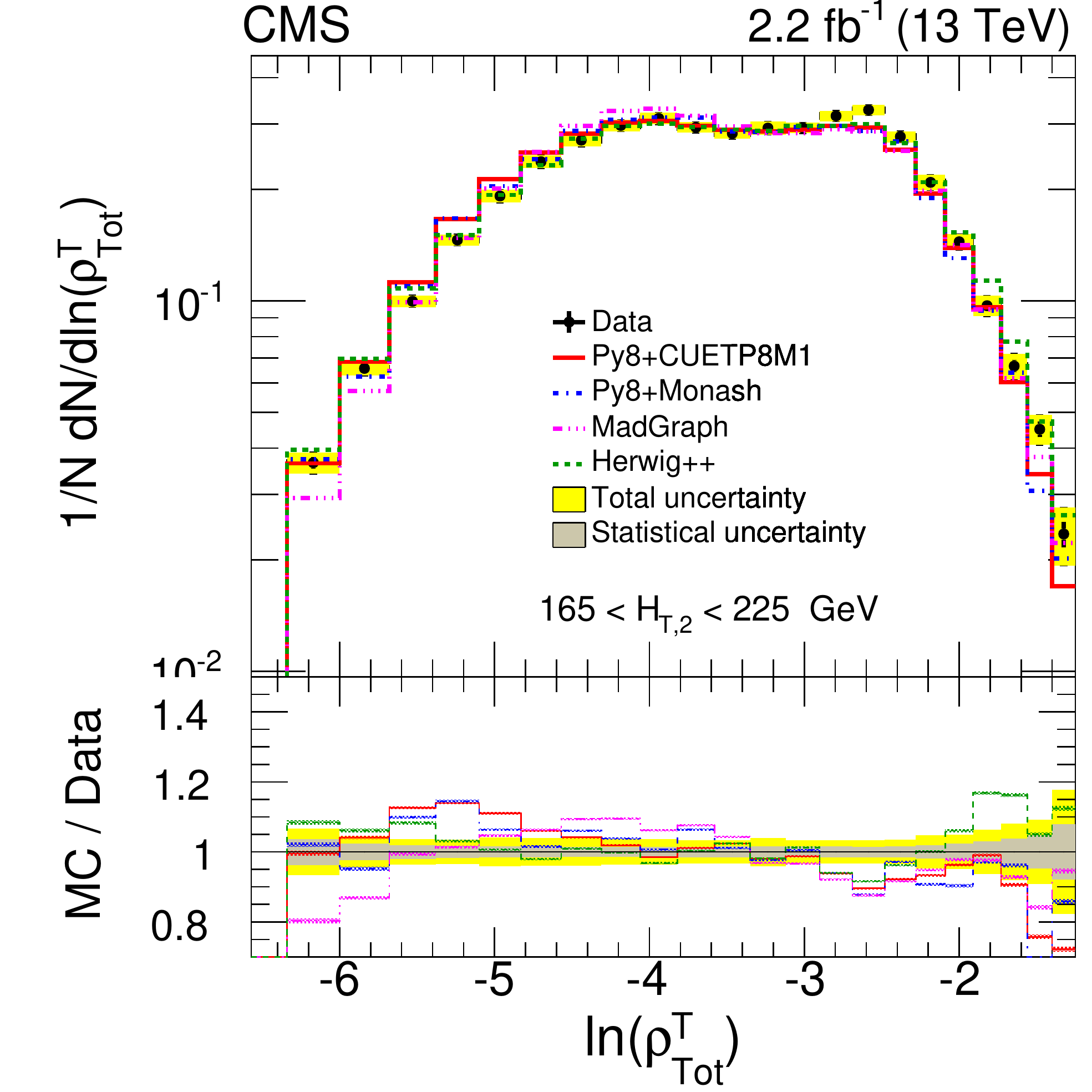}
  \caption{Normalized differential distributions of unfolded data compared with theoretical (MC)
	predictions of \pythiaC (red line), \pythiaM (blue dash-dotted line),
	\MGvATNLO (pink dash-dot-dotted line) and
        \HERWIGpp (brown dash-dot-dotted line)  as a function of ESV:
        complement of transverse thrust (\taup) (upper left),
        total jet broadening (\bt) (upper right),
        total jet mass (\rhoTot) (lower left) and
        total transverse jet mass (\rhoPerp) (lower right) for $\HTtwo\GeV$.
        In each ratio plot, the inner gray band represents statistical uncertainty and the yellow band represents
	the total uncertainty (systematic and statistical components added in quadrature) on data and the MC
        predictions include only statistical uncertainty.}
  \label{fig:Comparison-DataMC-two}
\end{figure}

\begin{figure}[hbtp]
\centering
  \includegraphics[width=0.49\textwidth]{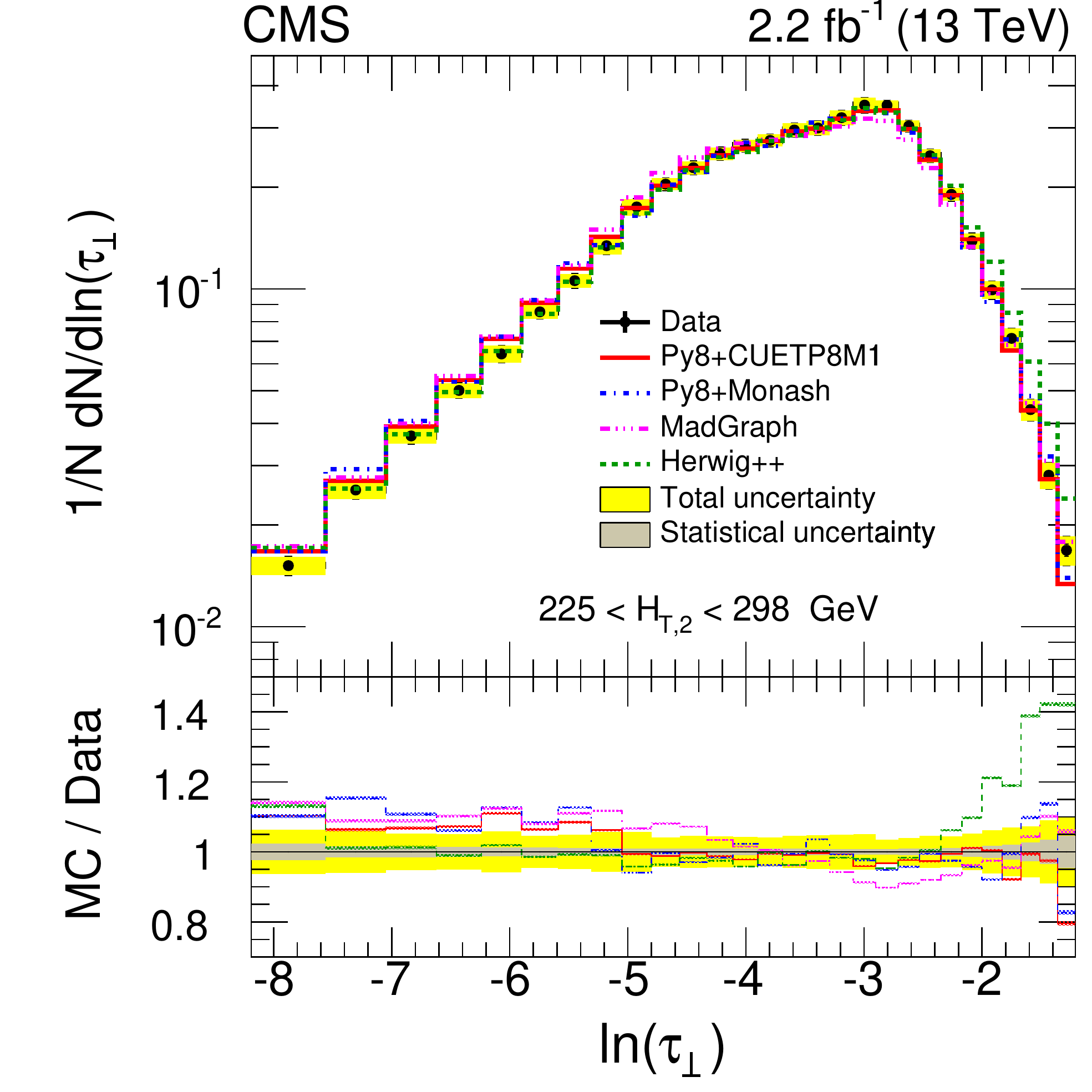}
  \includegraphics[width=0.49\textwidth]{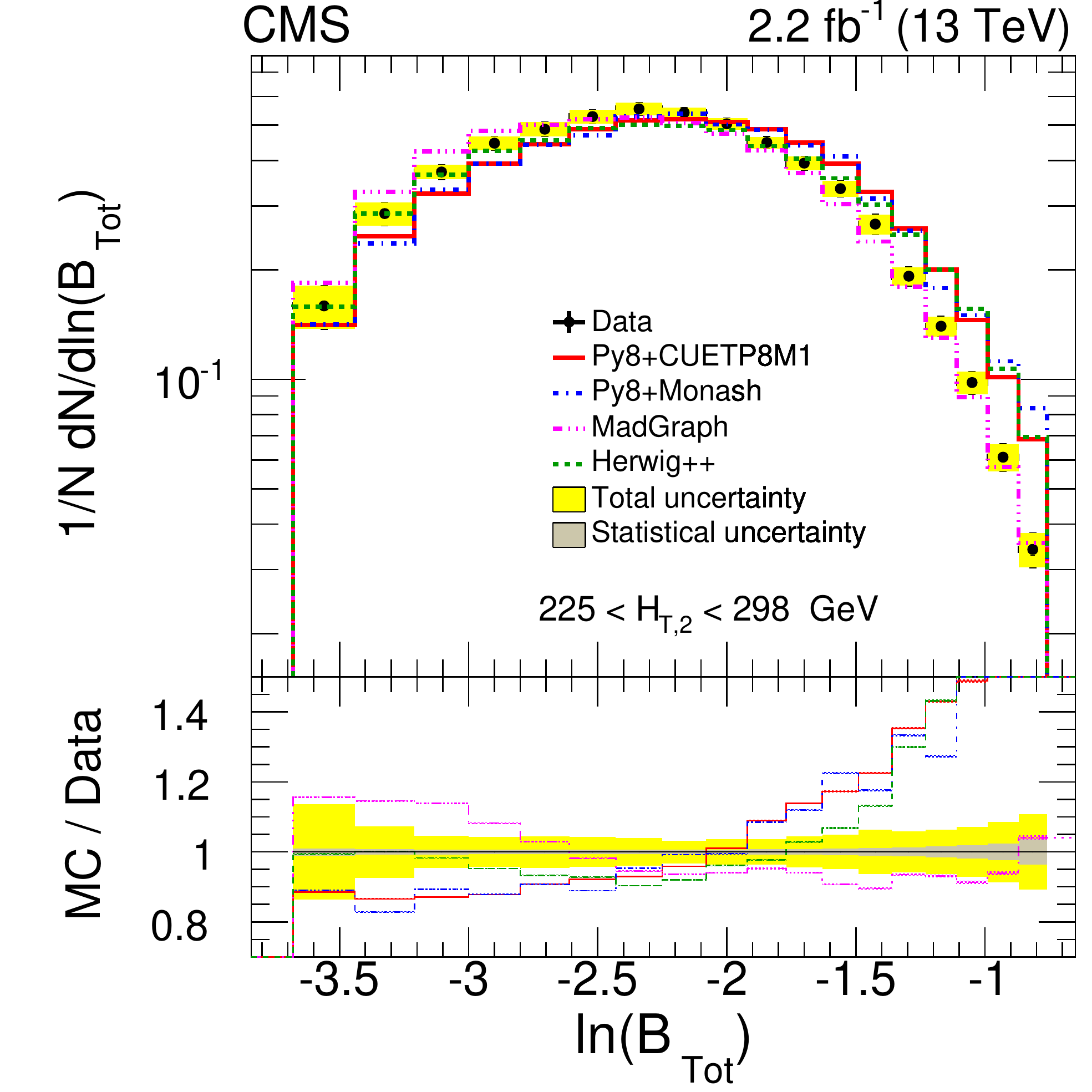}
  \includegraphics[width=0.49\textwidth]{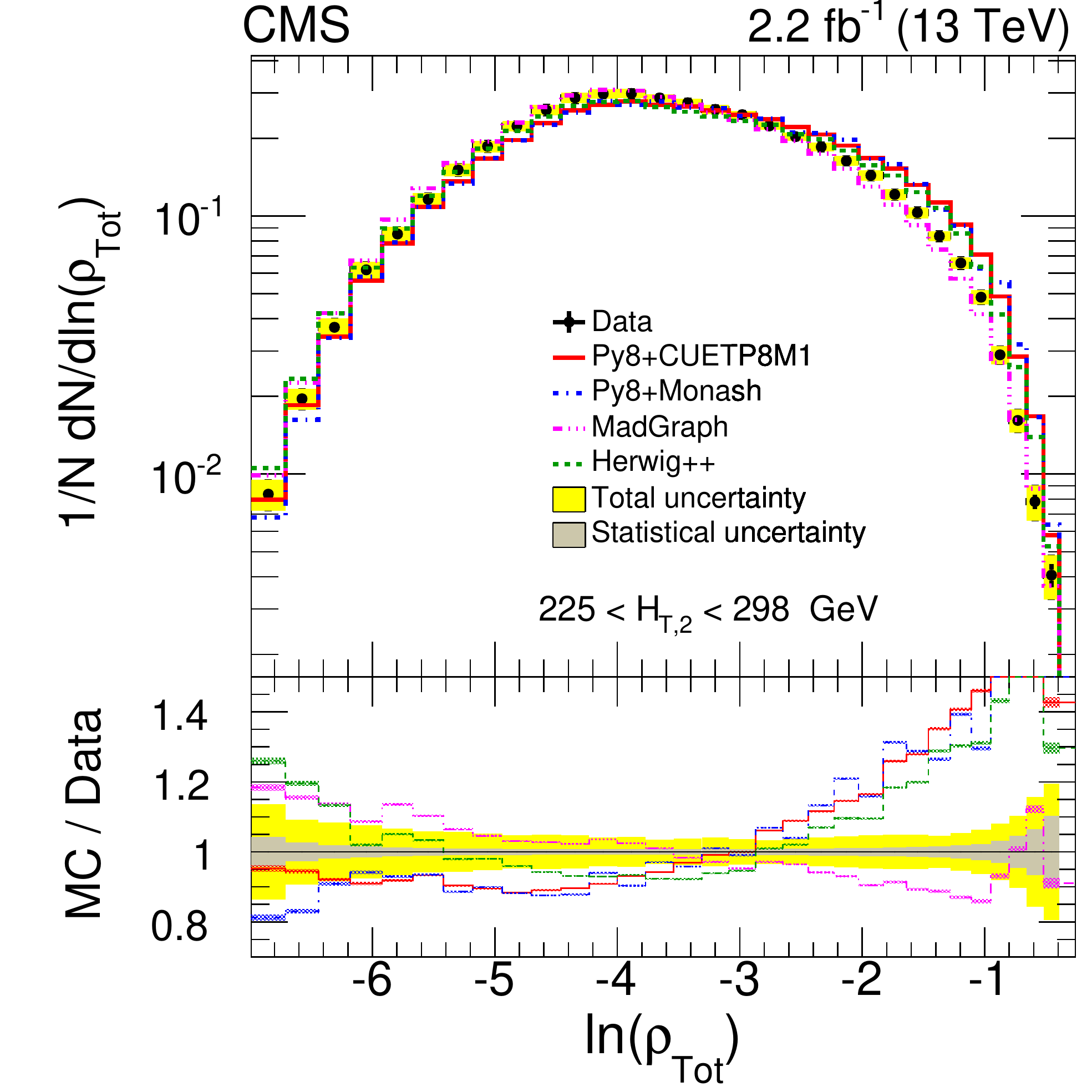}
  \includegraphics[width=0.49\textwidth]{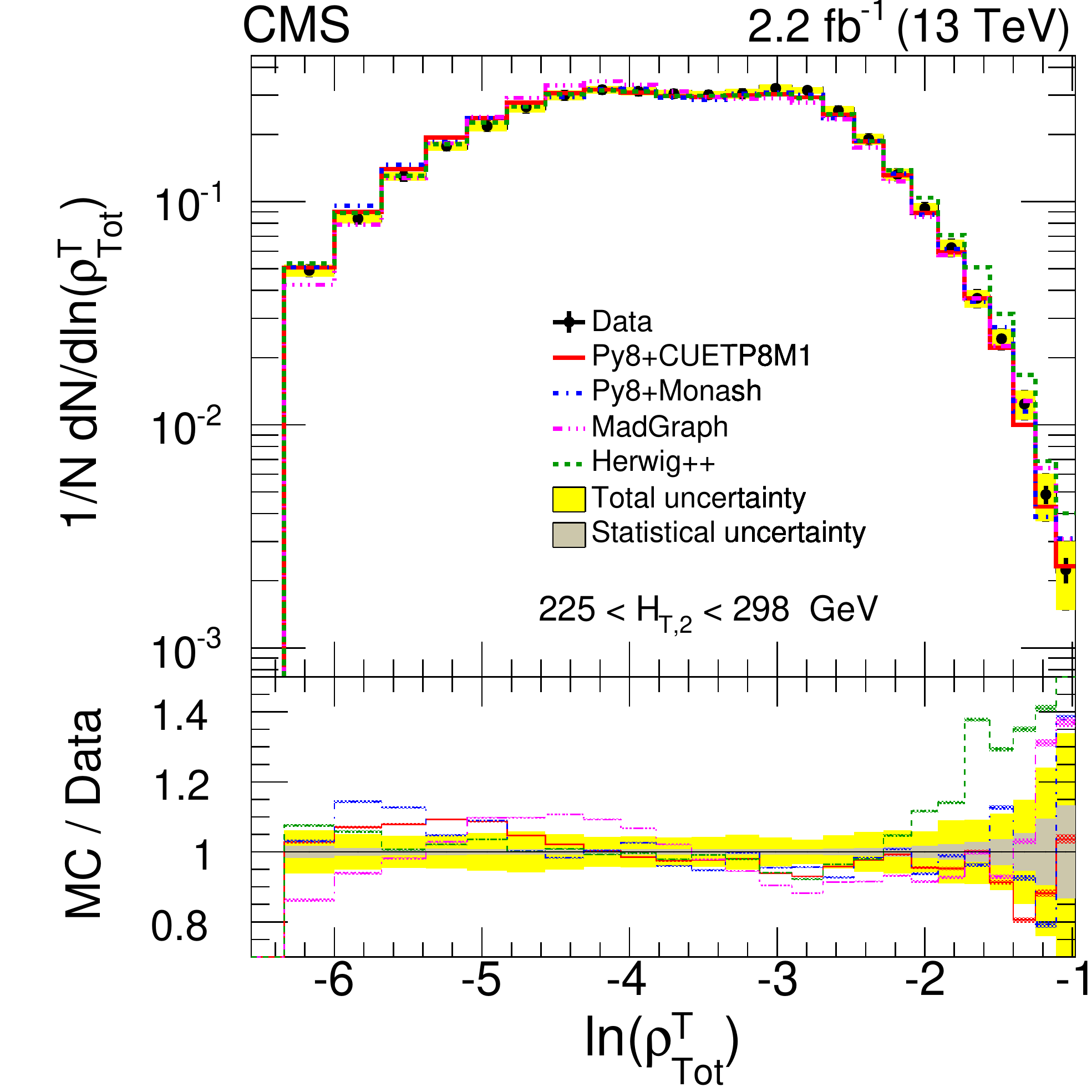}
  \caption{Normalized differential distributions of unfolded data compared with theoretical (MC)
	predictions of \pythiaC (red line), \pythiaM (blue dash-dotted line),
	\MGvATNLO (pink dash-dot-dotted line) and
        \HERWIGpp (brown dash-dot-dotted line)  as a function of ESV:
        complement of transverse thrust (\taup) (upper left),
        total jet broadening (\bt) (upper right),
        total jet mass (\rhoTot) (lower left) and
        total transverse jet mass (\rhoPerp) (lower right) for $\HTthr\GeV$.
        In each ratio plot, the inner gray band represents statistical uncertainty and the yellow band represents
	the total uncertainty (systematic and statistical components added in quadrature) on data and the MC
        predictions include only statistical uncertainty.}
  \label{fig:Comparison-DataMC-thr}
\end{figure}

\begin{figure}[hbtp]
\centering
  \includegraphics[width=0.49\textwidth]{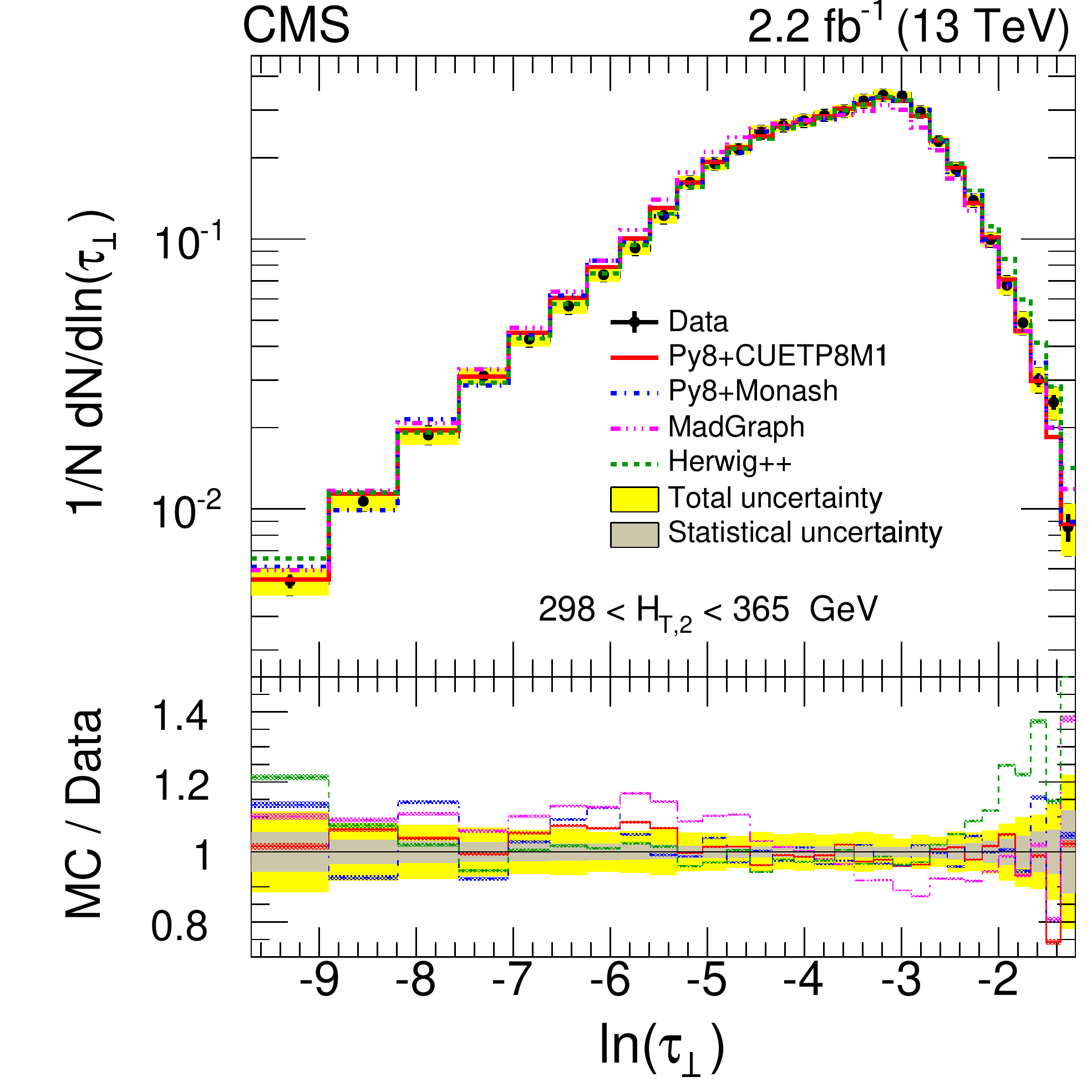}
  \includegraphics[width=0.49\textwidth]{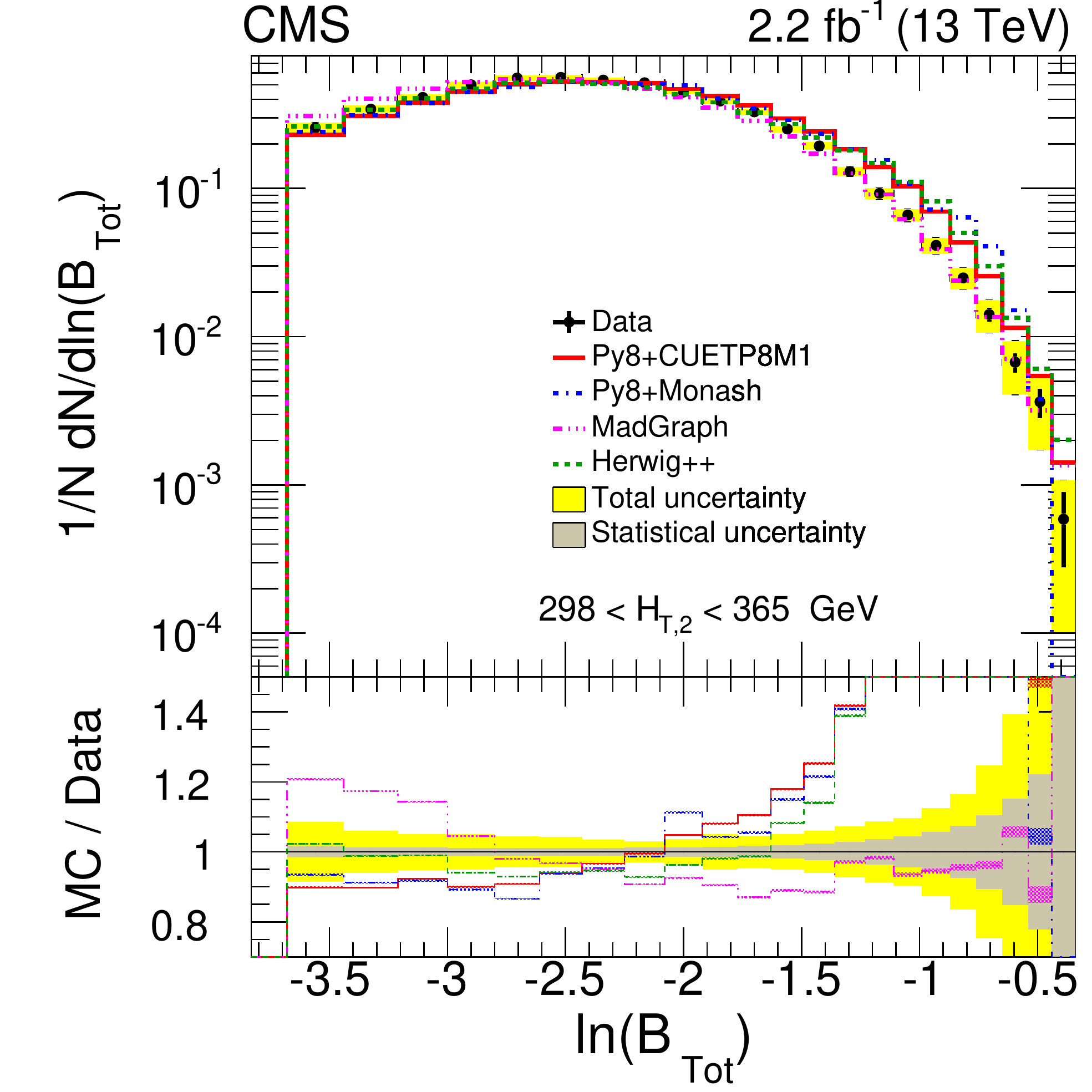}
  \includegraphics[width=0.49\textwidth]{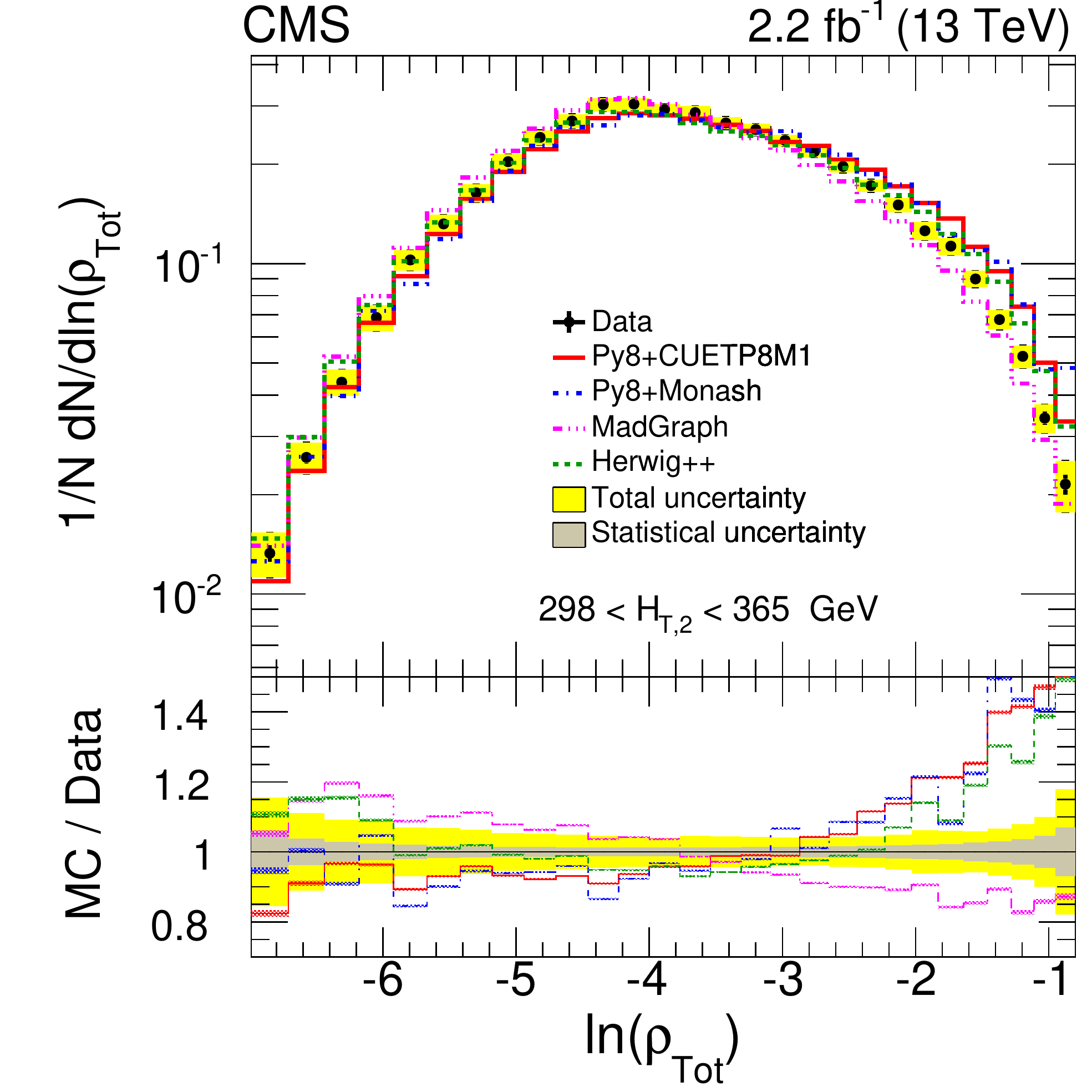}
  \includegraphics[width=0.49\textwidth]{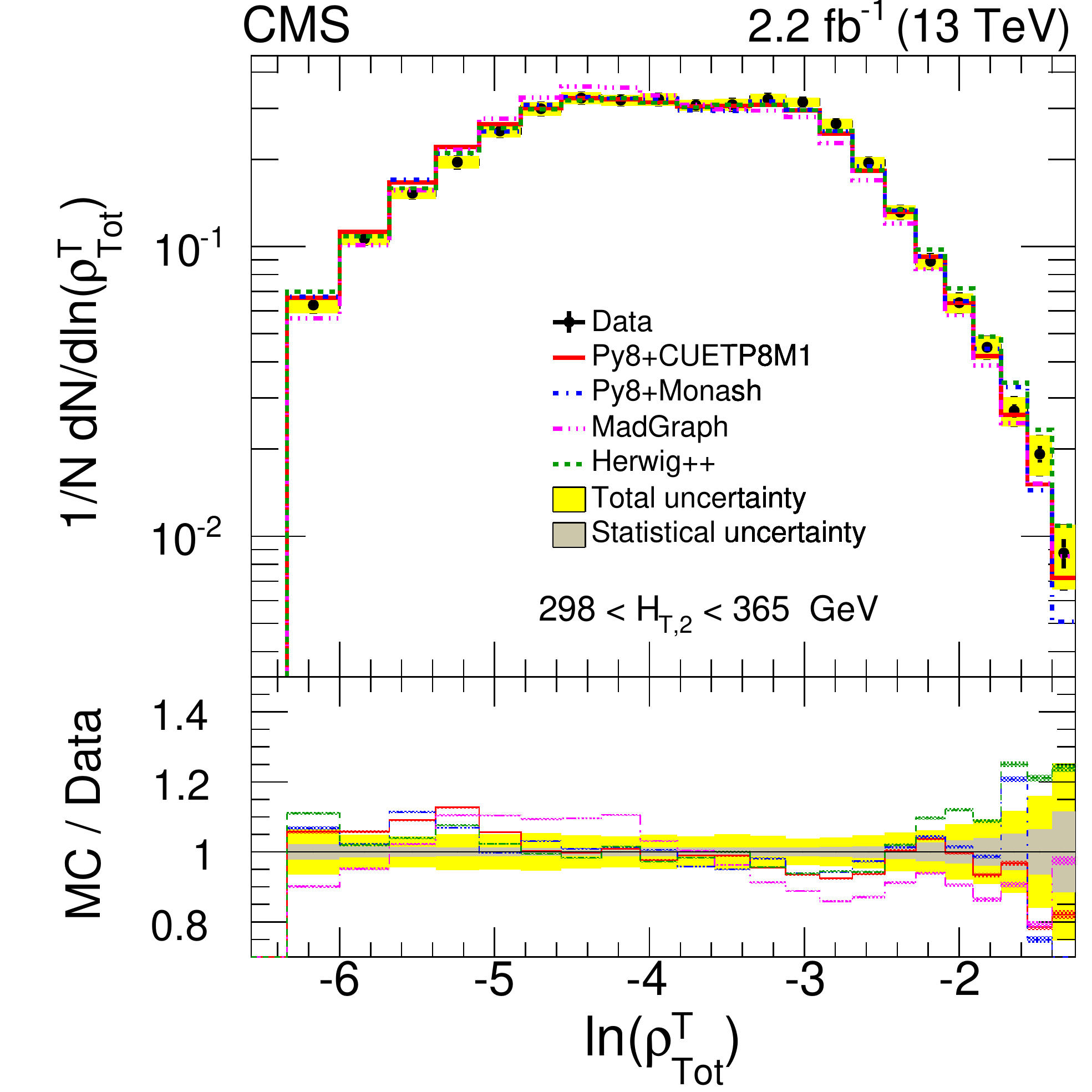}
  \caption{Normalized differential distributions of unfolded data compared with theoretical (MC)
	predictions of \pythiaC (red line), \pythiaM (blue dash-dotted line),
	\MGvATNLO (pink dash-dot-dotted line) and
        \HERWIGpp (brown dash-dot-dotted line)  as a function of ESV:
        complement of transverse thrust (\taup) (upper left),
        total jet broadening (\bt) (upper right),
        total jet mass (\rhoTot) (lower left) and
        total transverse jet mass (\rhoPerp) (lower right) for $\HTfou\GeV$.
        In each ratio plot, the inner gray band represents statistical uncertainty and the yellow band represents
	the total uncertainty (systematic and statistical components added in quadrature) on data and the MC
        predictions include only statistical uncertainty.}
  \label{fig:Comparison-DataMC-fou}
\end{figure}

\begin{figure}[hbtp]
\centering
  \includegraphics[width=0.49\textwidth]{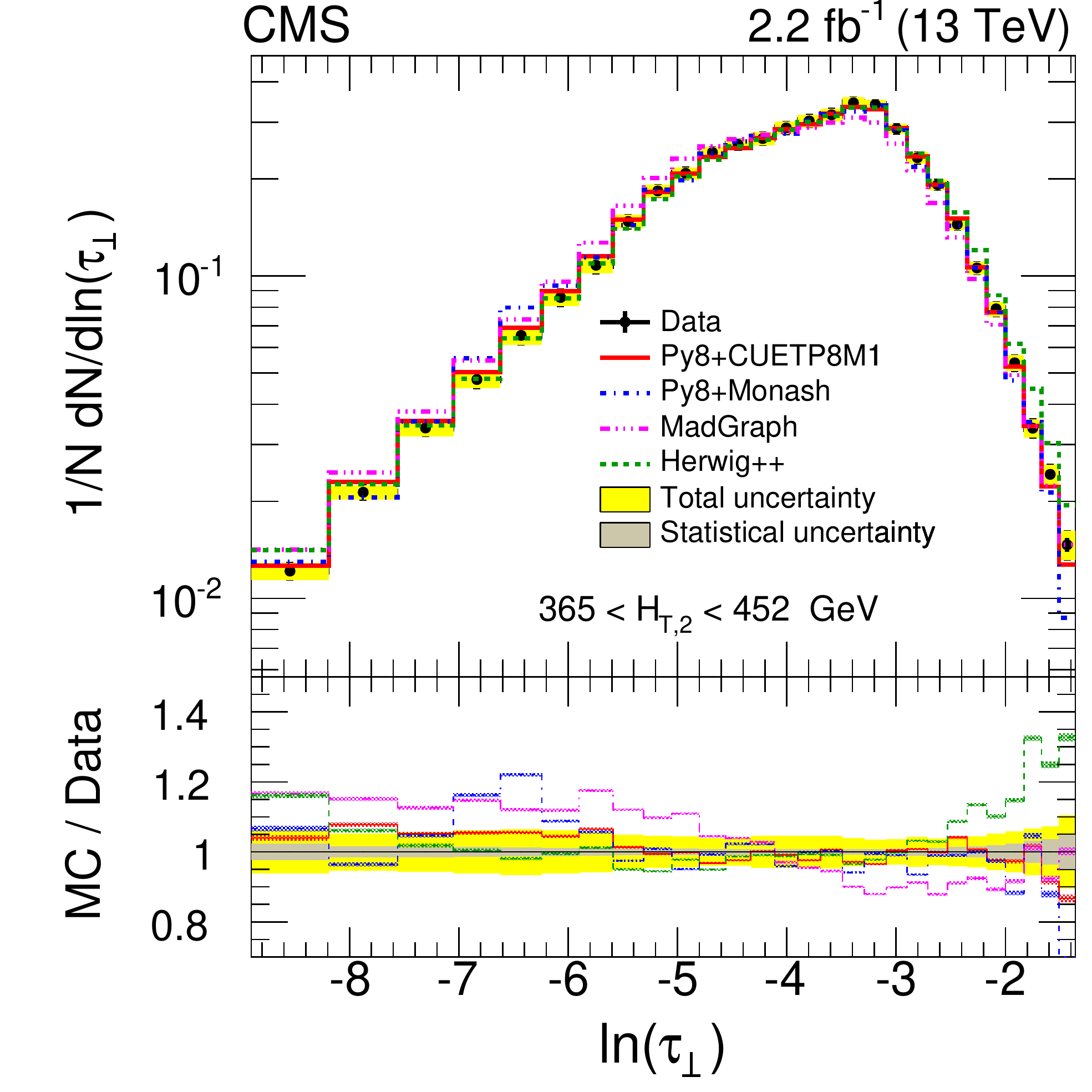}
  \includegraphics[width=0.49\textwidth]{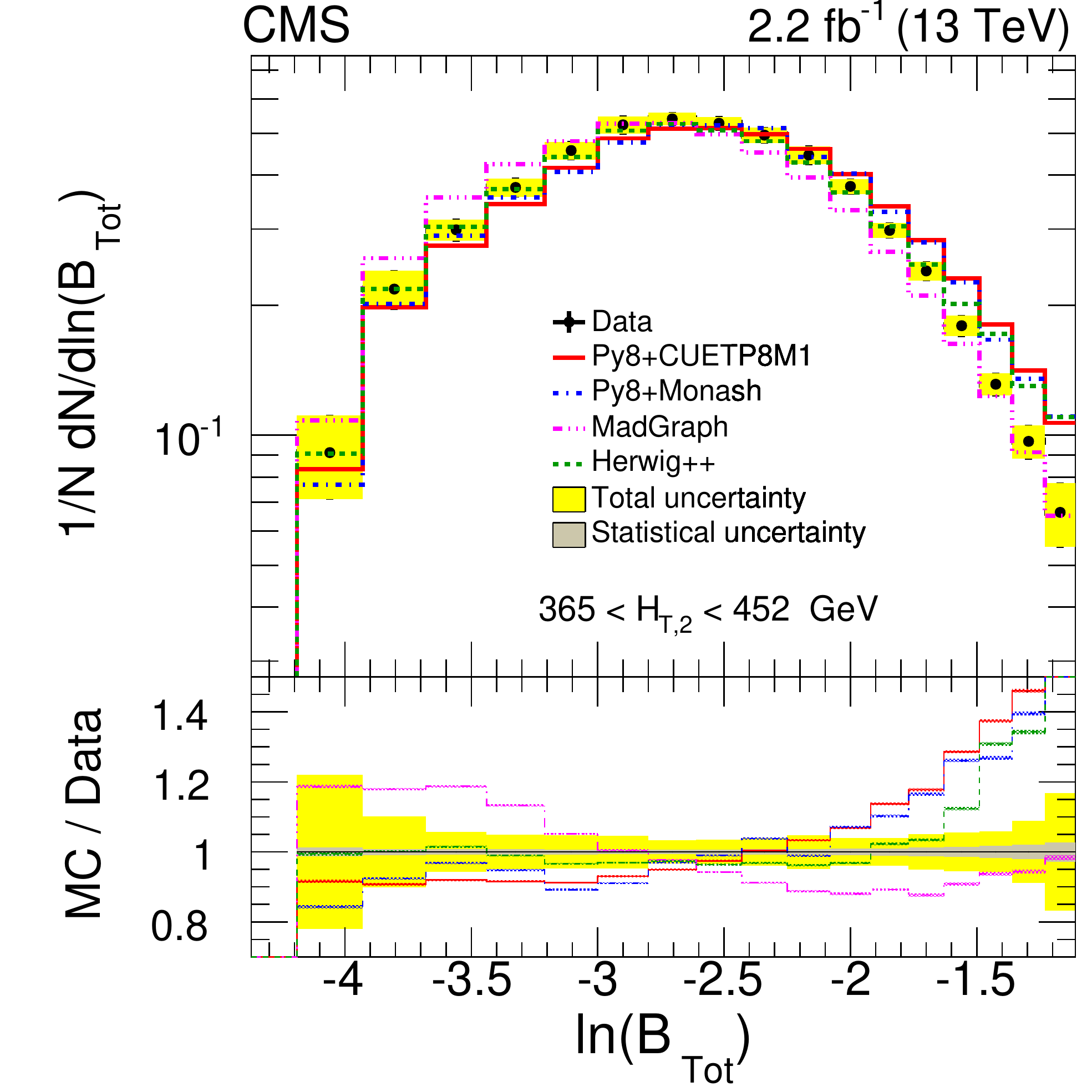}
  \includegraphics[width=0.49\textwidth]{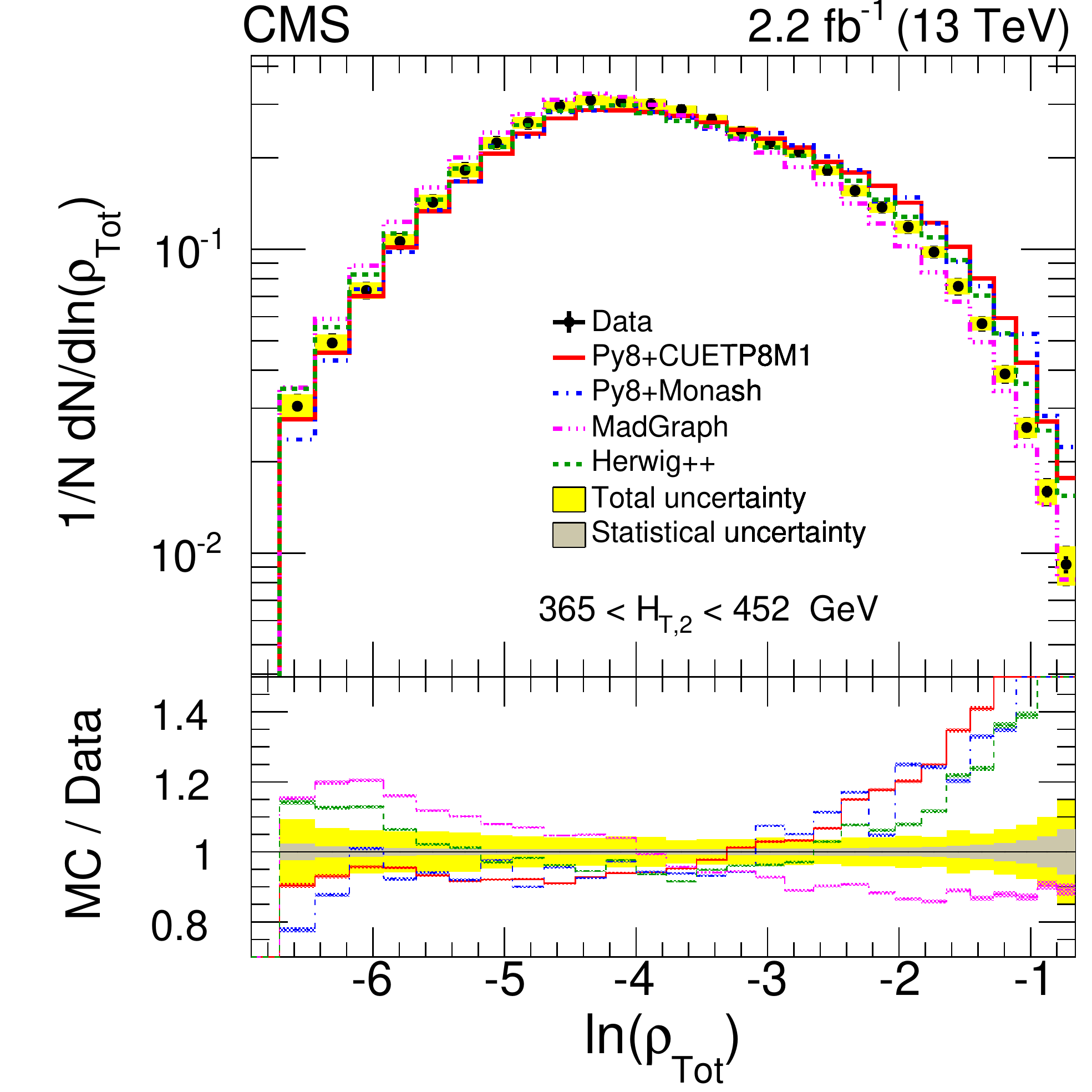}
  \includegraphics[width=0.49\textwidth]{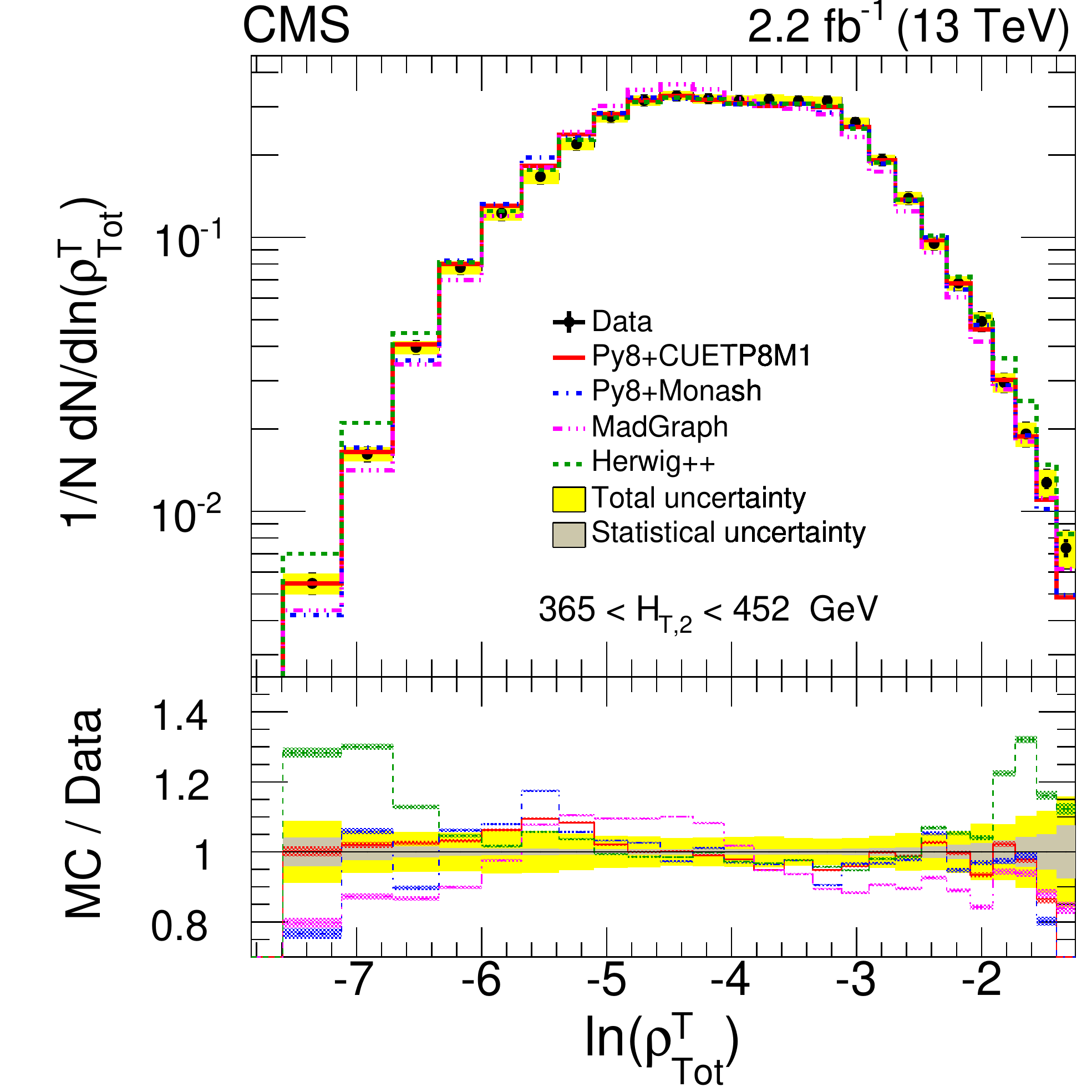}
  \caption{Normalized differential distributions of unfolded data compared with theoretical (MC)
	predictions of \pythiaC (red line), \pythiaM (blue dash-dotted line),
	\MGvATNLO (pink dash-dot-dotted line) and
        \HERWIGpp (brown dash-dot-dotted line)  as a function of ESV:
        complement of transverse thrust (\taup) (upper left),
        total jet broadening (\bt) (upper right),
        total jet mass (\rhoTot) (lower left) and
        total transverse jet mass (\rhoPerp) (lower right) for $\HTfiv\GeV$.
        In each ratio plot, the inner gray band represents statistical uncertainty and the yellow band represents
	the total uncertainty (systematic and statistical components added in quadrature) on data and the MC
        predictions include only statistical uncertainty.}
  \label{fig:Comparison-DataMC-fiv}
\end{figure}

\begin{figure}[hbtp]
\centering
  \includegraphics[width=0.49\textwidth]{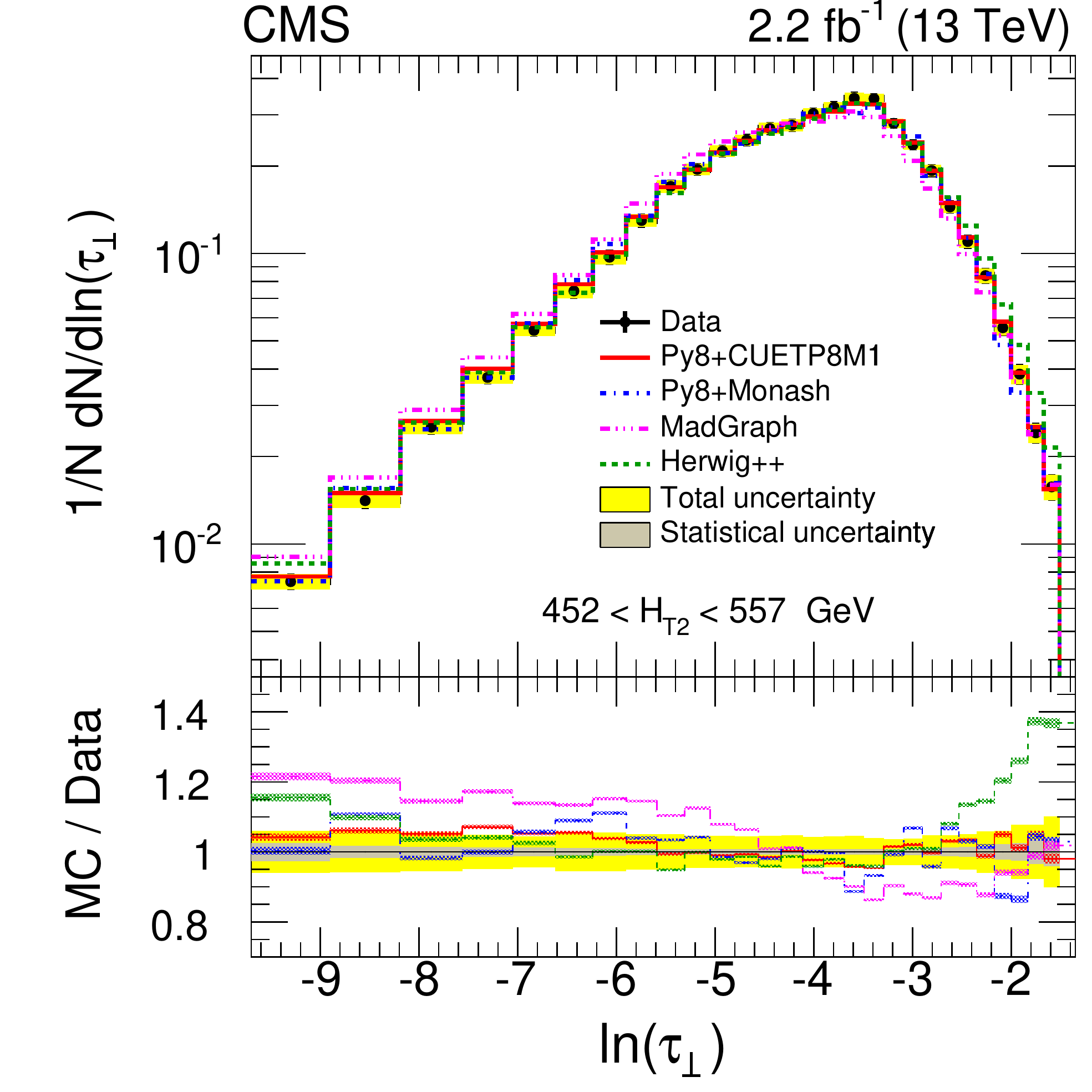}
  \includegraphics[width=0.49\textwidth]{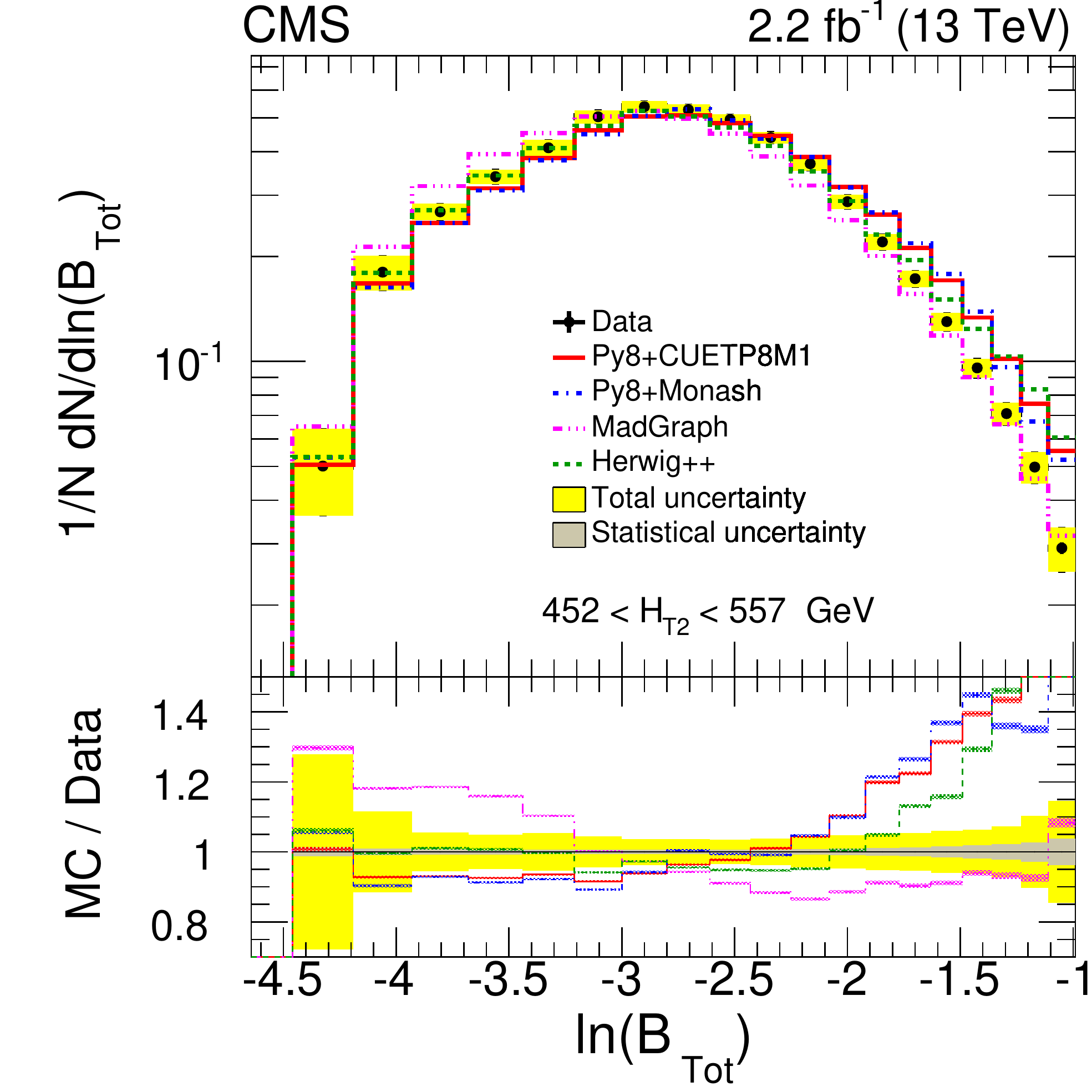}
  \includegraphics[width=0.49\textwidth]{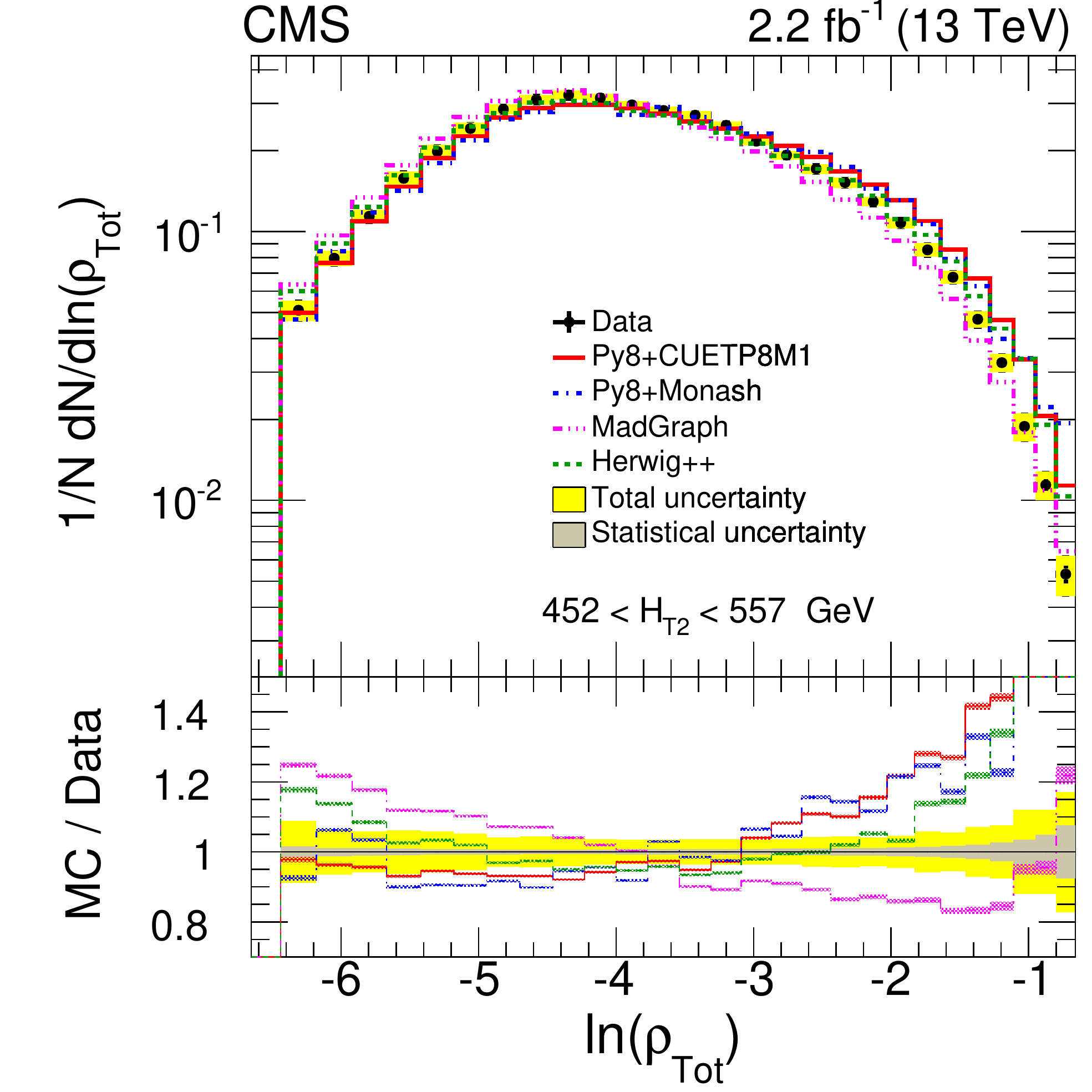}
  \includegraphics[width=0.49\textwidth]{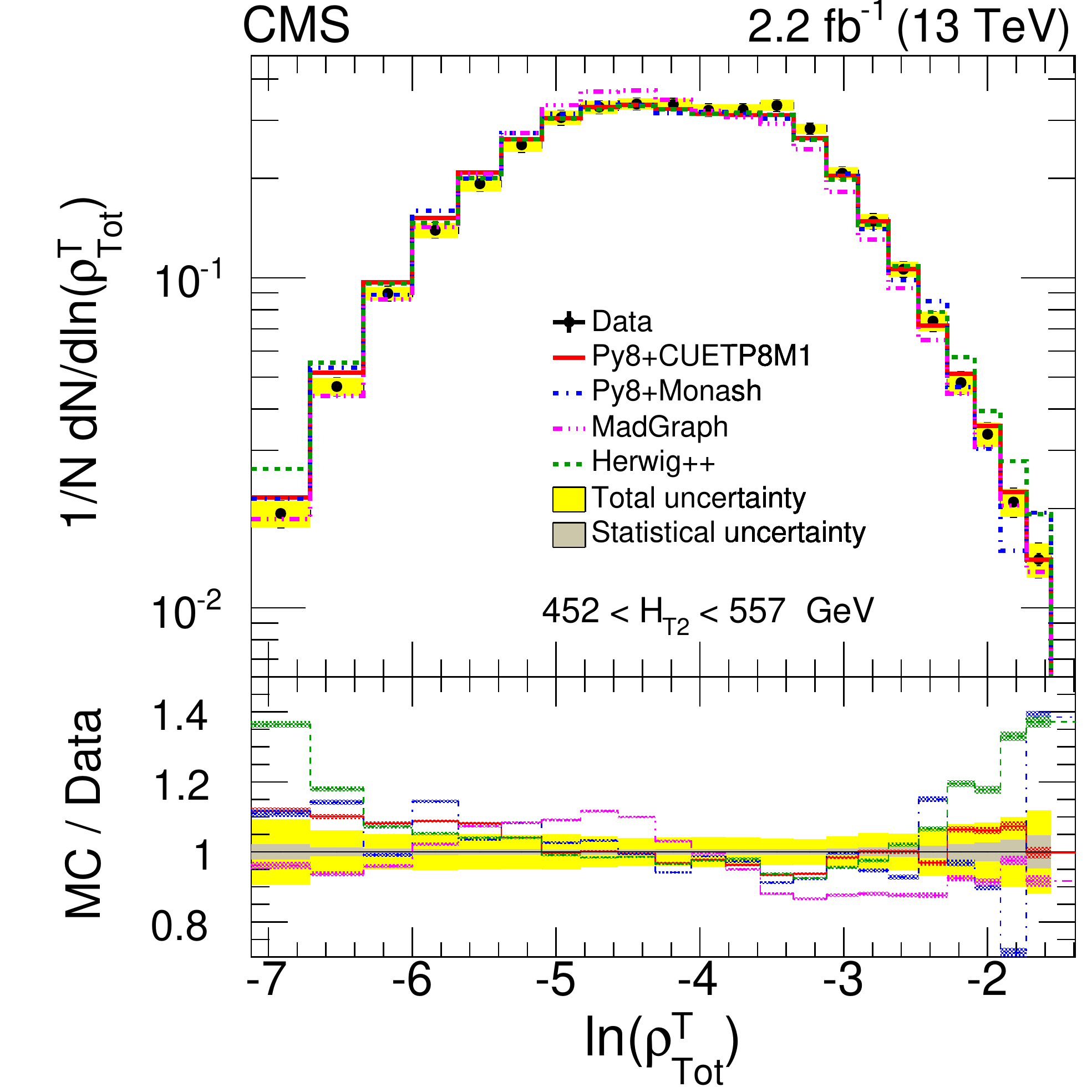}
  \caption{Normalized differential distributions of unfolded data compared with theoretical (MC)
	predictions of \pythiaC (red line), \pythiaM (blue dash-dotted line),
	\MGvATNLO (pink dash-dot-dotted line) and
        \HERWIGpp (brown dash-dot-dotted line)  as a function of ESV:
        complement of transverse thrust (\taup) (upper left),
        total jet broadening (\bt) (upper right),
        total jet mass (\rhoTot) (lower left) and
        total transverse jet mass (\rhoPerp) (lower right) for $\HTsix\GeV$.
        In each ratio plot, the inner gray band represents statistical uncertainty and the yellow band represents
	the total uncertainty (systematic and statistical components added in quadrature) on data and the MC
        predictions include only statistical uncertainty.}
  \label{fig:Comparison-DataMC-six}
\end{figure}

\begin{figure}[hbtp]
\centering
  \includegraphics[width=0.49\textwidth]{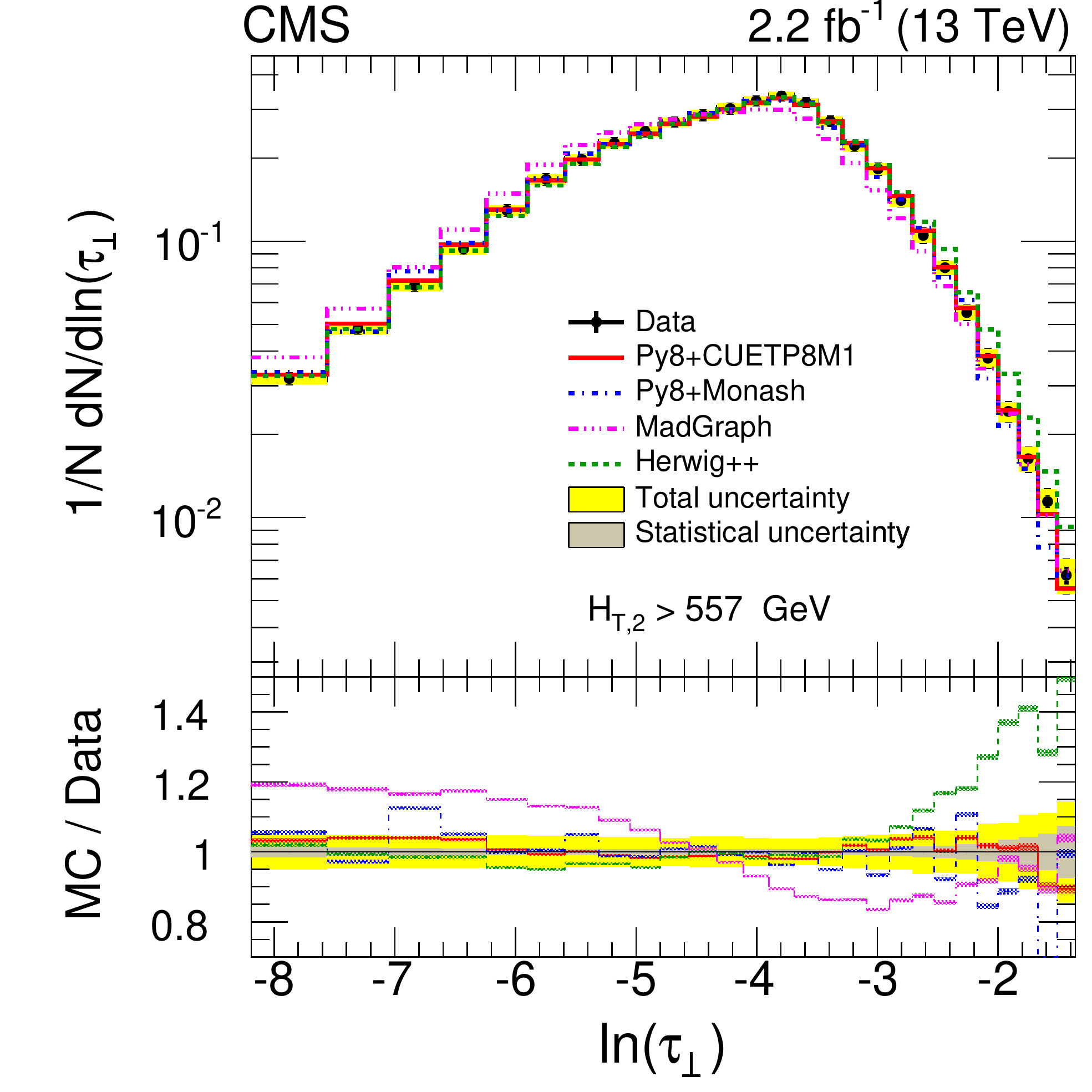}
  \includegraphics[width=0.49\textwidth]{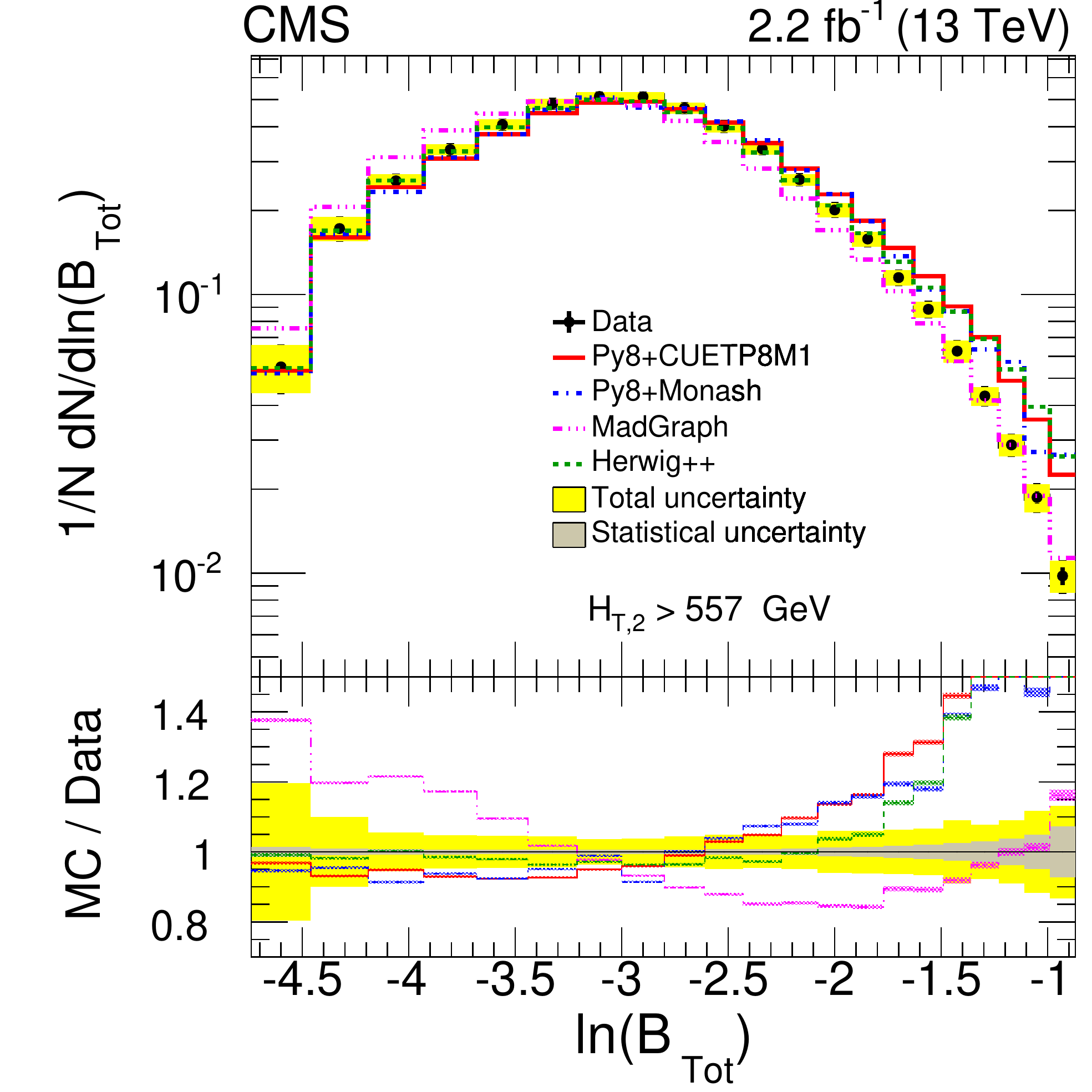}
  \includegraphics[width=0.49\textwidth]{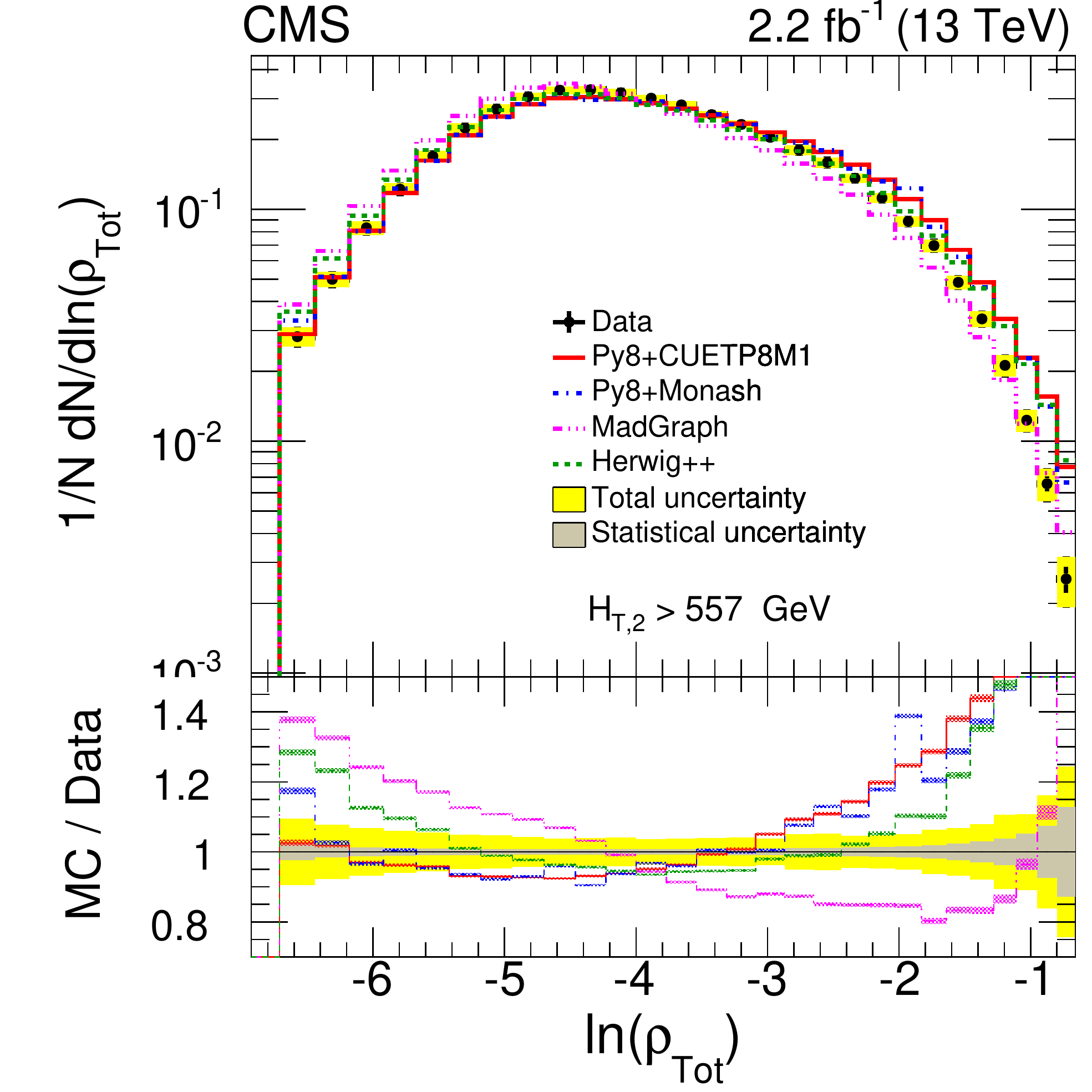}
  \includegraphics[width=0.49\textwidth]{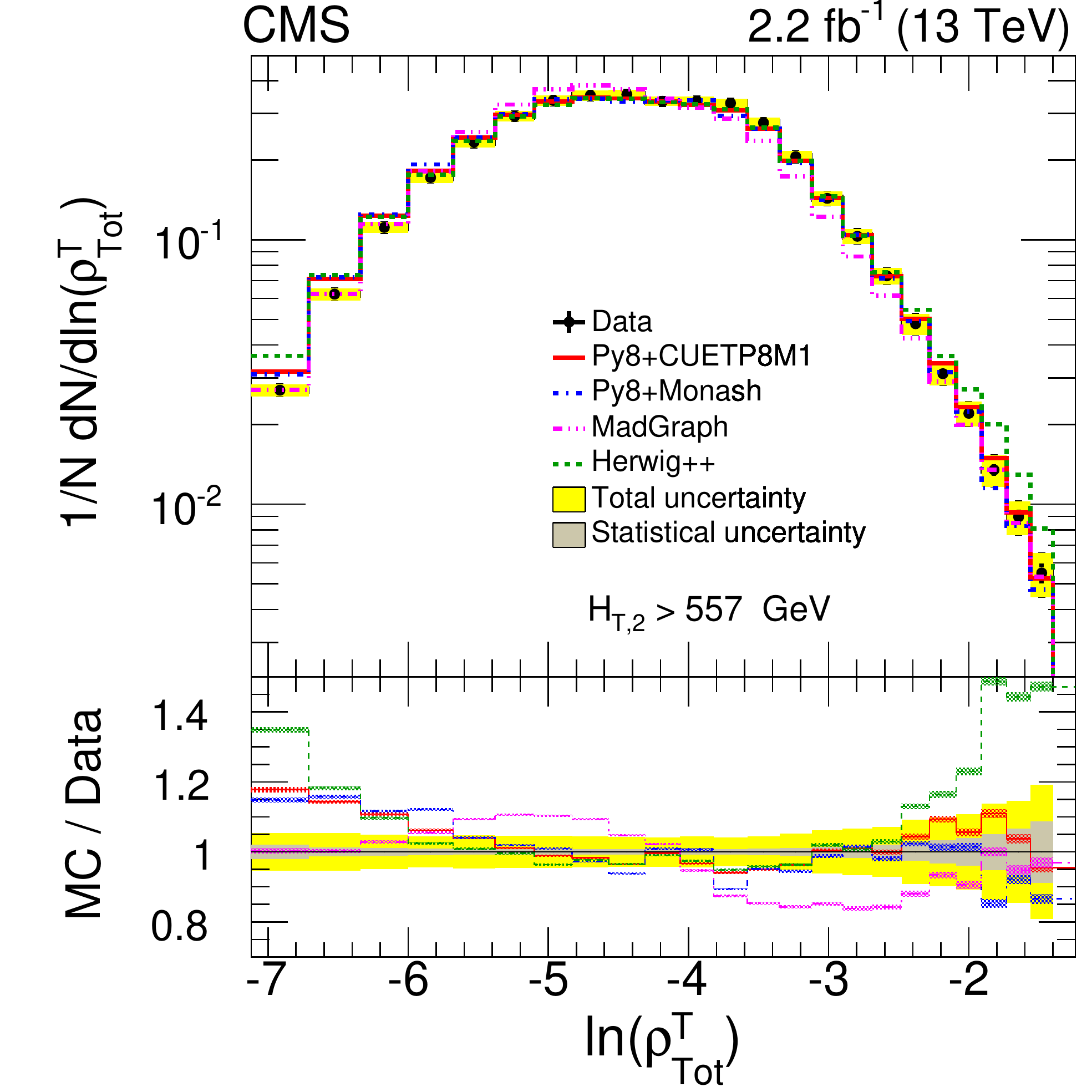}
  \caption{Normalized differential distributions of unfolded data compared with theoretical (MC)
	predictions of \pythiaC (red line), \pythiaM (blue dash-dotted line),
	\MGvATNLO (pink dash-dot-dotted line) and
        \HERWIGpp (brown dash-dot-dotted line)  as a function of ESV:
        complement of transverse thrust (\taup) (upper left),
        total jet broadening (\bt) (upper right),
        total jet mass (\rhoTot) transverse jet mass (\rhoPerp) (lower left) and
        total transverse jet mass (\rhoPerp) (lower right) for $\HTsev\GeV$.
        In each ratio plot, the inner gray band represents statistical uncertainty and the yellow band represents
	the total uncertainty (systematic and statistical components added in quadrature) on data and the MC
        predictions include only statistical uncertainty.}
\label{fig:Comparison-DataMC-sev}
\end{figure}

The unfolded distributions for the ESVs obtained from data are compared with the particle-level
predictions of various MC generators, as shown in
Figures~\ref{fig:Comparison-DataMC-zer}--\ref{fig:Comparison-DataMC-sev} for various \h ranges.
Comparisons are made to the central predictions of the event generators only. Each figure
presents the variables \taup (upper left), \bt (upper right), \rhoTot (lower left), and \rhoPerp
(lower right) for a range of $\h$. The ratios of individual MC predictions to  that of data are shown
in the lower panel of each plot.

The MPI parameters in the \PYTHIAeight Monash and \textsc{CUETP8M1} tunes are very similar.
The predictions of these two tunes agree well for the four ESVs studied. In general, the agreement
between them improves with increasing $\h$. Both tunes show good agreement with data for the \taup
and \rhoPerp variables, except for the two lowest ranges of $\h$, and both overestimate the
multijet contribution to \rhoTot and \bt. We note that \taup and \rhoPerp variables are evaluated in
the transverse plane, whereas \bt and \rhoTot are evaluated using both longitudinal and transverse
components of the jets. This indicates that the treatment of the energy flow in the transverse
plane is modelled well in the Monash and \textsc{CUETP8M1} tunes of \PYTHIAeight,
whereas the energy flow out of the transverse plane is not.

The \HERWIGpp generator shows good agreement with data for all four ESVs studied, and it is better
than the \textsc{CUETP8M1} and Monash tunes of \PYTHIAeight
in predicting $\rhoTot$ and $\bt$. This implies its
better treatment of energy flow out of the transverse plane. Although both
\PYTHIAeight and \HERWIGpp use a PS approach to generate multijet events and hadronization,
the former uses string fragmentation and a \pt -ordered shower, whereas the
latter uses cluster fragmentation and angular-ordered shower.

The \MGvATNLO generator shows good agreement with data for $\taup$ and $\rhoPerp$ and its
agreement with data for $\rhoTot$ and $\bt$ is much better compared to the
\textsc{CUETP8M1} and Monash
tunes of \PYTHIAeight. The ME approach for generating multiparton hard scattering processes models the
transverse as well as longitudinal flows of energy better than PYTHIA8.

\begin{figure}[hbtp]
\centering
  \includegraphics[width=0.49\textwidth]{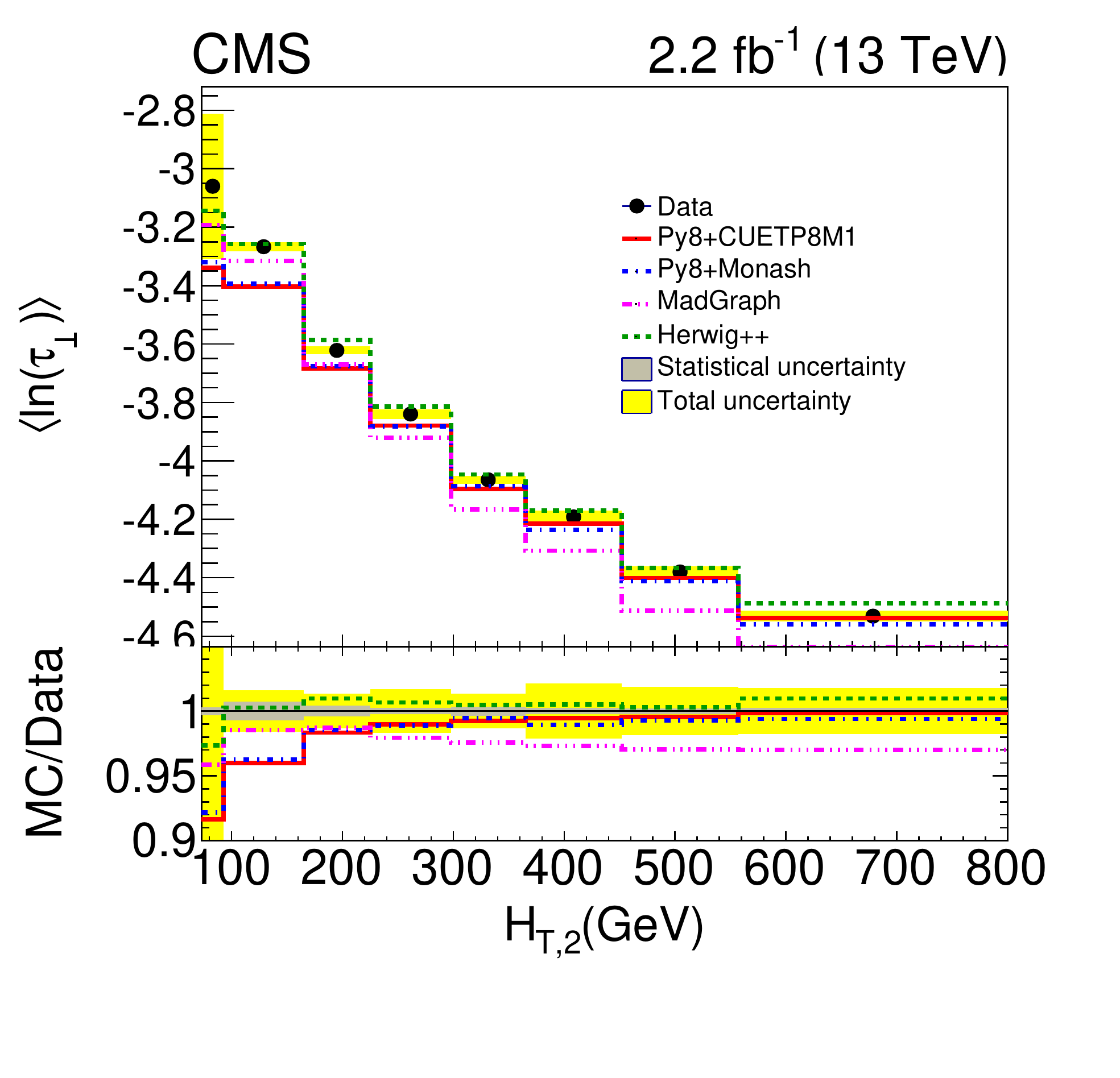}
  \includegraphics[width=0.49\textwidth]{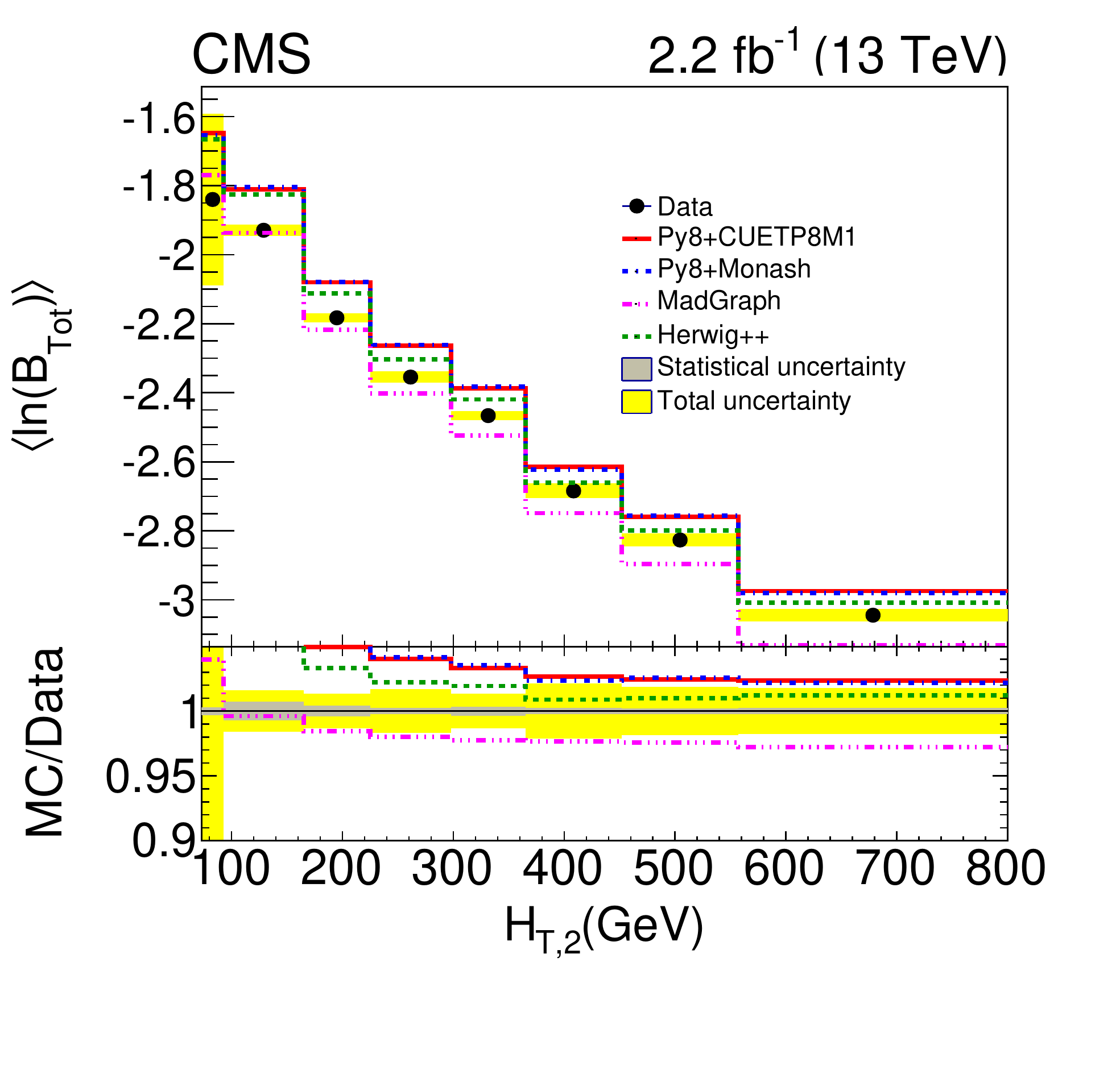}\hfill
  \includegraphics[width=0.49\textwidth]{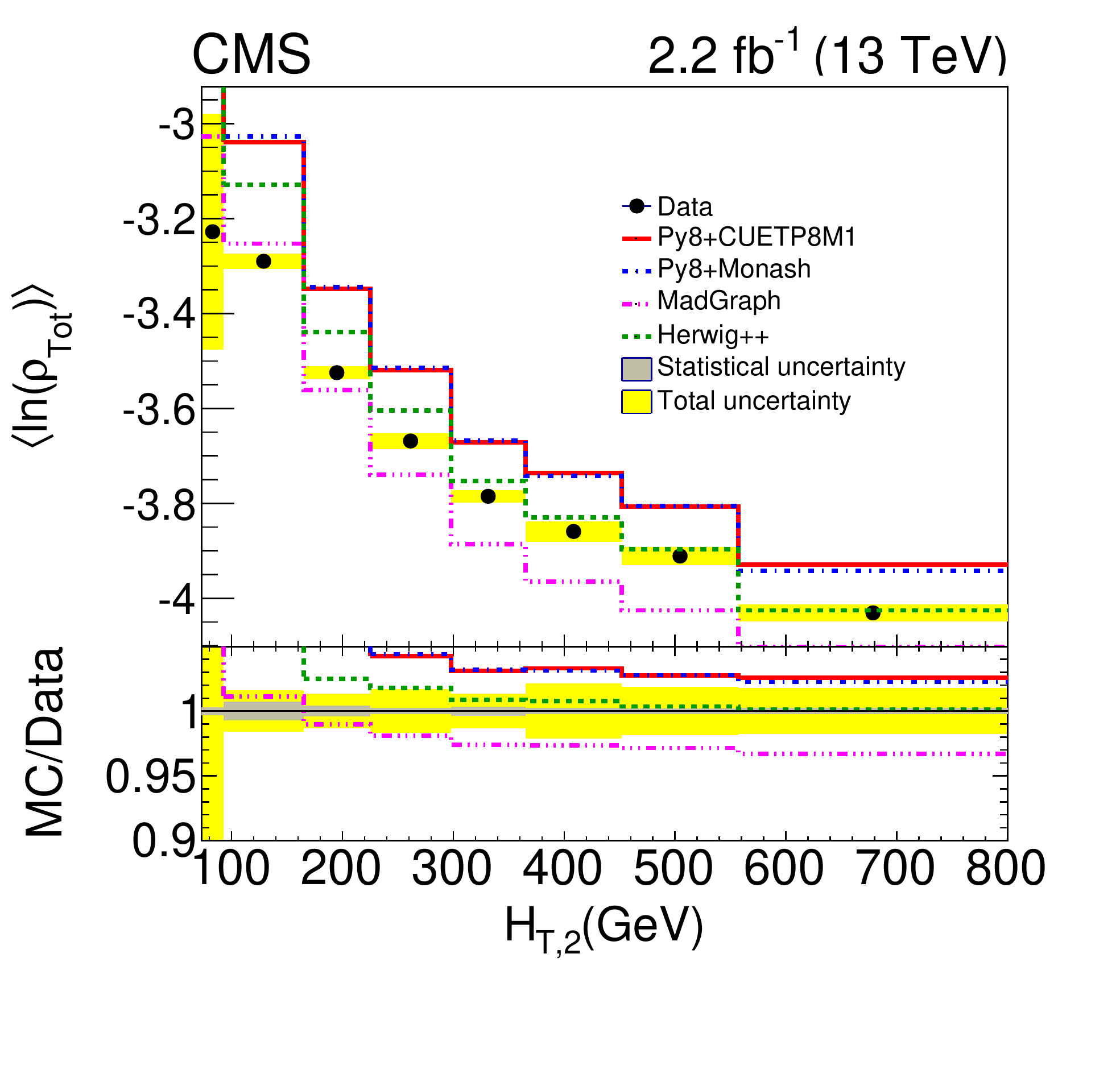}
  \includegraphics[width=0.49\textwidth]{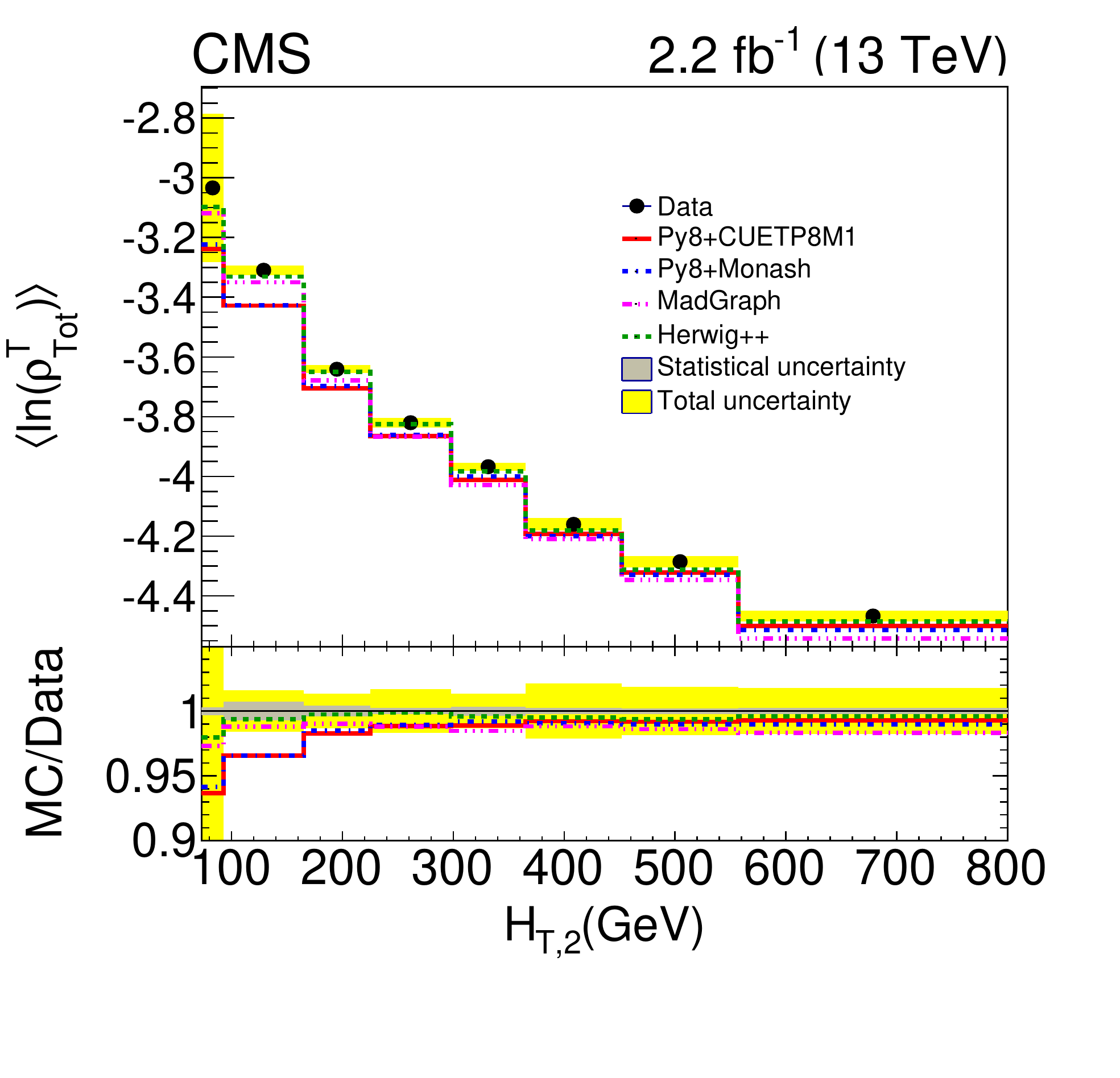}
  \caption{The evolution of the mean of \taup (upper left), \bt (upper right),
	\rhoTot (lower left) and \rhoPerp (lower right) and
	with increasing $\h$. The ratio plots with respect to data are presented
	in the bottom panel to compare predictions of
	\pythiaC (red line), \pythiaM (blue dash-dotted line),
	\MGvATNLO (pink dash-dot-dotted line) and \HERWIGpp (brown dash-dot-dotted line).
	The yellow band represents the total uncertainty (systematic and statistical
	components added in quadrature).}
  \label{fig:Mean}
\end{figure}

The following features emerge from the comparison plots of the four ESVs.. Agreement between
data and benchmark event generators improves with $\h$. Figure~\ref{fig:Mean} shows the evolution
of the mean value of each ESV with $\h$ and confirms the above observations. With
higher $\h$, the initial partons are more boosted, and hence the event tends to be less spherical.
Also, $\alpha_S$ decreases with $\h$, resulting in less emission of hard gluons, which further
spoils the multijet, spherical nature of the event. Thus, the mean value of each ESV decreases
with increasing $\h$.


\section{Summary}\label{sec:summary}
This paper presents the first measurement at $\sqrt{s} = 13\TeV$ of four event shape variables:
complement of transverse thrust (\taup), total jet broadening (\bt), total jet mass (\rhoTot), and total
transverse jet mass (\rhoPerp) using proton-proton collision data. It also covers a wider range of
energy than the analysis at $\sqrt{s} = 7\TeV$~\cite{Khachatryan:2011dx, Khachatryan:2014ika}.
Data are compared with theoretical predictions from event generators \PYTHIAeight, \HERWIGpp, and \MGvATNLO{}+\PYTHIAeight. The \PYTHIAeight generator describes the flow of energy in the transverse plane well as seen
in the \taup and \rhoPerp distributions. \HERWIGpp and \MGvATNLO show good agreement with the data for all
the four event shape variables and are better than \PYTHIAeight in predicting \rhoTot and \bt. A
study of the effects of initial state radiation, final state radiation, and multiple parton interactions
in \PYTHIAeight is also presented.

\begin{acknowledgments}
We congratulate our colleagues in the CERN accelerator departments for the excellent performance of the LHC and thank the technical and administrative staffs at CERN and at other CMS institutes for their contributions to the success of the CMS effort. In addition, we gratefully acknowledge the computing centres and personnel of the Worldwide LHC Computing Grid for delivering so effectively the computing infrastructure essential to our analyses. Finally, we acknowledge the enduring support for the construction and operation of the LHC and the CMS detector provided by the following funding agencies: BMBWF and FWF (Austria); FNRS and FWO (Belgium); CNPq, CAPES, FAPERJ, FAPERGS, and FAPESP (Brazil); MES (Bulgaria); CERN; CAS, MoST, and NSFC (China); COLCIENCIAS (Colombia); MSES and CSF (Croatia); RPF (Cyprus); SENESCYT (Ecuador); MoER, ERC IUT, and ERDF (Estonia); Academy of Finland, MEC, and HIP (Finland); CEA and CNRS/IN2P3 (France); BMBF, DFG, and HGF (Germany); GSRT (Greece); NKFIA (Hungary); DAE and DST (India); IPM (Iran); SFI (Ireland); INFN (Italy); MSIP and NRF (Republic of Korea); MES (Latvia); LAS (Lithuania); MOE and UM (Malaysia); BUAP, CINVESTAV, CONACYT, LNS, SEP, and UASLP-FAI (Mexico); MOS (Montenegro); MBIE (New Zealand); PAEC (Pakistan); MSHE and NSC (Poland); FCT (Portugal); JINR (Dubna); MON, RosAtom, RAS, RFBR, and NRC KI (Russia); MESTD (Serbia); SEIDI, CPAN, PCTI, and FEDER (Spain); MOSTR (Sri Lanka); Swiss Funding Agencies (Switzerland); MST (Taipei); ThEPCenter, IPST, STAR, and NSTDA (Thailand); TUBITAK and TAEK (Turkey); NASU and SFFR (Ukraine); STFC (United Kingdom); DOE and NSF (USA).

\hyphenation{Rachada-pisek} Individuals have received support from the Marie-Curie programme and the European Research Council and Horizon 2020 Grant, contract No. 675440 (European Union); the Leventis Foundation; the A. P. Sloan Foundation; the Alexander von Humboldt Foundation; the Belgian Federal Science Policy Office; the Fonds pour la Formation \`a la Recherche dans l'Industrie et dans l'Agriculture (FRIA-Belgium); the Agentschap voor Innovatie door Wetenschap en Technologie (IWT-Belgium); the F.R.S.-FNRS and FWO (Belgium) under the ``Excellence of Science - EOS" - be.h project n. 30820817; the Ministry of Education, Youth and Sports (MEYS) of the Czech Republic; the Lend\"ulet (``Momentum") Programme and the J\'anos Bolyai Research Scholarship of the Hungarian Academy of Sciences, the New National Excellence Program \'UNKP, the NKFIA research grants 123842, 123959, 124845, 124850 and 125105 (Hungary); the Council of Science and Industrial Research, India; the HOMING PLUS programme of the Foundation for Polish Science, cofinanced from European Union, Regional Development Fund, the Mobility Plus programme of the Ministry of Science and Higher Education, the National Science Center (Poland), contracts Harmonia 2014/14/M/ST2/00428, Opus 2014/13/B/ST2/02543, 2014/15/B/ST2/03998, and 2015/19/B/ST2/02861, Sonata-bis 2012/07/E/ST2/01406; the National Priorities Research Program by Qatar National Research Fund; the Programa Estatal de Fomento de la Investigaci{\'o}n Cient{\'i}fica y T{\'e}cnica de Excelencia Mar\'{\i}a de Maeztu, grant MDM-2015-0509 and the Programa Severo Ochoa del Principado de Asturias; the Thalis and Aristeia programmes cofinanced by EU-ESF and the Greek NSRF; the Rachadapisek Sompot Fund for Postdoctoral Fellowship, Chulalongkorn University and the Chulalongkorn Academic into Its 2nd Century Project Advancement Project (Thailand); the Welch Foundation, contract C-1845; and the Weston Havens Foundation (USA).
\end{acknowledgments}

\bibliography{auto_generated}

\providecommand{\href}[2]{#2}\begingroup\raggedright\begin{thebibliography}{10}%
\makeatletter
\providecommand{\hrefCMSnoop }[0]{\@secondoftwo}%
\makeatother
\providecommand{\doi}{\texttt{doi:}\begingroup \urlstyle{tt}\Url}

\bibitem{Banfi:2004nk}
\hrefCMSnoop {}{A.~Banfi, G.~P. Salam, and G.~Zanderighi, ``{Resummed event
  shapes at hadron-hadron colliders}'',} \textit{ JHEP} \textbf{ 08} (2004)
  062,
  \href{http://dx.doi.org/10.1088/1126-6708/2004/08/062}{\doi{10.1088/1126-6708/2004/08/062}},
\href{http://www.arXiv.org/abs/hep-ph/0407287}{\texttt{arXiv:hep-ph/0407287}}.
%%CITATION = HEP-PH/0407287;%%.

\bibitem{Banfi:2010xy}
\hrefCMSnoop {}{A.~Banfi, G.~P. Salam, and G.~Zanderighi, ``{Phenomenology of
  event shapes at hadron colliders}'',} \textit{ JHEP} \textbf{ 06} (2010) 038,
  \href{http://dx.doi.org/10.1007/JHEP06(2010)038}{\doi{10.1007/JHEP06(2010)038}},
\href{http://www.arXiv.org/abs/1001.4082}{\texttt{arXiv:1001.4082}}.
%%CITATION = ARXIV:1001.4082;%%.

\bibitem{Dasgupta:2003iq}
\hrefCMSnoop {}{M.~Dasgupta and G.~P. Salam, ``{Event shapes in
  $\mathrm{e^{+}e^{-}}$ annihilation and deep inelastic scattering}'',}
  \textit{ J. Phys. G} \textbf{ 30} (2004) R143,
  \href{http://dx.doi.org/10.1088/0954-3899/30/5/R01}{\doi{10.1088/0954-3899/30/5/R01}},
\href{http://www.arXiv.org/abs/hep-ph/0312283}{\texttt{arXiv:hep-ph/0312283}}.
%%CITATION = HEP-PH/0312283;%%.

\bibitem{Rubin:2010xp}
\hrefCMSnoop {}{M.~Rubin, G.~P. Salam, and S.~Sapeta, ``{Giant QCD K-factors
  beyond NLO}'',} \textit{ JHEP} \textbf{ 09} (2010) 084,
  \href{http://dx.doi.org/10.1007/JHEP09(2010)084}{\doi{10.1007/JHEP09(2010)084}},
\href{http://www.arXiv.org/abs/1006.2144}{\texttt{arXiv:1006.2144}}.
%%CITATION = ARXIV:1006.2144;%%.

\bibitem{Jones:2003yv}
R.~W.~L. Jones\hrefCMSnoop {}{ {et~al.}, ``{Theoretical uncertainties on
  $\alpha{_s}$ from event shape variables in $\mathrm{e^{+}e^{-}}$
  annihilations}'',} \textit{ JHEP} \textbf{ 12} (2003) 007,
  \href{http://dx.doi.org/10.1088/1126-6708/2003/12/007}{\doi{10.1088/1126-6708/2003/12/007}},
\href{http://www.arXiv.org/abs/hep-ph/0312016}{\texttt{arXiv:hep-ph/0312016}}.
%%CITATION = HEP-PH/0312016;%%.

\bibitem{Dissertori:2007xa}
G.~Dissertori\hrefCMSnoop {}{ {et~al.}, ``{First determination of the strong
  coupling constant using NNLO predictions for hadronic event shapes in
  $\mathrm{e^{+} e^{-}}$ annihilations}'',} \textit{ JHEP} \textbf{ 02} (2008)
  040,
  \href{http://dx.doi.org/10.1088/1126-6708/2008/02/040}{\doi{10.1088/1126-6708/2008/02/040}},
\href{http://www.arXiv.org/abs/0712.0327}{\texttt{arXiv:0712.0327}}.
%%CITATION = ARXIV:0712.0327;%%.

\bibitem{Dissertori:2009ik}
G.~Dissertori\hrefCMSnoop {}{ {et~al.}, ``{Determination of the strong coupling
  constant using matched NNLO+NLLA predictions for hadronic event shapes in
  $\mathrm{e^{+}e^{-}}$ annihilations}'',} \textit{ JHEP} \textbf{ 08} (2009)
  036,
  \href{http://dx.doi.org/10.1088/1126-6708/2009/08/036}{\doi{10.1088/1126-6708/2009/08/036}},
\href{http://www.arXiv.org/abs/0906.3436}{\texttt{arXiv:0906.3436}}.
%%CITATION = ARXIV:0906.3436;%%.

\bibitem{Chatterjee:2012qt}
\hrefCMSnoop {}{R.~M. Chatterjee, M.~Guchait, and D.~Sengupta, ``{Probing
  supersymmetry using event shape variables at 8 TeV LHC}'',} \textit{ Phys.
  Rev. D} \textbf{ 86} (2012) 075014,
  \href{http://dx.doi.org/10.1103/PhysRevD.86.075014}{\doi{10.1103/PhysRevD.86.075014}},
\href{http://www.arXiv.org/abs/1206.5770}{\texttt{arXiv:1206.5770}}.
%%CITATION = ARXIV:1206.5770;%%.

\bibitem{Datta:2011vg}
\hrefCMSnoop {}{A.~Datta, A.~Datta, and S.~Poddar, ``{Enriching the exploration
  of the mUED model with event shape variables at the CERN LHC}'',} \textit{
  Phys. Lett. B} \textbf{ 712} (2012) 219,
  \href{http://dx.doi.org/10.1016/j.physletb.2012.03.012}{\doi{10.1016/j.physletb.2012.03.012}},
\href{http://www.arXiv.org/abs/1111.2912}{\texttt{arXiv:1111.2912}}.
%%CITATION = ARXIV:1111.2912;%%.

\bibitem{Konar:2005bd}
\hrefCMSnoop {}{P.~Konar and P.~Roy, ``{Event shape discrimination of
  supersymmetry from large extra dimensions at a linear collider}'',} \textit{
  Phys. Lett. B} \textbf{ 634} (2006) 295,
  \href{http://dx.doi.org/10.1016/j.physletb.2006.01.056}{\doi{10.1016/j.physletb.2006.01.056}},
\href{http://www.arXiv.org/abs/hep-ph/0509161}{\texttt{arXiv:hep-ph/0509161}}.
%%CITATION = HEP-PH/0509161;%%.

\bibitem{Heister:2003aj}
\hrefCMSnoop {}{{ALEPH} Collaboration, ``{Studies of QCD at
  $\mathrm{e^{+}e^{-}}$ centre-of-mass energies between 91 GeV and 209 GeV}'',}
  \textit{ Eur. Phys. J. C} \textbf{ 35} (2004) 457,
\href{http://dx.doi.org/10.1140/epjc/s2004-01891-4}{\doi{10.1140/epjc/s2004-01891-4}}.
%%CITATION = EPHJA,C35,457;%%.

\bibitem{Abreu:1996na}
\hrefCMSnoop {}{{DELPHI} Collaboration, ``{Tuning and test of fragmentation
  models based on identified particles and precision event shape data}'',}
  \textit{ Z. Phys. C} \textbf{ 73} (1996) 11,
\href{http://dx.doi.org/10.1007/s002880050295}{\doi{10.1007/s002880050295}}.
%%CITATION = ZEPYA,C73,11;%%.

\bibitem{Acciarri:1997dn}
\hrefCMSnoop {}{{L3} Collaboration, ``{Study of hadronic events and
  measurements of $\alpha_{s}$ between 30 GeV and 91 GeV}'',} \textit{ Phys.
  Lett. B} \textbf{ 411} (1997) 339,
\href{http://dx.doi.org/10.1016/S0370-2693(97)01000-9}{\doi{10.1016/S0370-2693(97)01000-9}}.
%%CITATION = PHLTA,B411,339;%%.

\bibitem{Achard:2004sv}
\hrefCMSnoop {}{{L3} Collaboration, ``{Studies of hadronic event structure in
  $\mathrm{e^{+}e^{-}}$ annihilation from 30 GeV to 209 GeV with the L3
  detector}'',} \textit{ Phys. Rept.} \textbf{ 399} (2004) 71,
  \href{http://dx.doi.org/10.1016/j.physrep.2004.07.002}{\doi{10.1016/j.physrep.2004.07.002}},
\href{http://www.arXiv.org/abs/hep-ex/0406049}{\texttt{arXiv:hep-ex/0406049}}.
%%CITATION = HEP-EX/0406049;%%.

\bibitem{Acton:1993zh}
\hrefCMSnoop {}{{OPAL} Collaboration, ``{A determination of $\alpha_{s}$ (M
  (Z0)) at LEP using resummed QCD calculations}'',} \textit{ Z. Phys. C}
  \textbf{ 59} (1993) 1,
\href{http://dx.doi.org/10.1007/BF01555834}{\doi{10.1007/BF01555834}}.
%%CITATION = ZEPYA,C59,1;%%.

\bibitem{Aktas:2005tz}
\hrefCMSnoop {}{{H1} Collaboration, ``{Measurement of event shape variables in
  deep-inelastic scattering at HERA}'',} \textit{ Eur. Phys. J. C} \textbf{ 46}
  (2006) 343,
  \href{http://dx.doi.org/10.1140/epjc/s2006-02493-x}{\doi{10.1140/epjc/s2006-02493-x}},
\href{http://www.arXiv.org/abs/hep-ex/0512014}{\texttt{arXiv:hep-ex/0512014}}.
%%CITATION = HEP-EX/0512014;%%.

\bibitem{Aaltonen:2011et}
\hrefCMSnoop {}{{CDF} Collaboration, ``{Measurement of event ehapes in
  proton-antiproton collisions at center-of-mass energy 1.96 TeV}'',} \textit{
  Phys. Rev. D} \textbf{ 83} (2011) 112007,
  \href{http://dx.doi.org/10.1103/PhysRevD.83.112007}{\doi{10.1103/PhysRevD.83.112007}},
\href{http://www.arXiv.org/abs/1103.5143}{\texttt{arXiv:1103.5143}}.
%%CITATION = ARXIV:1103.5143;%%.

\bibitem{Sjostrand:2006za}
\hrefCMSnoop {}{T.~Sj{\"o}strand, S.~Mrenna, and P.~Z. Skands, ``{PYTHIA 6.4
  Physics and manual}'',} \textit{ JHEP} \textbf{ 05} (2006) 026,
  \href{http://dx.doi.org/10.1088/1126-6708/2006/05/026}{\doi{10.1088/1126-6708/2006/05/026}},
\href{http://www.arXiv.org/abs/hep-ph/0603175}{\texttt{arXiv:hep-ph/0603175}}.
%%CITATION = HEP-PH/0603175;%%.

\bibitem{Khachatryan:2011dx}
\hrefCMSnoop {}{{CMS Collaboration}, ``{First measurement of hadronic event
  shapes in pp collisions at $\sqrt{s} = 7$ TeV}'',} \textit{ Phys. Lett. B}
  \textbf{ 699} (2011) 48,
  \href{http://dx.doi.org/10.1016/j.physletb.2011.03.060}{\doi{10.1016/j.physletb.2011.03.060}},
\href{http://www.arXiv.org/abs/1102.0068}{\texttt{arXiv:1102.0068}}.
%%CITATION = ARXIV:1102.0068;%%.

\bibitem{Chatrchyan:2013tna}
\hrefCMSnoop {}{{CMS Collaboration}, ``{Event shapes and azimuthal correlations
  in Z+jets events in pp collisions at $\sqrt{s} = 7$ TeV}'',} \textit{ Phys.
  Lett. B} \textbf{ 722} (2013) 238,
  \href{http://dx.doi.org/10.1016/j.physletb.2013.04.025}{\doi{10.1016/j.physletb.2013.04.025}},
\href{http://www.arXiv.org/abs/1301.1646}{\texttt{arXiv:1301.1646}}.
%%CITATION = ARXIV:1301.1646;%%.

\bibitem{Chatrchyan:2013ala}
\hrefCMSnoop {}{{CMS Collaboration}, ``{Jet and underlying event properties as
  a function of charged-particle multiplicity in proton–proton collisions at
  $\sqrt{s} = 7$ TeV}'',} \textit{ Eur. Phys. J. C} \textbf{ 73} (2013) 2674,
  \href{http://dx.doi.org/10.1140/epjc/s10052-013-2674-5}{\doi{10.1140/epjc/s10052-013-2674-5}},
\href{http://www.arXiv.org/abs/1310.4554}{\texttt{arXiv:1310.4554}}.
%%CITATION = ARXIV:1310.4554;%%.

\bibitem{Khachatryan:2014ika}
\hrefCMSnoop {}{{CMS Collaboration}, ``{Study of hadronic event-shape variables
  in multijet final states in pp collisions at $\sqrt{s} = 7$ TeV}'',} \textit{
  JHEP} \textbf{ {\bf 10}} (2014) 087,
  \href{http://dx.doi.org/10.1007/JHEP10(2014)087}{\doi{10.1007/JHEP10(2014)087}},
\href{http://www.arXiv.org/abs/1407.2856}{\texttt{arXiv:1407.2856}}.
%%CITATION = ARXIV:1407.2856;%%.

\bibitem{Aad:2012np}
\hrefCMSnoop {}{{ATLAS Collaboration}, ``{Measurement of event shapes at large
  momentum transfer with the ATLAS detector in $pp$ collisions at $\sqrt{s} =
  7$ TeV}'',} \textit{ Eur. Phys. J. C} \textbf{ 72} (2012) 2211,
  \href{http://dx.doi.org/10.1140/epjc/s10052-012-2211-y}{\doi{10.1140/epjc/s10052-012-2211-y}},
\href{http://www.arXiv.org/abs/1206.2135}{\texttt{arXiv:1206.2135}}.
%%CITATION = ARXIV:1206.2135;%%.

\bibitem{Aad:2012fza}
\hrefCMSnoop {}{{ATLAS Collaboration}, ``{Measurement of charged-particle event
  shape variables in $\sqrt{s} = 7$ TeV proton-proton interactions with the
  ATLAS detector}'',} \textit{ Phys. Rev. D} \textbf{ 88} (2013) 032004,
  \href{http://dx.doi.org/10.1103/PhysRevD.88.032004}{\doi{10.1103/PhysRevD.88.032004}},
\href{http://www.arXiv.org/abs/1207.6915}{\texttt{arXiv:1207.6915}}.
%%CITATION = ARXIV:1207.6915;%%.

\bibitem{Aad:2016ria}
\hrefCMSnoop {}{{ATLAS Collaboration}, ``{Measurement of event-shape
  observables in $Z \rightarrow \ell^{+} \ell^{-}$ events in $pp$ collisions at
  $\sqrt{s} = 7$ TeV with the ATLAS detector at the LHC}'',} \textit{ Eur.
  Phys. J. C} \textbf{ 76} (2016) 375,
  \href{http://dx.doi.org/10.1140/epjc/s10052-016-4176-8}{\doi{10.1140/epjc/s10052-016-4176-8}},
\href{http://www.arXiv.org/abs/1602.08980}{\texttt{arXiv:1602.08980}}.
%%CITATION = ARXIV:1602.08980;%%.

\bibitem{Abelev:2012sk}
\hrefCMSnoop {}{{ALICE Collaboration}, ``{Transverse sphericity of primary
  charged particles in minimum bias proton-proton collisions at $\sqrt{s} =
  0.9$, 2.76 and 7 TeV}'',} \textit{ Eur. Phys. J. C} \textbf{ 72} (2012) 2124,
  \href{http://dx.doi.org/10.1140/epjc/s10052-012-2124-9}{\doi{10.1140/epjc/s10052-012-2124-9}},
\href{http://www.arXiv.org/abs/1205.3963}{\texttt{arXiv:1205.3963}}.
%%CITATION = ARXIV:1205.3963;%%.

\bibitem{Sjostrand:2014zea}
T.~Sj{\"o}strand\hrefCMSnoop {}{ {et~al.}, ``{An Introduction to PYTHIA
  8.2}'',} \textit{ Comput. Phys. Commun.} \textbf{ 191} (2015) 159,
  \href{http://dx.doi.org/10.1016/j.cpc.2015.01.024}{\doi{10.1016/j.cpc.2015.01.024}},
\href{http://www.arXiv.org/abs/1410.3012}{\texttt{arXiv:1410.3012}}.
%%CITATION = ARXIV:1410.3012;%%.

\bibitem{Alwall:2014hca}
J.~Alwall\hrefCMSnoop {}{ {et~al.}, ``{The automated computation of tree-level
  and next-to-leading order differential cross sections, and their matching to
  parton shower simulations}'',} \textit{ JHEP} \textbf{ 07} (2014) 079,
  \href{http://dx.doi.org/10.1007/JHEP07(2014)079}{\doi{10.1007/JHEP07(2014)079}},
\href{http://www.arXiv.org/abs/1405.0301}{\texttt{arXiv:1405.0301}}.
%%CITATION = ARXIV:1405.0301;%%.

\bibitem{Bellm:2015jjp}
\hrefCMSnoop {}{J.~Bellm {et~al.}, ``{Herwig 7.0/Herwig++ 3.0 release note}'',}
  \textit{ Eur. Phys. J. C} \textbf{ 76} (2016) 196,
  \href{http://dx.doi.org/10.1140/epjc/s10052-016-4018-8}{\doi{10.1140/epjc/s10052-016-4018-8}},
\href{http://www.arXiv.org/abs/1512.01178}{\texttt{arXiv:1512.01178}}.
%%CITATION = ARXIV:1512.01178;%%.

\bibitem{Chatrchyan:2008zzk}
\hrefCMSnoop {}{{CMS Collaboration}, ``The {CMS} experiment at the {CERN}
  {LHC}'',} \textit{ JINST} \textbf{ 3} (2008) S08004,
  \href{http://dx.doi.org/10.1088/1748-0221/3/08/S08004}{\doi{10.1088/1748-0221/3/08/S08004}}.

\bibitem{Sirunyan:2017ulk}
\hrefCMSnoop {}{{CMS Collaboration}, ``{Particle-flow reconstruction and global
  event description with the CMS detector}'',} \textit{ JINST} \textbf{ 12}
  (2017) P10003,
  \href{http://dx.doi.org/10.1088/1748-0221/12/10/P10003}{\doi{10.1088/1748-0221/12/10/P10003}},
\href{http://www.arXiv.org/abs/1706.04965}{\texttt{arXiv:1706.04965}}.
%%CITATION = ARXIV:1706.04965;%%.

\bibitem{Cacciari:2008gp}
\hrefCMSnoop {}{M.~Cacciari, G.~P. Salam, and G.~Soyez, ``{The anti-\kt jet
  clustering algorithm}'',} \textit{ JHEP} \textbf{ 04} (2008) 063,
  \href{http://dx.doi.org/10.1088/1126-6708/2008/04/063}{\doi{10.1088/1126-6708/2008/04/063}},
\href{http://www.arXiv.org/abs/0802.1189}{\texttt{arXiv:0802.1189}}.
%%CITATION = ARXIV:0802.1189;%%.

\bibitem{Cacciari:2011ma}
\hrefCMSnoop {}{M.~Cacciari, G.~P. Salam, and G.~Soyez, ``{FastJet User
  Manual}'',} \textit{ Eur. Phys. J. C} \textbf{ 72} (2012) 1896,
  \href{http://dx.doi.org/10.1140/epjc/s10052-012-1896-2}{\doi{10.1140/epjc/s10052-012-1896-2}},
\href{http://www.arXiv.org/abs/1111.6097}{\texttt{arXiv:1111.6097}}.
%%CITATION = ARXIV:1111.6097;%%.

\bibitem{Chatrchyan:2011ds}
\hrefCMSnoop {}{{CMS Collaboration}, ``{Determination of jet energy calibration
  and transverse momentum resolution in CMS}'',} \textit{ JINST} \textbf{ 6}
  (2011) P11002,
  \href{http://dx.doi.org/10.1088/1748-0221/6/11/P11002}{\doi{10.1088/1748-0221/6/11/P11002}},
\href{http://www.arXiv.org/abs/1107.4277}{\texttt{arXiv:1107.4277}}.
%%CITATION = ARXIV:1107.4277;%%.

\bibitem{Ball:2014uwa}
\hrefCMSnoop {}{{NNPDF} Collaboration, ``{Parton distributions for the LHC Run
  II}'',} \textit{ JHEP} \textbf{ 04} (2015) 040,
  \href{http://dx.doi.org/10.1007/JHEP04(2015)040}{\doi{10.1007/JHEP04(2015)040}},
\href{http://www.arXiv.org/abs/1410.8849}{\texttt{arXiv:1410.8849}}.
%%CITATION = ARXIV:1410.8849;%%.

\bibitem{Khachatryan:2015pea}
\hrefCMSnoop {}{{CMS Collaboration}, ``{Event generator tunes obtained from
  underlying event and multiparton scattering measurements}'',} \textit{ Eur.
  Phys. J. C} \textbf{ 76} (2016) 155,
  \href{http://dx.doi.org/10.1140/epjc/s10052-016-3988-x}{\doi{10.1140/epjc/s10052-016-3988-x}},
\href{http://www.arXiv.org/abs/1512.00815}{\texttt{arXiv:1512.00815}}.
%%CITATION = ARXIV:1512.00815;%%.

\bibitem{Skands:2014pea}
\hrefCMSnoop {}{P.~Skands, S.~Carrazza, and J.~Rojo, ``{Tuning PYTHIA 8.1: the
  Monash 2013 Tune}'',} \textit{ Eur. Phys. J. C} \textbf{ 74} (2014) 3024,
  \href{http://dx.doi.org/10.1140/epjc/s10052-014-3024-y}{\doi{10.1140/epjc/s10052-014-3024-y}},
\href{http://www.arXiv.org/abs/1404.5630}{\texttt{arXiv:1404.5630}}.
%%CITATION = ARXIV:1404.5630;%%.

\bibitem{Alwall:2007fs}
\hrefCMSnoop {}{J.~Alwall {et~al.}, ``{Comparative study of various algorithms
  for the merging of parton showers and matrix elements in hadronic
  collisions}'',} \textit{ Eur. Phys. J. C} \textbf{ 53} (2008) 473,
  \href{http://dx.doi.org/10.1140/epjc/s10052-007-0490-5}{\doi{10.1140/epjc/s10052-007-0490-5}},
\href{http://www.arXiv.org/abs/0706.2569}{\texttt{arXiv:0706.2569}}.
%%CITATION = ARXIV:0706.2569;%%.

\bibitem{Adye:2011gm}
\hrefCMSnoop {}{T.~Adye, ``{Unfolding algorithms and tests using RooUnfold}'',}
  in \textit{ {Proceedings, PHYSTAT 2011 Workshop on Statistical Issues Related
  to Discovery Claims in Search Experiments and Unfolding, CERN, Geneva,
  Switzerland 17-20 January 2011}}, p.~313, CERN.
\newblock 2011.
\newblock \href{http://www.arXiv.org/abs/1105.1160}{\texttt{arXiv:1105.1160}}.
\newblock
\href{http://dx.doi.org/10.5170/CERN-2011-006.313}{\doi{10.5170/CERN-2011-006.313}}.
%%CITATION = ARXIV:1105.1160;%%.

\bibitem{DAgostini:1994fjx}
\hrefCMSnoop {}{G.~D'Agostini, ``{A multidimensional unfolding method based on
  Bayes' theorem}'',} \textit{ Nucl. Instrum. Meth. A} \textbf{ 362} (1995)
  487,
\href{http://dx.doi.org/10.1016/0168-9002(95)00274-X}{\doi{10.1016/0168-9002(95)00274-X}}.
%%CITATION = NUIMA,A362,487;%%.

\bibitem{Hocker:1995kb}
\hrefCMSnoop {}{A.~Hocker and V.~Kartvelishvili, ``{SVD approach to data
  unfolding}'',} \textit{ Nucl. Instrum. Meth. A} \textbf{ 372} (1996) 469,
  \href{http://dx.doi.org/10.1016/0168-9002(95)01478-0}{\doi{10.1016/0168-9002(95)01478-0}},
\href{http://www.arXiv.org/abs/hep-ph/9509307}{\texttt{arXiv:hep-ph/9509307}}.
%%CITATION = HEP-PH/9509307;%%.

\bibitem{Khachatryan:2016kdb}
\hrefCMSnoop {}{{CMS Collaboration}, ``{Jet energy scale and resolution in the
  CMS experiment in pp collisions at 8 TeV}'',} \textit{ JINST} \textbf{ 12}
  (2017) P02014,
  \href{http://dx.doi.org/10.1088/1748-0221/12/02/P02014}{\doi{10.1088/1748-0221/12/02/P02014}},
\href{http://www.arXiv.org/abs/1607.03663}{\texttt{arXiv:1607.03663}}.
%%CITATION = ARXIV:1607.03663;%%.

\end{thebibliography}\endgroup

\cleardoublepage \appendix\section{The CMS Collaboration \label{app:collab}}\begin{sloppypar}\hyphenpenalty=5000\widowpenalty=500\clubpenalty=5000\vskip\cmsinstskip
\textbf{Yerevan Physics Institute, Yerevan, Armenia}\\*[0pt]
A.M.~Sirunyan, A.~Tumasyan
\vskip\cmsinstskip
\textbf{Institut f\"{u}r Hochenergiephysik, Wien, Austria}\\*[0pt]
W.~Adam, F.~Ambrogi, E.~Asilar, T.~Bergauer, J.~Brandstetter, M.~Dragicevic, J.~Er\"{o}, A.~Escalante~Del~Valle, M.~Flechl, R.~Fr\"{u}hwirth\cmsAuthorMark{1}, V.M.~Ghete, J.~Hrubec, M.~Jeitler\cmsAuthorMark{1}, N.~Krammer, I.~Kr\"{a}tschmer, D.~Liko, T.~Madlener, I.~Mikulec, N.~Rad, H.~Rohringer, J.~Schieck\cmsAuthorMark{1}, R.~Sch\"{o}fbeck, M.~Spanring, D.~Spitzbart, A.~Taurok, W.~Waltenberger, J.~Wittmann, C.-E.~Wulz\cmsAuthorMark{1}, M.~Zarucki
% ...
\vskip\cmsinstskip
% ...
1:  Also at Vienna University of Technology, Vienna, Austria\\
% ...
\end{sloppypar}
\end{document}